%% file: JAIF_2023_V7.tex
\documentclass[a4paper,twocolumn, 10pt]{IEEEtran} 

\newlength{\figureheight} 
\newlength{\figurewidth} 

\newcommand{\exportFigures}{true}
\newcommand{\arxiv}{true} 

\input{./definitions.tex}
\input{./pgfplot.tex}

\graphicspath{{./plots/}}
\newcommand{\tikzfolder}{./compiledPlots/}

\tcbset{colback=red!10,colframe=red!10,arc=0.5mm}

\newcommand{\legendref}[1]{\tikzexternaldisable\pgfplotslegendfromname{#1}\tikzexternaldisable}
\newcommand{\sref}[1]{{\protect\subref{#1}}}

\input{\folder/acronyms.tex}

\IEEEoverridecommandlockouts

\begin{document}
\frenchspacing

\title{\LARGE 
Multipath-based SLAM for Non-Ideal Reflective Surfaces Exploiting Multiple-Measurement Data Association}
\author{
Lukas Wielandner, 
Alexander Venus,
Thomas Wilding, and
Erik Leitinger
\thanks{L.\ Wielandner, A.\ Venus, T.\ Wilding, and E.\ Leitinger are with the Signal Processing and Speech Communication Laboratory, Graz University of Technology, Graz, Austria (e-mail: (lukas.wielandner, a.venus, thomas.wilding, erik.leitinger)@tugraz.at). Submission: 28.03.2023, revision: 24.08.2023, associate editor: Florian Meyer. This work was supported in part by the Christian Doppler Research Association; the Austrian Federal Ministry for Digital and Economic Affairs; the National Foundation for Research, Technology, and Development;  the European Union's Horizon 2020 research and innovation programme under grant agreement No 101013425 (Project ``REINDEER'').}
\vspace*{-4mm}}
\maketitle

\begin{abstract}
  \input{\folder/abstract.tex}
\end{abstract}  



\acresetall

\vspace*{-3mm}

\section{Introduction}\label{sec:introduction}

\input{\folder/introduction.tex}
\input{\folder/geom_model.tex}

\input{\folder/signal_model.tex}
\input{\folder/system_model.tex}

\input{\folder/SPA.tex}

\input{\folder/simulation_results.tex}

\vspace*{-2mm}
\section{Conclusions}\label{sec:conclusion}
\input{\folder/conclusions.tex}

\appendices
\input{\folder/appendix_DA.tex}


\bibliographystyle{IEEEtran}
\bibliography{IEEEabrv,references}
\input{biography}

\end{document}

%% file: definitions.tex
\usepackage{amsthm,amsmath,array}   
\usepackage{amssymb}
\usepackage{graphicx}
\usepackage{color}
\usepackage{xcolor}
\usepackage{cite}
\usepackage{bm}
\usepackage{epsfig,psfrag}
\usepackage{tabularx}
\usepackage{acronym}
\usepackage[yyyymmdd,hhmmss]{datetime}
\usepackage{bbm}
\usepackage{notation}
\usepackage{mathrsfs} 
\usepackage{bardiag}
\usepackage{tabularx}
\usepackage{multirow}
\usepackage{colortbl,pgfplotstable}
\usepackage{dsfont}
\usepackage{times}
\usepackage{rotating}
\usepackage{enumitem}

\usepackage[ruled,vlined,linesnumbered]{algorithm2e}
\SetKwRepeat{Do}{do}{while}

\usepackage[caption=false,font=footnotesize]{subfig}

\usepackage{pgfplots}
\usepackage{tikzscale}
\usepackage{tikz}

\usetikzlibrary{shapes.misc}
\usetikzlibrary{arrows}
\usetikzlibrary{arrows.meta}
\usetikzlibrary{angles}
\usetikzlibrary{quotes}
\usetikzlibrary{fit}
\usetikzlibrary{graphs}
\usetikzlibrary{patterns}
\usetikzlibrary{petri}
\usetikzlibrary{plotmarks}
\usetikzlibrary{positioning}
\usetikzlibrary{decorations.pathmorphing}
\usetikzlibrary{decorations.markings}
\usetikzlibrary{decorations.pathreplacing}
\usetikzlibrary{decorations.shapes}
\usetikzlibrary{decorations.text}
\usetikzlibrary{backgrounds}
\usetikzlibrary{calc}
\usetikzlibrary{spy}
\usetikzlibrary{shapes,arrows}

\usepackage{tikzscale}

\usepackage{pgfplots}
\pgfplotsset{compat=newest}
\usepgfplotslibrary{colormaps}
\usepgfplotslibrary{patchplots}
\usepgfplotslibrary{units}

\usepackage[most,theorems]{tcolorbox}
\tcbuselibrary{most}
\tcbset{every box/.style={enhanced,breakable},shield externalize}

\definecolor{darkgreen}{rgb}{0.0, 0.5, 0.0}
\definecolor{magenta_}{rgb}{1.0, 0, 1.0}

\definecolor{darkgrey}{rgb}{0.3, 0.3, 0.3}

\newcommand{\be}{\begin{equation}}
\newcommand{\ee}{\end{equation}}
\newcommand{\ist}{\hspace*{.3mm}}
\newcommand{\rmv}{\hspace*{-.3mm}}
\newcommand{\iist}{\hspace*{1mm}}
\newcommand{\rrmv}{\hspace*{-1mm}}
\newcommand{\rreq}{\hspace*{-1mm}= \hspace*{-1mm}}

\newcommand{\nn}{\nonumber}

\makeatletter
\newcommand*\bigcdot{\mathpalette\bigcdot@{.5}}
\newcommand*\bigcdot@[2]{\mathbin{\vcenter{\hbox{\scalebox{#2}{$\m@th#1\bullet$}}}}}
\makeatother

\allowdisplaybreaks
\newcommand{\ticked}{$\text{\rlap{$\checkmark$}}\square$}
\newcommand{\unticked}{{$\square$}}
\newcommand{\tick}[1]{\ifthenelse{#1=1}{\ticked}{\unticked}}

\newcolumntype{L}[1]{>{\raggedright\arraybackslash}p{#1}}
\newcolumntype{C}[1]{>{\centering\arraybackslash}p{#1}}
\newcolumntype{R}[1]{>{\raggedleft\arraybackslash}p{#1}}

\providecommand{\ist}{\hspace*{.3mm}}

\providecommand{\rmv}{\hspace*{-.3mm}}
\providecommand{\iist}{\hspace*{1mm}}
\providecommand{\rrmv}{\hspace*{-1mm}}
\providecommand{\nn}{\nonumber}

\newcommand{\T}{\text{T}}

\definecolor{colA}{rgb}{0,0,1}
\definecolor{colB}{rgb}{0.5,0.0,0.5}
\definecolor{colC}{rgb}{0,0.5,0}
\definecolor{colD}{rgb}{0.1333,0.5451,0.1333}
\definecolor{colE}{rgb}{1.00000,0.55,0.00000}
\definecolor{colF}{rgb}{1.00000,0.0,0.00000}
\definecolor{col_mag}{rgb}{1.00000,0.00000,1.00000}
\definecolor{col_cyan}{rgb}{0,1,1}
\definecolor{col_grey}{rgb}{0.4,0.4,0.4}
\definecolor{col_grey2}{rgb}{0.6,0.6,0.6}
\definecolor{lightgray}{rgb}{0.9,0.9,0.9}  
\definecolor{Gray}{rgb}{225, 225, 225} 
\definecolor{matlabBlue}{rgb}{0.00000,0.44700,0.74100}%
\definecolor{matlabOrange}{rgb}{0.85000,0.32500,0.09800}%
\definecolor{matlabYellow}{rgb}{0.92900,0.69400,0.12500}%
\definecolor{matlabLila}{rgb}{0.49400,0.18400,0.55600}%
\definecolor{matlabGreen}{rgb}{0.46600,0.67400,0.18800}%
\definecolor{col_PAgreen}{rgb}{0.00000,0.49800,0.00000}%

\def\linewidthA{0.5}

\def\mylinewidth{0.7}
\def\mylinewidth2{1}

\def\figureW{6cm}
\def\figureW2{5cm}

\def\marksizeA{1.5}

\pgfplotsset{styleA/.style={color=colA, line width=\linewidthA, mark=\markA, mark size=\marksizeA, mark repeat=1, mark phase = 0, mark optiodvipsnamesns={solid}}}
\pgfplotsset{styleB/.style={color=colB, line width=\linewidthA, mark=\markB, mark size=\marksizeA, mark repeat=1, mark phase = 0, mark options={solid}}}
\pgfplotsset{styleC/.style={color=colC, line width=\linewidthA, mark=\markC, mark size=\marksizeA, mark repeat=1, mark phase = 0, mark options={solid}}}
\pgfplotsset{styleD/.style={color=col_cyan, line width=\linewidthA, mark=\markD, mark size=\marksizeA, mark repeat=1, mark phase = 0, mark options={solid}}}
\pgfplotsset{styleE/.style={color=colE, line width=\linewidthA, mark=\markE, mark size=\marksizeA, mark repeat=1, mark phase = 0, mark options={solid}}}
\pgfplotsset{styleF/.style={color=colF, line width=\linewidthA, mark=\markF, mark size=\marksizeA, mark repeat=1, mark phase = 0, mark options={solid}}}

\pgfplotsset{styleSLAM0p0/.style={mymarkfixednumber={\markA}{mark size=\marksizeA,solid}{12},color=matlabBlue, line width=\linewidthA, mark repeat=1, mark phase = 0, mark options={solid,matlabBlue}}}
\pgfplotsset{styleSLAM0p03/.style={mymarkfixednumber={\markB}{mark size=\marksizeA,solid}{6},color=matlabBlue,dashdotted, line width=\linewidthA, mark repeat=1, mark phase = 0, mark options={solid,matlabBlue}}}
\pgfplotsset{styleSLAM0p15/.style={mymarkfixednumber={\markC}{mark size=\marksizeA,solid}{10},color=matlabBlue,densely dashed, line width=\linewidthA, mark repeat=1, mark phase = 0, mark options={solid,matlabBlue}}}
\pgfplotsset{styleSLAM0p3/.style={mymarkfixednumber={\markD}{mark size=\marksizeA,solid}{8},color=matlabBlue,densely dotted, line width=\linewidthA, mark repeat=1, mark phase = 0, mark options={solid,matlabBlue}}}

\pgfplotsset{styleEFSLAM0p0/.style={mymarkfixednumber={\markAa}{mark size=\marksizeA,solid}{10},color=matlabOrange, line width=\linewidthA, mark repeat=1, mark phase = 0, mark options={solid,matlabOrange}}}
\pgfplotsset{styleEFSLAM0p03/.style={mymarkfixednumber={\markBb}{mark size=\marksizeA,solid}{13},color=matlabOrange,dashdotted, line width=\linewidthA, mark repeat=1, mark phase = 0, mark options={solid,matlabOrange}}}
\pgfplotsset{styleEFSLAM0p15/.style={mymarkfixednumber={\markCc}{mark size=\marksizeA,solid}{7},color=matlabOrange,densely dashed, line width=\linewidthA, mark repeat=1, mark phase = 0, mark options={solid,matlabOrange}}}
\pgfplotsset{styleEFSLAM0p3/.style={mymarkfixednumber={\markDd}{mark size=\marksizeA,solid}{8},color=matlabOrange,densely dotted, line width=\linewidthA, mark repeat=1, mark phase = 0, mark options={solid,matlabOrange}}}

\pgfplotsset{CDFstyleSLAM0p0/.style={mymarkfixednumber={\markA}{mark size=\marksizeA,solid}{3},color=matlabBlue, line width=\linewidthA, mark repeat=1, mark phase = 0, mark options={solid,matlabBlue}}}
\pgfplotsset{CDFstyleSLAM0p03/.style={mymarkfixednumber={\markB}{mark size=\marksizeA,solid}{100},color=matlabBlue,dashdotted, line width=\linewidthA, mark repeat=1, mark phase = 0, mark options={solid,matlabBlue}}}
\pgfplotsset{CDFstyleSLAM0p15/.style={mymarkfixednumber={\markC}{mark size=\marksizeA,solid}{40},color=matlabBlue,densely dashed, line width=\linewidthA, mark repeat=1, mark phase = 0, mark options={solid,matlabBlue}}}
\pgfplotsset{CDFstyleSLAM0p3/.style={mymarkfixednumber={\markD}{mark size=\marksizeA,solid}{40},color=matlabBlue,densely dotted, line width=\linewidthA, mark repeat=1, mark phase = 0, mark options={solid,matlabBlue}}}

\pgfplotsset{CDFstyleEFSLAM0p0/.style={mymarkfixednumber={\markAa}{mark size=\marksizeA,solid}{2},color=matlabOrange, line width=\linewidthA, mark repeat=1, mark phase = 0, mark options={solid,matlabOrange}}}
\pgfplotsset{CDFstyleEFSLAM0p03/.style={mymarkfixednumber={\markBb}{mark size=\marksizeA,solid}{4},color=matlabOrange,dashdotted, line width=\linewidthA, mark repeat=1, mark phase = 0, mark options={solid,matlabOrange}}}
\pgfplotsset{CDFstyleEFSLAM0p15/.style={mymarkfixednumber={\markCc}{mark size=\marksizeA,solid}{7},color=matlabOrange,densely dashed, line width=\linewidthA, mark repeat=1, mark phase = 0, mark options={solid,matlabOrange}}}
\pgfplotsset{CDFstyleEFSLAM0p3/.style={mymarkfixednumber={\markDd}{mark size=\marksizeA,solid}{15},color=matlabOrange,densely dotted, line width=\linewidthA, mark repeat=1, mark phase = 0, mark options={solid,matlabOrange}}}

\pgfdeclarelayer{backback1}
\pgfdeclarelayer{back1}
\pgfdeclarelayer{lay1}
\pgfdeclarelayer{backback2}
\pgfdeclarelayer{back2}
\pgfdeclarelayer{lay2}
\pgfsetlayers{backback2,back2,lay2,backback1,back1,lay1,main}

\definecolor{FGgreen}{RGB}{34,139,34}
\definecolor{FGblue}{RGB}{80,120,255}
\definecolor{FGred}{RGB}{255,110,110}

\colorlet{fgColor_node}{colA}
\colorlet{fgColorUp1}{colB}
\colorlet{fgColor_coop}{colB}
\colorlet{fgColorBG}{colC}
\colorlet{fgColor_facAlpha}{colE}
\colorlet{fgColorPost3}{colF}
\colorlet{fgColorComb}{colE}
\colorlet{fgColorBox}{colE}

\newcommand{\mumN}[2]{\ensuremath{\mu_\text{m}\big(\tilde{\V{x}}_{#2} , \overline{\V{x}}^{(j)}_{#1 , n}\big)}}
\newcommand{\mumL}[2]{\ensuremath{\mu_\text{m}\big(\tilde{\V{x}}_{#2} , \underline{\V{x}}^{(j)}_{#1 , n}\big)}}

\newcommand{\x}{\tilde{\V{x}}_n}
\newcommand{\xn}[1]{\tilde{\V{x}}_{#1}}

\newcommand{\al}{\underline{a}^{(j)}_{kl,n}}
\newcommand{\an}[2]{\overline{a}^{(j)}_{#1 #2,n}}

\newcommand{\Lxn}[1]{\underline{\V{x}}^{(j)}_{k,#1}}
\newcommand{\Nx}[1]{\overline{\V{x}}^{(j)}_{#1,n}}

\newcommand{\norm}[2]{\ensuremath{\lVert #1 \rVert^{#2}}}

\newcommand{\minus}{\rmv - \rmv}

\newcommand{\s}{\hspace*{0.5pt}}

\providecommand{\norm}[1]{\lVert#1\rVert}

\newcommand{\zd}{\ensuremath{{z_\tau^{(j)}}_{\rmv\rmv\rmv\rmv\rmv\rmv\rmv  m,n}}}
\newcommand{\zu}{\ensuremath{{z_\mathrm{u}^{(j)}}_{\rmv\rmv\rmv\rmv\rmv\rmv\rmv m,n}}}
\newcommand{\zdr}{\ensuremath{{\rv{z}_\tau^{(j)}}_{\rmv\rmv\rmv\rmv\rmv\rmv\rmv  m,n}}}
\newcommand{\zur}{\ensuremath{{\rv{z}_\mathrm{u}^{(j)}}_{\rmv\rmv\rmv\rmv\rmv\rmv\rmv m,n}}}

\newcommand{\scale}{0.6}
\newcommand{\bkm}{\scalebox{\scale}{$b^{(j)}_{l,n} = \underline{K} + m$}}
\newcommand{\scb}{\scalebox{\scale}{$b^{(j)}_{l,n}$}}
\newcommand{\bi}{\scalebox{\scale}{$b^{(j)}_{l,n} = i$}}

\newcommand{\noise}{\ensuremath{n}}

%% file: pgfplot.tex
\usetikzlibrary{shapes.misc}
\usetikzlibrary{arrows}
\usetikzlibrary{arrows.meta}
\usetikzlibrary{angles}
\usetikzlibrary{quotes}
\usetikzlibrary{fit}
\usetikzlibrary{graphs}
\usetikzlibrary{patterns}
\usetikzlibrary{petri}
\usetikzlibrary{plotmarks}
\usetikzlibrary{positioning}
\usetikzlibrary{decorations.pathmorphing}
\usetikzlibrary{decorations.markings}
\usetikzlibrary{decorations.pathreplacing}
\usetikzlibrary{decorations.shapes}
\usetikzlibrary{decorations.text}
\usetikzlibrary{backgrounds}
\usetikzlibrary{calc}
\usetikzlibrary{spy}

\usepackage{tikzscale}

\usepackage{pgfplots}
\pgfplotsset{compat=newest}
\usepgfplotslibrary{colormaps}
\usepgfplotslibrary{patchplots}
\usepgfplotslibrary{units}

\def\figHresults{3cm}
\def\figWresults{4.5cm}

\usepackage{ifluatex}

\ifluatex 
    \usepackage{fontspec}
    \defaultfontfeatures{Ligatures={TeX}}

\else
    \usepackage[latin1]{inputenc}
    \usepackage[T2A,T1]{fontenc}
\fi

\ifthenelse{\equal{\exportFigures}{true}}
{
  \usepgfplotslibrary{external}\tikzexternalize[prefix=\tikzfolder]
  \tikzset{external/system call={pdflatex \tikzexternalcheckshellescape -halt-on-error -interaction=batchmode -jobname "\image" "\texsource"}}
  \tikzset{external/up to date check=diff}
}
{}

\def\addlegendimage{\csname pgfplots@addlegendimage\endcsname}
\def\addlegendentry{\csname pgfplots@addlegendentry\endcsname}

\newcommand{\pgfref}[1]
{\ifthenelse{\equal{\exportFigures}{true}}
{\tikzexternalenable\ref{#1}\tikzexternalenable}
{\ref{#1}}}

\pgfplotsset{every axis/.append style={
  label style={font=\normalsize},
  tick label style={font=\scriptsize},
  xticklabel={
   \ifdim \tick pt < 0pt
    \pgfmathparse{abs(\tick)}%
    \llap{$-{}$}\pgfmathprintnumber{\pgfmathresult}
   \else
    \pgfmathprintnumber{\tick}
   \fi}
   }}
 
\usetikzlibrary{decorations.markings}
\makeatletter
\tikzset{
  nomorepostactions/.code={\let\tikz@postactions=\pgfutil@empty},
  mymarkfixednumber/.style n args={3}{decoration={markings,
    mark= between positions 0.1 and 1 step (1/#3)*\pgfdecoratedpathlength with{%
        \tikzset{#2,every mark}\tikz@options
        \pgftransformresetnontranslations
        \pgfuseplotmark{#1}%
      },  
    },
    postaction={decorate},
    /pgfplots/legend image post style={
        mark=#1,mark options={#2},every path/.append style={nomorepostactions}
    },
  },
}
\tikzset{
  nomorepostactions/.code={\let\tikz@postactions=\pgfutil@empty},
  mymarkfixeddistance/.style n args={3}{decoration={markings,
    mark= between positions 0 and 1 step #3cm with{%
        \tikzset{#2,every mark}\tikz@options
        \pgftransformresetnontranslations%
        \pgfuseplotmark{#1}%
      },  
    },
    postaction={decorate},
    /pgfplots/legend image post style={
        mark=#1,mark options={#2},every path/.append style={nomorepostactions}
    },
  },
}
\tikzset{
  nomorepostactions/.code={\let\tikz@postactions=\pgfutil@empty},
  mymark/.style n args={3}{decoration={markings,
    mark= between positions 0 and 1 step 0.75cm with{%
        \tikzset{#2,every mark}\tikz@options
        \pgftransformresetnontranslations%
        \pgfuseplotmark{#1}%
      },  
    },
    postaction={decorate},
    /pgfplots/legend image post style={
        mark=#1,mark options={#2},every path/.append style={nomorepostactions}
    },
  },
}
\makeatother

\def\mylinewidth{0.5}

\def\mymarksize{1.5}
\def\mymarksize2{2.5}
\def\markA{o}
\def\markAa{*}
\def\markB{square}
\def\markBb{square*}
\def\markC{triangle}
\def\markCc{triangle*}
\def\markD{diamond}
\def\markDd{diamond*}
\def\markE{pentagon}
\def\markF{asterisk}

\pgfdeclareplotmark{markarrow}
{%
  \pgfpathmoveto{\pgfqpoint{\pgfplotmarksize}{0pt}}
  \pgfpathlineto{\pgfqpointpolar{130}{1.5\pgfplotmarksize}}
  \pgfpathmoveto{\pgfqpoint{\pgfplotmarksize}{0pt}}
  \pgfpathlineto{\pgfqpointpolar{-130}{1.5\pgfplotmarksize}}
  \pgfusepathqstroke
}

%% file: inputFiles/acronyms.tex
\acrodefplural{esl}[ESLs]{electronic shelf labels}
\acrodef{aoa}[AOA]{angle-of-arrival}
\acrodef{aod}[AOD]{angle-of-departure}
\acrodef{awgn}[AWGN]{additive white Gaussian noise}
\acrodef{blt}[BLT]{bluetooth}
\acrodef{bp}[BP]{belief propagation}
\acrodef{bw}[BW]{bandwidth}
\acrodef{cdf}[CDF]{cumulative distribution function}
\acrodef{ceda}[CEDA]{channel estimation and detection algorithm}
\acrodef{ci}[CI]{confidence interval}
\acrodef{cl}[CL]{confidence level}
\acrodef{crlb}[CRLB]{Cram\'er-Rao lower bound}
\acrodef{dmc}[DMC]{dense multipath component}
\acrodef{dm}[DM]{dense multipath}
\acrodef{dnr}[DNR]{dense-to-noise ratio}
\acrodef{dps}[DPS]{delay power spectrum}
\acrodef{dut}[DUT]{device under test}
\acrodef{eirp}[EIRP]{equivalent isotropic radiated power}
\acrodef{eot}[EOT]{extended object tracking}
\acrodef{fa}[FA]{false alarm}
\acrodef{glrt}[GLRT]{generalized likelihood ratio test}
\acrodef{iid}[iid]{independent and identically distributed}
\acrodef{lhf}[LHF]{likelihood function}
\acrodef{los}[LOS]{line-of-sight}
\acrodef{lpbo}[LPBO]{legacy \ac{pbo}}
\acrodef{mc}[MC]{Monte Carlo}
\acrodef{mf}[MF]{matched filter}
\acrodef{mie}[MIE]{method of interval estimation}
\acrodef{mimo}[MIMO]{multiple input multiple output}
\acrodef{ml}[ML]{maximum likelihood}
\acrodef{mmse}[MMSE]{minimum mean-square error}
\acrodef{mpc}[MPC]{multipath component}
\acrodef{mp}[MP]{multipath}
\acrodef{MP}[MP]{message passing}
\acrodef{mpslam}[MP-SLAM]{multipath-based simultaneous localization and mapping}
\acrodef{Mpslam}[MP-SLAM]{Multipath-based simultaneous localization and mapping}
\acrodef{mse}[MSE]{mean squared error}
\acrodef{mse}[MSE]{mean squared error}
\acrodef{mva}[MVA]{master \ac{va}}
\acrodef{nlike}[NLIKE]{normalized likelihood}
\acrodef{nlos}[NLOS]{non-\ac{los}}
\acrodef{nnlike}[NNLIKE]{normalized noise-free likelihood}
\acrodef{npbo}[NPBO]{new \ac{pbo}}
\acrodef{olos}[OLOS]{obstructed \ac{los}}
\acrodef{ospa}[OSPA]{optimal subpattern assignment}
\acrodef{mospa}[MOSPA]{mean \ac{ospa}}
\acrodef{pa}[PA]{physical anchor}
\acrodef{pbo}[PBO]{potential bias object}
\acrodef{pcb}[PCB]{printed circuit board}
\acrodef{pcrlb}[PCRLB]{posterior Cram\'er-Rao lower bound}
\acrodef{pdaai}[AIPDA]{amplitude-information \ac{pda}}
\acrodef{pdaf}[PDAF]{probabilistic data association filter}
\acrodef{pda}[PDA]{probabilistic data association}
\acrodef{pdf}[PDF]{probability density function}
\acrodef{pdp}[PDP]{power delay profile}
\acrodef{pmf}[PMF]{probability mass function}
\acrodef{pva}[PVA]{potential \ac{va}}
\acrodef{reb}[REB]{ranging error bound}
\acrodef{rmse}[RMSE]{root mean squared error}
\acrodef{rss}[RSS]{received signal strength}
\acrodef{rv}[RV]{random variable}
\acrodef{sinr}[SINR]{signal-to-interference-plus-noise-ratio}
\acrodef{slam}[SLAM]{simultaneous localization and mapping}
\acrodef{smc}[SMC]{specular multipath component}
\acrodef{snr}[SNR]{signal-to-noise-ratio}
\acrodef{spa}[SPA]{sum-product algorithm}
\acrodef{stdv}[STDV]{standard deviation}
\acrodef{tdoa}[TDOA]{time difference of arrival}
\acrodef{tka}[TKA]{trusted keyless access}
\acrodef{toa}[TOA]{time-of-arrival}
\acrodef{ut}[UT]{unscented transform}
\acrodef{uwb}[UWB]{ultra wide band}
\acrodef{va}[VA]{virtual anchor}
\acrodef{zzb}[ZZB]{Ziv-Zakai bound}

%% file: inputFiles/abstract.tex
\ac{Mpslam} is a promising approach to obtain position information of transmitters and receivers as well as information regarding the propagation environments in future mobile communication systems. Usually, specular reflections of the radio signals occurring at flat surfaces are modeled by \acp{va} that are mirror images of the \acp{pa}.  In existing methods for \ac{Mpslam}, each \ac{va} is assumed to generate only a single measurement. However, due to imperfections of the measurement equipment such as non-calibrated antennas or model mismatch due to roughness of the reflective surfaces, there are potentially multiple \acp{mpc} that are associated to one single \ac{va}. In this paper, we introduce a Bayesian particle-based \ac{spa} for \ac{Mpslam} that can cope with multiple-measurements being associated to a single \ac{va}. Furthermore, we introduce a novel statistical measurement model that is strongly related to the radio signal. It introduces additional dispersion parameters into the likelihood function to capture additional \ac{mpc}-related measurements. We demonstrate that the proposed SLAM method can robustly fuse multiple measurements per \ac{va} based on numerical simulations.

%% file: inputFiles/introduction.tex

\ac{Mpslam} is a promising approach to obtain position information of transmitters and receivers as well as information regarding their propagation environments in future mobile communication systems. 
Usually, specular reflections of radio signals at flat surfaces are modeled by \acp{va} that are mirror images of the \acp{pa} \cite{LeitingerJSAC2015,WitrisalSPM2016,LeitMeyHlaWitTufWin:TWC2019,MenMeyBauWin:JSTSP2019}. The positions of these \acp{va} are unknown. \ac{Mpslam} algorithms can detect and localize \acp{va} and jointly estimate the time-varying position of mobile agents \cite{GentnerTWC2016,LeitMeyHlaWitTufWin:TWC2019,MenMeyBauWin:JSTSP2019}. 
The availability of \ac{va} location information makes it possible to leverage multiple propagation paths of radio signals for agent localization and can thus significantly improve localization accuracy and robustness. 
In non-ideal scenarios with rough reflective surfaces \cite{KulmerPIMRC2018, WenKulWitWym:TWC2021} and limitations in the measurement equipment, such as non-calibrated antennas \cite{PohZhaStaCaiDamHoe:IEEEAcess2022}, those standard methods are prone to fail since multiple measurements can originate from the same PA or VA. 
This shows the need for developing new methods to cope with these limitations.

\subsection{State of the Art}

The proposed algorithm follows the feature-based \ac{slam} approach \cite{DurrantWhyte2006, Dissanayake2001}, i.e., the map is represented by an unknown number of \textit{features}, whose unknown positions are estimated in a sequential (time-recursive) manner. 
Existing \ac{Mpslam} algorithms consider \acp{va} \cite{LeitMeyHlaWitTufWin:TWC2019,LeiGreWit:ICC2019,MenMeyBauWin:JSTSP2019,KimGraGaoBatKimWym:TWC2020,KimGranSveKimWym:TVT2022} or \acp{mva} \cite{LeiMey:Asilomar2020_DataFusion,LeiTeaZhaLiaMey:Fusion2022,LeiVenTeaMey:TSP2023} as features to be mapped. 
Most of these methods use estimated parameters related to \acp{mpc} contained in the radio signal, such as distances (which are proportional to delays), \acp{aoa}, or \acp{aod} \cite{RichterPhD2005}. 
These parameters are estimated from the signal in a preprocessing stage \cite{RichterPhD2005,ShutWanJos:CSTA2013,HanBadFleRao:SAM2014, BadHanFle:TSP2017, HanFleuRao:TSP2018, LiLeiVenTuf:TWC2022, GreLeiFleWit:Arxiv2023} and are used as ``measurements'' available to the \ac{slam} algorithm. 
A complicating factor in feature-based \ac{slam} is measurement origin uncertainty, i.e., the unknown association of measurements with features \cite{LeitMeyHlaWitTufWin:TWC2019, LeiGreWit:ICC2019, MenMeyBauWin:JSTSP2019, MeyWilJ21, LiLeiVenTuf:TWC2022}. 
In particular, (i) it is not known which map feature was generated by which measurement, (ii) there are missed detections due to low \ac{snr} or occlusion of features, and (iii) there are false positive measurements due to clutter. 
Thus, an important aspect of \ac{Mpslam} is \emph{data association} between these measurements and the \acp{va} or the \acp{mva}. 
Probabilistic data association can increase the robustness and accuracy of \ac{Mpslam} but introduces additional unknown parameters. 
State-of-the-art methods for multipath-based \ac{slam} are Bayesian estimators that perform the \ac{spa} on a factor graph  \cite{LeitMeyHlaWitTufWin:TWC2019, LeiGreWit:ICC2019, MenMeyBauWin:JSTSP2019} to avoid the curse of dimensionality related to the high-dimensional estimation problems. 

In these existing methods for \ac{Mpslam}, each feature is assumed to generate only a single measurement \cite{WilliamsLauTAE2014, MeyerProc2018}. 
However, due to imperfections of the measurement equipment or model mismatch due to non-ideal reflective surfaces (such as rough surfaces characterized by diffuse multipath \cite{KulmerPIMRC2018, WenKulWitWym:TWC2021}), there are potentially multiple \acp{mpc} that need to be associated to a single feature (\acp{va} or \acp{mva}) to accurately represent the environment. 
This is related to the multiple-measurement-to-object data association in \ac{eot} \cite{Koc:TAES2008_EOT, GranstroemFatemiSvenssonTAES2020_PMBMFilterExtendedTargets, MeyerICASSP2020,MeyWilJ21}. 
In \ac{eot}, the point object assumption is no longer valid, hence one single object can potentially generate more than one measurement resulting in a particularly challenging data association due to the large number of possible association events \cite{GranstroemTAES2012, GraBau:JAIF2017, GranstroemFatemiSvenssonTAES2020_PMBMFilterExtendedTargets}. 
In \cite{MeyerICASSP2020, MeyWilJ21}, an innovative approach to this multiple-measurements-to-object data association problem is presented. 
It is based on the framework of graphical models \cite{KolFri:PGM2009}. 
In particular, a \ac{spa} was proposed with computational complexity that scales only quadratically in the number of objects and the number of measurements avoiding suboptimal clustering of spatially close measurements.

\subsection{Contributions}

In this paper, we introduce a Bayesian particle-based \ac{spa} for \ac{Mpslam} that can cope with multiple-measurements associated to a single \ac{va}. 
The proposed method is based on a factor graph designed for scalable probabilistic multiple-measurement-to-feature association proposed in \cite{MeyerICASSP2020,MeyWilJ21}. 
We also introduce a novel statistical measurement model that is strongly related to the radio signal. 
It introduces additional dispersion parameters into the likelihood function to capture additional \ac{mpc}-related measurements. The key contributions of this paper are as follows.
\begin{itemize}
	\item We introduce the multiple-measurement-to-feature data association proposed in \cite{MeyWilJ21} to \ac{Mpslam} \cite{LeiGreWit:ICC2019,LeitMeyHlaWitTufWin:TWC2019}.
	\item We use this multiple-measurement data association to incorporate additional \ac{mpc}-related measurements originating from non-ideal effects such as rough reflective surfaces or non-calibrated antennas.
	\item We introduce a novel likelihood function model that is augmented with dispersion parameters to capture these additional \ac{mpc}-related measurements that are associated to a single \ac{va}.
	\item We demonstrate based on synthetically generated measurements that the proposed \ac{slam} method robustly associates multiple measurements per \ac{va} and that it is able to significantly outperform state-of-the-art \ac{Mpslam} methods \cite{LeiGreWit:ICC2019, LeitMeyHlaWitTufWin:TWC2019} in case additional \ac{mpc}-related measurements occur.
\end{itemize}
This paper advances over the preliminary account of our method provided in the conference publication \cite{WieVenWilLei:Fusion2023} by (i) presenting a detailed derivation of the factor graph, (ii) providing additional simulation results, and (iii) demonstrating performance advantages compared to the classical \ac{Mpslam} \cite{LeitMeyHlaWitTufWin:TWC2019,LeiGreWit:ICC2019}.

\subsection{Notation}
\label{sec:notation}
Random variables are displayed in sans serif, upright fonts; their realizations in serif, italic fonts. 
Vectors and matrices are denoted by bold lowercase and uppercase letters, respectively. For example, a random variable and its realization are denoted by $\rv x$ and $x$, respectively, and a random vector and its realization 
by $\RV x$ and $\V x$, respectively. 
Furthermore, $\|\V{x}\|$ and ${\V{x}}^{\text T}$ denote the Euclidean norm and the transpose of vector $\V x$, respectively; $\propto$ indicates equality up to a normalization factor;
$f(\V x)$ denotes the \ac{pdf} of random vector $\RV x$ (this is a short notation for  $f_{\RV x}(\V x)$); 
$f(\V x | \V y)$ denotes the conditional \ac{pdf} of random vector $\RV x$ conditioned on random vector  $\RV y$  (this is a short notation for  $f_{\RV x | \RV y}(\V x | \V y)$).
The cardinality of a set $\Set{X}$ is denoted as $ \vert\Set{X}\vert $. $\delta(\cdot)$ denotes the Dirac delta function.
Furthermore, ${1}_{\mathbb{A}}(\V{x})$ denotes the indicator function that is ${1}_{\mathbb{A}}(\V{x}) = 1$ if $\V{x} \in \mathbb{A}$ and 0 otherwise, for $\mathbb{A}$ being an arbitrary set and $\mathbb{R}^{\text{+}}$ is the set of positive real numbers.
Finally, $ \delta_{e}$ denotes the indicator function of the event $ e \rmv=\rmv 0 $ (i.e., $ \delta_{e} \rmv=\rmv 1 $ if $ e \rmv=\rmv 0 $ and $0$ otherwise).
We define the following \acp{pdf} with respect to $\rv{x}$: 
The Gaussian \ac{pdf} is
\vspace*{-1mm}
\begin{align} \label{eq:eq:truncated_gaussian_pdf}
	f_\text{N}(x; \mu , \sigma) = \frac{1}{\sqrt{2\pi} \sigma} e^{\frac{-(x-\mu)^2}{2\,\sigma^2}} \\[-7mm]\nn
\end{align}
with mean $\mu$ and standard deviation $\sigma$ \cite{Kay1998}. 
The truncated Rician \ac{pdf} is \cite[Ch. 1.6.7]{BarShalom11} 
\vspace*{-1mm}
\begin{align} \label{eq:truncated_rice_pdf}
	f_\text{TRice}(x;\rmv s ,\rmv u , \lambda) = \frac{1}{Q_1(\frac{u}{s}, \frac{\lambda}{s})}\frac{x}{s^2} e^{\frac{-(x^2+u^2)}{2\,s^2}} I_0(\frac{x\, u}{s^2}) {1}_{\mathbb{R}^{\text{+}}}(x \minus \lambda)\\[-7mm]\nn
\end{align}
with non-centrality parameter $u$, scale parameter $s$ and truncation threshold $\lambda$. 
$I_0(\cdot)$ is the 0th-order modified first-kind Bessel function and $Q_1(\cdot,\cdot)$ denotes the Marcum Q-function \cite{Kay1998}. 
The truncated Rayleigh \ac{pdf} is \cite[Ch. 1.6.7]{BarShalom11}
\vspace*{-1mm}
\begin{align} \label{eq:truncated_rayleigh_pdf}
	f_\text{TRayl}(x; s , \lambda) = \frac{x}{s^2}\, e^{\frac{-(x^2-\lambda^2)}{2\, s^2}}  {1}_{\mathbb{R}^{\text{+}}}(x - \lambda)\\[-7mm]\nn
\end{align}
with scale parameter $s$ and truncation threshold $\lambda$. 
This formula corresponds to the so-called Swerling I model\cite{BarShalom11}. 
The Gamma \ac{pdf} is denoted as 
\vspace*{-1mm}
\begin{align}
\mathcal{G}(x;\alpha,\beta) =& \frac{1}{\beta^\alpha \Gamma(\alpha)} x^{k-1} e^{-\frac{x}{\beta}}
\label{eq:gammapdf}\\[-7mm]\nn
\end{align}
where $\alpha$ is the shape parameter, $\beta$ is the scale parameter and $\Gamma(\cdot)$ is the gamma-function. 
Finally, we define the uniform \ac{pdf} $f_\mathrm{U}(x;a,b) = 1/(b-a) {1}_{[a,b]}(x)$.

%% file: inputFiles/geom_model.tex

\section{Geometrical Relations}\label{sec:geometricRel} 

At each time $n$, we consider a mobile agent at position $\V{p}_n$ equipped with a single antenna and $J$ base stations, called \acp{pa}, equipped with a single antenna and at known positions $\V{p}_{\mathrm{pa}}^{(j)} = \big[ {p}_{1,\mathrm{pa}}^{(j)} \ist\ist\ist {p}_{2,\mathrm{pa}}^{(j)} \big]^{\mathrm{T}} \rmv\rmv\in \mathbb{R}^2\rmv$, $j \rmv\in\rmv \{1,\ldots,J\}$, where $J$ is assumed to be known, in an environment described by reflective surfaces. 
Specular reflections of radio signals at flat surfaces are modeled by \acp{va} that are mirror images of \acp{pa}. 
In particular, \ac{va} positions associated to single-bounce reflections are given by
\vspace*{-1mm}
\begin{align}
	\V{p}^{(j)}_{l,\mathrm{va}} &= \V{p}^{(j)}_{\mathrm{pa}} + 2\big( \V{u}_l^{\T}\V{e}_l - \V{u}_l^{\T}\V{p}^{(j)}_{\mathrm{pa}}\big)\V{u}_l \label{eq:VA1BPAequation}
	\\[-7mm]
	\nn
\end{align}
where $\V{u}_l$ is the normal vector of the according reflective surface, and $\V{e}_l$ is an arbitrary point on this surface. 
The second summand in \eqref{eq:VA1BPAequation} represents the normal vector w.r.t. this reflective surface in direction $\V{u}_l$ with the length of two times the distance between \ac{pa} $j$ at position $\V{p}^{(j)}_{\mathrm{pa}}$ and the normal-point at the reflective surface, i.e., $2\big( \V{u}_l^{\T}\V{e}_l - \V{u}_l^{\T}\V{p}^{(j)}_{\mathrm{pa}}\big)$. 
An example is shown in Fig.~\ref{fig:fp}. \ac{va} positions associated to multiple-bounce reflections are determined by applying \eqref{eq:VA1BPAequation} multiple times. The current number of \textit{visible} \acp{va}\footnote{A \ac{va} does not exist at time $n$, when the reflective surface corresponding to this \ac{va} is obstructed with respect to the agent.} within the scenario (associated with single-bounce and higher-order bounce reflections) is $L_n^{(j)}$ for each of the $J$ \acp{pa}.

\begin{figure}[t]
	\centering   
	\ifthenelse{\equal{\arxiv}{false}}
		{  
	\subfloat[]{
		\tikzsetnextfilename{indoor_environment_MPAs}
		\scalebox{1}{\input{./plots/indoor_environment_MPAs.tex}}
		\label{fig:fp}}
	\vspace*{2mm}	
	\subfloat[]{\centering
		\tikzsetnextfilename{environment_signal_example_seed3_log}
		\setlength{\figurewidth}{0.8\columnwidth}
		\setlength{\figureheight}{0.22\columnwidth}
		\def\datapath{./plots/signal_examples/environment_signal_example_seed3_log}
		\hspace{-18mm}{\input{\datapath/environment_signal_example_seed3_log.tex}}\vspace{-1mm}
		\label{fig:signal}}	}
		{ 	\subfloat[]{
		\scalebox{1}{\includegraphics{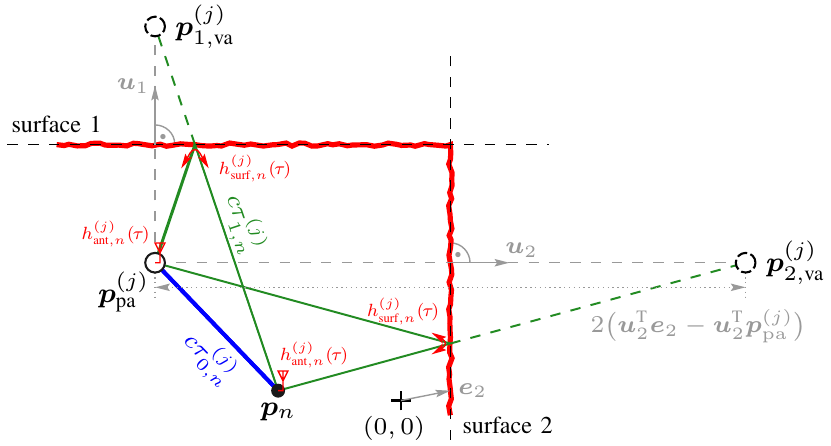}}
		\label{fig:fp}}
	\vspace*{2mm}	
	\subfloat[]{\centering
		\setlength{\figurewidth}{0.8\columnwidth}
		\setlength{\figureheight}{0.22\columnwidth}
		\hspace{-18mm}{\includegraphics{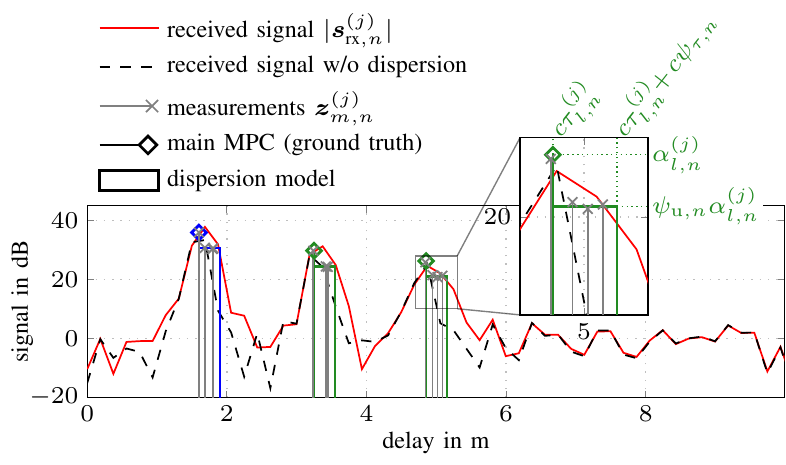}}\vspace{-1mm}
		\label{fig:signal}}	}
	\vspace*{-1.5mm}
	\caption{Exemplary indoor environment \sref{fig:fp} and representative realization of a received signal \sref{fig:signal}. The floor plan\vspace*{0.3mm} in \sref{fig:fp} includes an agent at position $\V{p}_n$ and a \ac{pa} at position \smash{$\V{p}^{(j)}_\text{pa}$} and two \acp{va} at positions \smash{$\V{p}^{(j)}_{l,\text{va}}$} for corresponding surfaces. The signal shown in \sref{fig:signal}\vspace*{0.3mm} is received by \ac{pa} at position \smash{$\V{p}^{(j)}_\text{pa}$}. Non-ideal antennas or reflective surfaces as indicated in \sref{fig:fp} by generic impulse responses $h_{\text{ant}, n}^{(j)}(\tau)$ and $h_{\text{surf}, n}^{(j)}(\tau)$ lead to the received signal {$\RV{s}_{\text{rx},n}^{(j)}$} shown in \sref{fig:signal} (c.f. received signal without dispersion). Resulting measurements (\ac{mpc} parameter estimates)\vspace*{0.3mm} $\V{z}^{(j)}_{m,n}$ are indicated in the received signal {$\RV{s}_{\text{rx},n}^{(j)}$} shown in \vspace*{0.3mm} \sref{fig:signal} alongside the proposed dispersion model.}
	\vspace*{-4mm}
	\label{fig:overview}
\end{figure}

%% file: inputFiles/signal_model.tex

\section{Radio Signal Model}\label{sec:signal_model}

At each time $n$, the mobile agent transmits a signal $s(t)$ from a single antenna and each \ac{pa} $j \rmv\rmv \in \rmv\rmv \{1,\ist\dots\ist,J\}$ acts as a receiver having a single antenna. 
The received complex baseband signal at the $j$th \ac{pa} is sampled $N_\text{s}$ times with sampling frequency $ f_{\text{s}} = 1/T_{\text{s}}$ yielding an observation period of $T =  N_{\text{s}} \, T_{\text{s}}$. 
By stacking the samples, we obtain the discrete-time received signal vector
\vspace*{-2mm}
\begin{align}
\hspace*{-3mm}\RV{s}_{\text{rx},n}^{(j)} \rmv\rmv=\rmv\rmv  \sum_{l = 1}^{{L}_n^{(j)}} {\alpha}_{l,n}^{(j)} \Big( \V{s}\big({\tau}_{l,n}^{(j)}\big) \rmv\rmv+ \rmv\rmv\rmv\rmv\sum_{i=1}^{S_l^{(j)}} \beta^{(j)}_{l,i,n} \V{s}\big({\tau}_{l,n}^{(j)}\rmv\rmv+\rmv\rmv\nu^{(j)}_{l,i,n}\big)\Big) \rmv\rmv + \rmv\rmv \RV{w}_{n}^{(j)}\rmv\rmv\rmv\rmv\rmv\rmv
\label{eq:signal_model_sampled}\\[-6mm]\nn
\end{align}
where $\bm{s}(\tau) \triangleq [s(-(N_\text{s}-1)/2\ist T_\text{s}-\tau) \,\,\, \cdots \,\,\, s((N_\text{s}-1)/2\ist T_\text{s} - \tau)]^\text{T}\in \mathbb{C}^{N_\text{s}\times 1}$ is the discrete-time transmit pulse. 
The first term contains the sum over the \ac{los} component ($l=1$) and the ${L}_n^{(j)}\rmv-\rmv 1$ specular \acp{mpc} (for $l \rmv\in\rmv \{ 2,\dots,L_n^{(j)}\}$) termed main components.
The $l$th main-component is characterized by its complex amplitude ${\alpha}_{l,n}^{(j)} \in \mathbb{C}$ and its delays ${\tau}_{l,n}^{(j)}$. The second term contains the sum over $S_l^{(j)}$ additional sub-components characterized by complex amplitudes ${\alpha}_{l,n}^{(j)} \beta^{(j)}_{l,i,n}$ and by (relative) delays ${\tau}_{l,n}^{(j)} + \nu^{(j)}_{l,i,n}$, where $\nu^{(j)}_{l,i,n}$ is the excess delay and $\beta^{(j)}_{l,i,n} \in \mathbb{R}$ is a relative dampening variable.
The delays $\tau_{l,n}^{(j)}$ are proportional to the distances (ranges) between the agent and either the $j$th \ac{pa} (for $l \!=\! 1$) or the corresponding \acp{va} (for $l \rmv\in\rmv \{ 2,\dots,L_n^{(j)}\}$). 
That is $\tau_{1,n}^{(j)} \rmv=\rmv \big\|\V{p}_n \!-\rmv \V{p}_{\text{pa}}^{(j)}\big\|/c$ and $\tau_{l,n}^{(j)} \rmv=\rmv \big\|\V{p}_n \! -\rmv \V{p}_{l,\text{va}}^{(j)}\big\|/c$ for $l \rmv\in\rmv \{ 2,\dots,L_n^{(j)}\}$, where $c$ is the speed of light.
The measurement noise vector $\RV{w}_{n}^{(j)} \in \mathbb{C}^{N_\text{s} \times 1} $ is a zero-mean, circularly-symmetric complex Gaussian random vector with covariance matrix ${\sigma}^{(j)\ist 2} \M{I}_{N_\text{s}}$ and noise variance ${\sigma}^{(j)\ist 2} = {N}_{0}^{(j)}/T_{\text{s}}$. 
The component \ac{snr} of \ac{mpc} $l$ is $ \mathrm{SNR}^{(j)}_{l,n} = |{\alpha}^{(j)}_{l,n}|^2 \|\V{s}({\tau}_{l,n}^{(j)})\|^2/ {\sigma}^{(j)\ist 2}$. 
The component \ac{snr} of the sub-components is given as $ \mathrm{SNR}^{(j)}_{l,i,n} = \beta^{(j)\, 2}_{l,i,n}\mathrm{SNR}^{(j)}_{l,n}$. 
The corresponding normalized amplitude is ${u}^{(j)}_{l,n} \triangleq \mathrm{SNR}^{(j)\ist\frac{1}{2}}_{l,n}$ and ${u}^{(j)}_{l,i,n} \triangleq \mathrm{SNR}^{(j)\ist\frac{1}{2}}_{l,i,n}$, respectively. 
Details about the signal model given in \eqref{eq:signal_model_sampled} are provided in Appendix~\ref{sec:app_signal_model}. 

\subsection{Signal Model Assumptions}\label{sec:assumption}

To capture effects such as non-calibrated antennas \cite[Section~VII-C]{LiLeiVenTuf:TWC2022}, the scattering from a user-body \cite{WilLeiMueWit:PIMRC2020, WilLeiMueWit:EuCAP2021}, rural environments \cite{SchuFle:EUCAP2010,SchuFle:EUCAP2012} as well as non-ideal reflective surfaces \cite{KulmerPIMRC2018}, we introduce the dispersion parameters $\psi^{(j)}_{\tau,l,n}$ and $\psi_{\text{u},l,n}^{(j)}$. 
In this work, we assume the \textit{following restrictions to this model:} (i) the additional sub-components with excess delays $\nu^{(j)}_{l,i,n} \in [0, \psi^{(j)}_{\tau,l,n}]$ after each \ac{mpc} $l$ have the same support, i.e., $ \psi^{(j)}_{\tau,l,n} \triangleq  \psi_{\tau,n}$ and (ii) the corresponding dampening variables are constant $\beta^{(j)}_{l,i,n} \triangleq \psi^{(j)}_{\text{u},l,n}$ with the same value for each \ac{mpc} $l$, i.e., $\psi^{(j)}_{\text{u},l,n} \triangleq \psi_{\text{u},n}$. 
This model can be applied to ultra-wideband systems with non-calibrated antennas \cite[Section~VII-C]{LiLeiVenTuf:TWC2022} that introduce delay dispersion or to environments containing moderate non-ideal reflective surfaces \cite{KulmerPIMRC2018,WenKulWitWym:TWC2021} that are approximately similar in behavior and do not change significantly over the explored area. An exemplary signal as well as the dispersion model is shown in Fig.~\ref{fig:signal}.\footnote{Note that the proposed algorithm can be reformulated in line with \cite{MeyWilJ21} to the general case with individual delay supports $\psi^{(j)}_{\tau,l,n}$ and to more complex amplitudes distributions for $\beta^{(j)}_{l,i,n}$, especially when multiple-antenna systems providing multiple \ac{mpc} parameters (delay, \ac{aoa}, \ac{aod}) \cite{LeiGreWit:ICC2019, MenMeyBauWin:JSTSP2019, LeiVenTeaMey:TSP2023}.}

\subsection{Parametric Channel Estimation}  \label{sec:channel_estimation}

By applying at each time $ n $, a \ac{ceda} \cite{ShutWanJos:CSTA2013, HanBadFleRao:SAM2014, BadHanFle:TSP2017, HanFleuRao:TSP2018, LiLeiVenTuf:TWC2022, GreLeiFleWit:Arxiv2023} to the {observed} discrete signal vector $\V{s}_{\text{rx},n}^{(j)}$, one obtains, for each anchor $j$, a number of $M_n^{(j)}$ measurements denoted by ${\V{z}^{(j)}_{m,n}}$ with $m \in  \Set{M}_n^{(j)} \triangleq \{1,\,\dots\,,M_n^{(j)}\} $.
Each  $\V{z}^{(j)}_{m,n} = [\zd \ \zu]^\text{T}$ representing a potential \ac{mpc} parameter estimate, contains a delay measurement $\zd \rmv\rmv\in\rmv\rmv [0, \tau_\text{max}]$ and a normalized amplitude measurement $\zu \rmv\rmv\in\rmv\rmv [\gamma, \infty )$, where $\gamma$ is the detection threshold. 
The \ac{ceda} decomposes the signal $\RV{s}_{\text{rx},n}^{(j)}$ into individual, decorrelated components according to \eqref{eq:signal_model_sampled}, reducing the number of dimensions (as ${M}_n^{(j)}$ is usually much smaller than $N_\text{s}$). It thus compresses the information contained in $\RV{s}_{\text{rx},n}^{(j)}$ into $\vspace*{-1mm}\V{z}^{(j)}_{n} = [{\bm{z}^{(j)\text{T}}_{1,n}}  \rmv \cdots  {\V{z}^{(j)\text{T}}_{M_n^{(j)},n}}]^\text{T}$. 
The stacked vector $\V{z}_n = [\V{z}^{(1)\, \text{T}}_{n} \rmv \cdots  \V{z}^{(J)\,\text{T}}_{n}]^\text{T}$ is used by the proposed algorithm as a noisy measurement.

%% file: inputFiles/system_model.tex

\section{System Model}\label{sec:system_model}

At each time $n$, the state $\RV{x}_n = [\RV{p}_n^\T\,\ist \RV{v}_n^\T]^\T$ of the agent consists of its position $\RV{p}_n$ and velocity $\RV{v}_n$. 
We also introduce the augmented agent state $\tilde{\RV{x}}_n = [\RV{x}_n^\T \,\ist \RV{\psi}_{n}^\T]^\T$ that contains the dispersion parameters $\RV{\psi}_{n} = [\rv{\psi}_{\tau,n}\iist \rv{\psi}_{\text{u},n}]^\T$. 
In line with \cite{MeyerProc2018, LiLeiVenTuf:TWC2022, LeiGreWit:ICC2019}, we account for the unknown number of \acp{va} by introducing for each \ac{pa} $j$ \acp{pva} $k \rmv\in\rmv \mathcal{K}^{(j)}_n \triangleq \{ 1,\dots,{K}_n^{(j)} \}$. 
The number of \acp{pva} $K_n^{(j)}$ is the maximum possible number of \acp{va} of \ac{pa} $j$ that produced measurements so far \cite{MeyerProc2018} (i.e., $K_n^{(j)}$ increases with time). 
The state of \ac{pva} $(j,k)$ is denoted as $\RV{y}_{k,n}^{(j)} \!\triangleq\rmv \big[\RV{x}_{k,n}^{(j)\text{T}}\,\ist \rv{r}_{k,n}^{(j)}\big]^\T$ with $\RV{x}_{k,n}^{(j)} = \big[\RV{p}_{k,\text{va}}^{(j)\text{T}} \,\ist \rv{u}_{k,n}^{(j)} \big]^\T$, which includes the normalized amplitude $\rv{u}_{k,n}^{(j)}$ \cite{LeiGreWit:ICC2019,LiLeiVenTuf:TWC2022}. 
The existence/nonexistence of \ac{pva} $k$ is modeled by the existence variable $\rv{r}^{(j)}_{k,n} \rmv\in \{0,1\}$ in the sense that \ac{pva} $k$ exists if and only if $r^{(j)}_{k,n} \!=\! 1$. 
The \ac{pva} state is considered formally also if \ac{pva} $k$ is nonexistent, i.e., if $r^{(j)}_{k,n} \!=\rmv 0$. 

Since a part of the \ac{pa} state is unknown, we also consider the \ac{pa} itself a \ac{pva}. 
Hence, we distinguish between the \ac{pva} $k=1$ that explicitly represents the \ac{pa}, which is a-priori existent and has known and fixed position $\V{p}_{1,\text{va}}^{(j)} = \V{p}_{\text{pa}}^{(j)} $, and all other \acp{pva} $k \in \{2,\ist\dots\ist, K_n^{(j)}\}$ whose existence and position are a-priori unknown. 
Note that the \acp{pva} state representing the \ac{pa} still considers the normalized amplitude $\rv{u}_{1,n}^{(j)}$ as well as the existence variable $\rv{r}^{(j)}_{1,n}$.
The states $\RV{x}^{(j)\ist\text{T}}_{k,n}$ of nonexistent \acp{pva} are obviously irrelevant. 
Therefore, all \acp{pdf} defined for \ac{pva} states, $f(\V{y}_{k,n}) =\rmv f(\V{x}_{k,n}, r_{k,n})$, are of the form $f(\V{x}^{(j)}_{k,n}, 0 )$ $=\rmv f^{(j)}_{k,n} f_{\text{d}}(\V{x}^{(j)}_{k,n})$, where $f_{\text{d}}(\V{x}^{(j)}_{k,n})$ is an arbitrary ``dummy'' \ac{pdf} and $f^{(j)}_{k,n} \!\rmv\in [0,1]$ is a constant. 
We also define the stacked vectors $\RV{y}_n^{(j)} \!\triangleq \big[\RV{y}_{1,n}^{(j)\text{T}} \rmv\cdots\ist \RV{y}_{K_n^{(j)}\rmv,n}^{(j)\text{T}} \big]^\T$ and $\RV{y}_n \!\triangleq \big[\RV{y}_n^{(1)\text{T}} \rmv\cdots\ist \RV{y}_n^{(J)\text{T}} \big]^\T\rmv$. 
Note that according to the model introduced in Section~\ref{sec:signal_model}, $\RV{\psi}_{n}$ is common for all \acp{pva}. 
However, this model can be extended to individual dispersion parameters for each \ac{pva} (see \cite{MeyWilJ21}).

\subsection{State Evolution}
\label{sec:state_statistics}

For each \ac{pva} with state $\RV{y}_{k,n-1}^{(j)}$ with $k \rmv\in\rmv \mathcal{K}^{(j)}_{n-1} \triangleq \{ 1,\dots,{K}_{n-1}^{(j)} \}$ at time $n-1$ and \ac{pa} $j$, there is one ``legacy'' \ac{pva} with state $\underline{\RV{y}}_{k,n}^{(j)} \!\triangleq \big[\underline{\RV{x}}_{k,n}^{(j)\text{T}}\,\ist \underline{\rv{r}}_{k,n}^{(j)}\big]^\T$ with $k \in \mathcal{K}^{(j)}_{n-1}$ at time $n$ and \ac{pa} $j$. 
We also define the joint states \vspace*{-0.8mm} $\underline{\RV{y}}_n^{(j)} \triangleq \big[\underline{\RV{y}}_{1,n}^{(j)\text{T}} \rmv\cdots\ist \underline{\RV{y}}_{K_{n-1}^{(j)}\rmv,n}^{(j)\text{T}} \big]^\T\rmv$ and $\underline{\RV{y}}_n \!\triangleq \big[\underline{\RV{y}}_n^{(1)\text{T}} \rmv\cdots\ist \underline{\RV{y}}_n^{(J)\text{T}} \big]^\T\rmv$. 
Assuming that the augmented agent state as well as the \ac{pva} states of all \acp{pa} evolve independently across $k$, $n$, and $j$, the joint state-transition \ac{pdf} factorizes as \cite{LeitMeyHlaWitTufWin:TWC2019,MeyerProc2018}
\vspace*{-1mm}
\begin{align}
	f\big(\tilde{\V{x}}_n,\underline{\V{y}}_n|\tilde{\V{x}}_{n-1},\V{y}_{n-1}\big) &= f(\V{x}_{n}|\V{x}_{n-1})f(\V{\psi}_{n}|\V{\psi}_{n-1})\nn \\[-1mm]
	& \hspace{5mm}\times  \prod_{j=1}^J \prod_{k=1}^{K_{n-1}^{(j)}}  \rmv\rmv f\big(\underline{\V{y}}_{k,n}^{(j)} \big| \V{y}_{k, n-1}^{(j)}\big) 
	\label{eq:state_space}\\[-6mm]\nn
\end{align}
where $f(\underline{\V{y}}_{k,n}^{(j)}| \V{y}_{k, n-1}^{(j)}) \triangleq f\big(\underline{\V{x}}_{k,n}^{(j)},\underline{r}_{k,n}^{(j)} \big| \V{x}_{k,n-1}^{(j)}, r_{k,n-1}^{(j)}\big)\vspace*{0.5mm}$ is the legacy \ac{pva} state-transition \ac{pdf}. 
If \ac{pva} did not exist at time $n \rmv-\! 1$, i.e., $r_{k,n-1}^{(j)} \!=\! 0$, it cannot exist as a legacy \ac{pva} at time $n$ either. 
Thus, 
\vspace*{-2mm}
\begin{align}
	f\big(&\underline{\V{x}}_{k,n}^{(j)},r_{k,n}^{(j)} \big| \V{x}_{k,n-1}^{(j)}, 0\big) = \begin{cases} 
		f_\text{d}\big(\underline{\V{x}}_{k,n}^{(j)}\big) , &\!\!\! \underline{r}_{k,n}^{(j)} \!=\rmv 0 \\[0mm]
		0 , &\!\!\! \underline{r}_{k,n}^{(j)} \!=\! 1.
	\end{cases} \!\!\!\!
	\label{eq:dummy_transition}\\[-7mm]\nn
\end{align}
If \ac{pva} existed at time $n \rmv-\! 1$, i.e., $r_{k,n-1}^{(j)} \!=\! 1$, it either dies, i.e., $\underline{r}_{k,n}^{(j)} \!=\rmv 0$, or survives, i.e., $\underline{r}_{k,n}^{(j)} \!=\! 1$ with survival probability denoted as $p_\text{s}$. 
If it does survive, its new state $\underline{\RV{y}}_{k,n}^{(j)}$ is distributed according to the state-transition PDF \vspace*{0.2mm}$f\big(\underline{\V{x}}_{k,n}^{(j)}\big| \V{x}_{k,n-1}^{(j)}\big) \triangleq  \delta\big(\underline{\V{p}}_{k,\text{va}}^{(j)} - \V{p}_{k,\text{va}}^{(j)}\big) f\big(u_{k,n}^{(j)} \big| u_{k,n-1}^{(j)}\big)$ \cite{LeiGreWit:ICC2019,LeitMeyHlaWitTufWin:TWC2019}. 
Thus,
\begin{align}
	&f\big(\underline{\V{x}}_{k,n}^{(j)},\underline{r}_{k,n}^{(j)} \big| \V{x}_{k,n-1}^{(j)}, 1\big)\nn\\
	& \hspace{4mm} =\begin{cases} 
		(1 \!-\rmv p_\text{s}) \ist f_\text{d}\big(\underline{\V{x}}_{k,n}^{(j)}\big) , &\!\!\! \underline{r}_{k,n}^{(j)} \!=\rmv 0 \\[0mm]
		p_\text{s} \ist \delta\big(\underline{\V{p}}_{k,\text{va}}^{(j)} - \V{p}_{k,\text{va}}^{(j)}\big) f\big(u_{k,n}^{(j)} \big| u_{k,n-1}^{(j)}\big) , &\!\!\! \underline{r}_{k,n}^{(j)} \!=\! 1
	\end{cases}\ist.
	\label{eq:survival_transition}\\[-6mm]\nn
\end{align}
The agent state $\RV{x}_n$ with state-transition \ac{pdf} $f(\V{x}_{n}|\V{x}_{n-1})$ is assumed to evolve in time according to a 2-dimensional, constant velocity and stochastic acceleration model \cite{BarShalom2002EstimationTracking} (linear movement) given as $\RV{x}_n = \bm{A}\, \RV{x}_{n\minus 1} + \bm{B}\, \RV{w}_{n}$, with the acceleration process $\RV{w}_n$ being \ac{iid} across $n$, zero mean, and Gaussian with covariance matrix ${\sigma_{\text{w}}^2}\, \bm{I}_2$, ${\sigma_{\text{w}}}$ is the acceleration standard deviation, and $\bm{A} \in \mathbb{R}^{4\times 4}$ and $\bm{B} \in \mathbb{R}^{4\times 2}$ are defined according to \cite[p.~273]{BarShalom2002EstimationTracking}, with observation period $\Delta T$. 
The state-transition \acp{pdf} of the dispersion parameter states $f(\V{\psi}_{n}|\V{\psi}_{n-1}) = f({\psi}_{\tau,n}|\psi_{\tau,n-1})f(\psi_{\text{u},n}|\psi_{\text{u},n-1})$ are assumed to evolve independently of each other across $n$. 
Since both dispersion parameters are strictly positive and independent, we model the individual state-transition \acp{pdf} by Gamma \acp{pdf} given respectively by $f(\psi_{\tau,n}|\psi_{\tau,n-1})= \mathcal{G}(\psi_{\tau,n};q_{\tau},\psi_{\tau,n-1}/q_{\tau})$ and $f(\psi_{\text{u},n}|\psi_{\text{u},n-1}) = \mathcal{G}(\psi_{\text{u},n};q_{\text{u}},\psi_{\text{u},n-1}/q_{\text{u}})$, where $q_{\tau}$ and $q_{\text{u}}$ represent the respective state noise parameters \cite{Koc:TAES2008_EOT,MeyWilJ21}. 
Note that a small $q$ implies a large state transition uncertainty.
The state-transition \ac{pdf} of the normalized amplitude $\underline{u}_{k,n}^{(j)}$ is modeled by a truncated Rician \ac{pdf}\vspace*{-0.4mm}, i.e., $f( \underline{u}_{k,n}^{(j)} | u_{k,n\minus 1}^{(j)}) = f_\text{TRice}(\underline{u}_{k,n}^{(j)}; \sigma_{\text{u},k} ,u^{(j)}_{k,n\minus 1}, 0)$ with state noise parameter $\sigma_{\text{u},k}$. 
The truncated Rician \ac{pdf} was found to be useful for the proposed amplitude model \cite{LiLeiVenTuf:TWC2022} (see \eqref{eq:los_amplitude} in Section~\ref{sec:measurementModel}).\footnote{In \cite{MerUlmKoc:TAES2016}, it is shown that for a Swerling model I and III a Gamma state-transition \ac{pdf} represents a conjugate prior making an analytical derivation possible.}

\subsection{Measurement Model}
\label{sec:measurementModel} 

At each time $n$ and for each anchor $j$, the \ac{ceda} provides the currently observed measurement vector $\V{z}_n^{(j)}$, with fixed ${M}^{(j)}_n$, according to Section~\ref{sec:channel_estimation}. Before the measurements 
are observed, they are random and represented by the vector  ${\RV{z}^{(j)}_{m,n}} = [\zdr~\zur]^\text{T}$. 
In line with Section~\ref{sec:channel_estimation} we define the nested random vectors $\RV{z}^{(j)}_{n} = [{\RV{z}^{(j)\text{T}}_{1,n}} \cdots\ist {\RV{z}^{(j)\text{T}}_{\rv{M}_n^{(j)},n}}]^\text{T}$, with length corresponding to the random number of measurements $\rv{M}^{(j)}_n$, and $\RV{z}_n = [\RV{z}^{(1)\, \text{T}}_{n}\cdots\ist \RV{z}^{(J)\,\text{T}}_{n}]^\text{T}$. 
The vector containing all numbers of measurements is defined as  $\RV{M}_n = [\rv{M}_n^{(1)}\ist\cdots\ist \rv{M}_n^{(J)}]^\text{T}$.

If \ac{pva} $k$ exists ($r^{(j)}_{k,n} = 1$), it gives rise to a random number of  
measurements. 
The mean number of measurements per (existing) \ac{pva} is modeled by a Poisson point process with mean $\mu_\text{m}\big({\V{\psi}}_n , {u}^{(j)}_{k,n}\big)$. 
%
The individual measurements $\RV{z}^{(j)}_{m,n} $ are assumed to be conditionally independent, i.e., the joint \ac{pdf} of all measurements factorizes as $f(\V{z}_{n}^{(j)}| {M}_n^{(j)}, {\V{x}}_n, {\nu}_{k,n}^{(j)}, {\beta}_{k,n}^{(j)}, \V{x}_{k,n}^{(j)})  = \prod_{m=1}^{{M}_n^{(j)}} f(\V{z}_{m,n}^{(j)}| {\V{x}}_n, {\nu}_{k,n}^{(j)}, {\beta}_{k,n}^{(j)}, \V{x}_{k,n}^{(j)}) $.

If ${\RV{z}^{(j)}_{m,n}}$ is generated by a \ac{pva}, i.e., it corresponds to a main-component (\ac{los} component or \ac{mpc}),
we assume that the single-measurement \acl{lhf} $f(\V{z}_{m,n}^{(j)}| {\V{x}}_n,  {\nu}_{k,n}^{(j)}, {\beta}_{k,n}^{(j)}, \V{x}_{k,n}^{(j)})$ is conditionally independent across $\zdr$ and $\zur$. 
Thus, it factorizes as
\begin{align}
	&f(\V{z}_{m,n}^{(j)}| {\V{x}}_n,  {\nu}_{k,n}^{(j)}, {\beta}_{k,n}^{(j)}, \V{x}_{k,n}^{(j)})  \nonumber\\ &\hspace{2mm}  = f(\zd| \V{p}_{n}, {\nu}_{k,n}^{(j)}, {\beta}_{k,n}^{(j)} , \V{x}_{k,n}^{(j)})
	f(\zu| {\beta}_{k,n}^{(j)}, {u}_{k,n}^{(j)}) .
	\label{eq:single_measurement_likelihood}
\end{align}
The \acl{lhf} of the corresponding delay measurement $\zdr$ is given by
\vspace*{-1mm}
\begin{align}
	&  f(\zd | \V{p}_{n}, {\nu}_{k,n}^{(j)}, {\beta}_{k,n}^{(j)}, \V{x}_{k,n}^{(j)}) \nonumber\\
	&\hspace{5mm} = f_\text{N}\Big(\zd;\, \tau(\bm{p}_{k,\text{va}}^{(j)},\bm{p}_n) + \nu_{k,n}^{(j)},\, \sigma_{\tau}^{2} ({\beta}_{k,n}^{(j)} u_{k,n}^{(j)}) \Big)\\[-6mm]\nn
\end{align}
with mean $\tau(\RV{p}_{k,\text{va}}^{(j)},\RV{p}_n)  + \rv{\nu}_{k,n}^{(j)}$ and variance $\sigma_{\tau}^{2} (\rv{\beta}_{k,n}^{(j)} \rv{u}_{k,n}^{(j)})$ where $\tau(\RV{p}_{k,\text{va}}^{(j)},\RV{p}_n) = \norm{\RV{p}_n - \RV{p}_{k,\text{va}}^{(j)}}{}/c$. 
The standard deviation is determined from the Fisher information given by $\sigma_{\tau}^{2} (u) = c^2 / ( 8\, \pi^2 \, \beta_\text{bw}^2 \, u^{2})$ with $\beta_\text{bw}$ being the root mean squared bandwidth \cite{WitrisalWCL2016,WilGreLeiMueWit:ACSSC2018} (see Section~\ref{sec:results}). 
The \acl{lhf} of the corresponding normalized amplitude measurement $\zu$ is obtained as\footnote{The proposed model describes the distribution of the amplitude estimates of the radio signal model given in \eqref{eq:signal_model_sampled} \cite{LepRabLeG:FUSION2013, LepRabLeG:TAES2016, LiLeiVenTuf:TWC2022, VenLeiTerWit:TWC2023}.}
\vspace*{-1mm}
\begin{align} \label{eq:los_amplitude}
	&f(\zu|  {\beta}_{k,n}^{(j)}, \rmv u_{k,n}^{(j)} ) \nn\\
	& \hspace*{8mm}\triangleq  \rmv f_\text{TRice}(\rmv\zu; \rmv \sigma_\mathrm{u} ( {\beta}_{k,n}^{(j)} u_{k,n}^{(j)} ) , {\beta}_{k,n}^{(j)} u_{k,n}^{(j)}  , \rmv\rmv\rmv \gamma)\\[-6mm]\nn
\end{align}
with scale parameter $\sigma_\mathrm{u} ( \rv{\beta}_{k,n}^{(j)} \rv{u}_{k,n}^{(j)} ) $, non-centrality parameter $\rv{\beta}_{k,n}^{(j)} \rv{u}_{k,n}^{(j)}$, and detection threshold $\gamma$ \cite{LiLeiVenTuf:TWC2022, VenLeiTerWit:TWC2023}. 
The scale parameter is similarly determined from the Fisher information given by
\vspace*{-1mm} 
\begin{align}\label{eq:scaleParam_amplitude}
	\sigma_{\mathrm{u}}^{2} (u) = 1/2 +u\, /(4 N_{\text{s}})\ist. \\[-6mm]\nn
\end{align}
Note that this expression reduces to $1/2$ if the \ac{awgn} noise variance $\sigma^{(j)\s 2}$ is assumed to be known or $N_{\text{s}}$ to grow indefinitely (see \cite[Appendix~D]{LiLeiVenTuf:TWC2022} for a detailed derivation). 
The probability of detection resulting from \eqref{eq:los_amplitude} is given by the Marcum Q-function, i.e., $p_{\text{D}}(  \rv{\beta}_{k,n}^{(j)} \rv{u}_{k,n}^{(j)} ) \triangleq  Q_1(u/{\sigma_\mathrm{u} ( {\beta}_{k,n}^{(j)} u_{k,n}^{(j)} )}, {\gamma}/{\sigma_\mathrm{u} ( {\beta}_{k,n}^{(j)} u_{k,n}^{(j)} )})$ \cite{LerroACC1990, LiLeiVenTuf:TWC2022} (see Section~\ref{sec:notation}).
Using the assumptions introduced in the Section~\ref{sec:assumption}, the joint \ac{pdf} of the dispersion variables can be constructed as follows
\begin{align} \label{eq:dispersion}
	f( \nu_{k,n}^{(j)}, {\beta}_{k,n}^{(j)} |  \V{\psi}_{n} ) =& \frac{1}{2} \left(\delta(\nu_{k,n}^{(j)})\, \delta({\beta}_{k,n}^{(j)} - 1)\right. \nonumber \\
	& \hspace{-2.7mm} +  \left. f_\text{U} (\nu_{k,n}^{(j)};0, {\psi}_{\tau,n}) \delta({\beta}_{k,n}^{(j)} - {\psi}_{\text{u},n})\right)\\[-7mm]\nn
\end{align}
where the according delay dispersion random variable is given as $\rv{\nu}_{k,n}^{(j)} \sim  f_\text{U}({\nu}_{k,n}^{(j)};0,{\psi}_{\tau,n})$ and the amplitude dispersion random variable is $\rv{\beta}_{k,n}^{(j)} \sim \delta({\beta}_{k,n}^{(j)} - {\psi}_{\text{u},n})$. The \ac{pdf} of a single measurement $\RV{z}_{m,n}^{(j)}$ can now be obtained by integrating out the dispersion variables as
\begin{align}
	&f(\V{z}_{m,n}^{(j)}| \tilde{\V{x}}_n,  \V{x}_{k,n}^{(j)}) =  f(\V{z}_{m,n}^{(j)}| {\V{x}}_n, \V{\psi}_{n},  \V{x}_{k,n}^{(j)}) \nonumber\\
	& = \int 	f(\V{z}_{m,n}^{(j)}| {\V{x}}_n,  {\nu}_{k,n}^{(j)}, {\beta}_{k,n}^{(j)}, \V{x}_{k,n}^{(j)}) \nonumber\\
	&\hspace{15mm}\times f( \nu_{k,n}^{(j)}, {\beta}_{k,n}^{(j)} |  \V{\psi}_{n} ) \mathrm{d}\nu_{k,n}^{(j)}\, \mathrm{d}{\beta}_{k,n}^{(j)} \nonumber \\
	& = 	f(\zd | \V{p}_{n}, \V{x}_{k,n}^{(j)})   	f(\zu| u_{k,n}^{(j)} )    \nonumber \\
	&\hspace{10mm}	+ 		f(\zd| \V{p}_{n},  \V{\psi}_{n}, \V{x}_{k,n}^{(j)})   	f(\zu| u_{k,n}^{(j)},  \psi_{\text{u},n} )  
\end{align}
with the main-component delay \ac{pdf}
\begin{equation}
\hspace*{-2mm}f(\zd | \V{p}_{n}, \V{x}_{k,n}^{(j)}) 	= f_\text{N}(\zd;\, \tau(\V{p}_{k,\text{va}}^{(j)},\bm{p}_n),\, \sigma_{\tau}^{2} (u_{k,n}^{(j)}) )\rmv\rmv
\end{equation}
and the main-component amplitude \ac{pdf}
\begin{equation}
	f(\zu| u_{k,n}^{(j)} )  =  
	f_\text{TRice}(\zu; \sigma_\mathrm{u} ( u_{k,n}^{(j)} ) ,  u_{k,n}^{(j)}  , \gamma)
\end{equation}
as well as the additional sub-component delay \ac{pdf}
\begin{align}
	& 	f(\zd | \V{p}_{n}, \V{\psi}_{n}, \V{x}_{k,n}^{(j)}) \nonumber \\
	&= \rmv\rmv\frac{1}{{\psi}_{\tau,n}} \rmv\rmv \int_{0}^{{\psi}_{\tau,n}} \rmv  \rmv \rmv \rmv \rmv \rmv \rmv\rmv\rmv\rmv\rmv\rmv\rmv\rmv\rmv\rmv\rmv
	f_\text{N}\Big(\zd; \tau(\V{p}_{k,\text{va}}^{(j)},\bm{p}_n) \rmv\rmv + \rmv\rmv \nu_{k,n}^{(j)}, \sigma_{\tau}^{2} ( {\psi}_{\text{u},n} u_{k,n}^{(j)}) \rmv\rmv \Big)   \s \mathrm{d} \nu_{\rmv  k,n}^{(j)} \nonumber \\	
	%
	&= \frac{1}{2 {\psi}_{\tau,n}} \Bigg( \text{erf}\left( \frac{\tau(\V{p}_{k,\text{va}}^{(j)},\bm{p}_n) + {\psi}_{\tau,n} - \zd }{  \sigma_{\tau} ( {\psi}_{\text{u},n} u_{k,n}^{(j)}) \sqrt{2}} \right) 
	\nonumber \\ &\hspace{25mm} - \text{erf}\Bigg( \frac{\tau(\V{p}_{k,\text{va}}^{(j)},\bm{p}_n) - \zd }{  \sigma_{\tau} ( {\psi}_{\text{u},n} u_{k,n}^{(j)}) \sqrt{2}} \Bigg) \Bigg)
\end{align}
and the additional sub-component amplitude \ac{pdf}
\vspace*{-2mm}
\begin{align}
	&f(\zu| {\psi}_{\text{u},n}, u_{k,n}^{(j)} ) \nn\\
	&\hspace*{10mm}= f_\text{TRice}(\rmv \zu; \rmv \sigma_\mathrm{u} ( {\psi}_{\text{u},n} u_{k,n}^{(j)} ) , {\psi}_{\text{u},n} u_{k,n}^{(j)}  , \rmv\rmv\rmv \gamma).\\[-7mm]\nn
\end{align}
The according probability of detection is given as $p_{\text{D}}( \rv{u}_{k,n}^{(j)} )$ for the main-component of each \ac{pva} or $p_{\text{D}}( \rv{\psi}_{\text{u},n}  \rv{u}_{k,n}^{(j)} )$ for the additional sub-components, respectively.

It is also possible that a measurement $\RV{z}_{m,n}^{(j)}$ did not originate from any \ac{pva} (\emph{false alarm}). 
False alarm measurements originating from the \ac{ceda} are assumed statistically independent of \ac{pva} states. 
They are modeled by a Poisson point process with mean $ {{\mu}_{\mathrm{fa}}} $ and \ac{pdf} $ f_{\mathrm{fa}}(\V{z}_{m,n}^{(j)}) $, which is assumed to factorize as $f_{\mathrm{fa}}(\V{z}_{m,n}^{(j)}) = {f_{\mathrm{fa}}}(\zd) {f_{\mathrm{fa}}}({z_\mathrm{u}}_{m,n}^{(j)})$. 
The false alarm \ac{pdf} for a single delay measurement is assumed to be uniformly distributed as $ {f_{\mathrm{fa}}}(\zd)\rmv\rmv= f_\mathrm{U}(\zd;0,\tau_\text{max})$.
In correspondence to \eqref{eq:los_amplitude} the false alarm \acl{lhf} of the normalized amplitude measurement is given as
$	f_\mathrm{fa}(\zu)\rmv \triangleq \rmv f_\text{TRayl}(\zu \,; \sqrt{1/2} \,, \gamma) $
with the scale parameter given as $\sqrt{1/2}$ and detection threshold $\gamma$.

Considering the measurement model for the normalized amplitudes in \eqref{eq:los_amplitude}, the mean number of \ac{pva}-related measurements $\mu_\text{m}\big(\tilde{\RV{x}}_n , \RV{x}^{(j)}_{k,n}\big) \triangleq \mu_\text{m}\big({\RV{\psi}}_n , \rv{u}^{(j)}_{k,n}\big) $ is well approximated as  
\vspace{-1mm}
\begin{align} \label{eq:mean_number_of_meas}
	\mu_\text{m}\big({\RV{\psi}}_n , \rv{u}^{(j)}_{k,n}\big)  = p_{\text{D}}(  \rv{u}_{k,n}^{(j)} ) +\frac{N_\text{ny}\, \rv{\psi}_{\tau,n} }{c \, {T}_\text{s}} \, p_{\text{D}}( \rv{\psi}_{\text{u},n}  \rv{u}_{k,n}^{(j)}) \ist .\\[-7mm]\nn
\end{align}
The right-hand side fraction denotes the average number of additional sub-components estimated by the \ac{ceda} at a detection threshold of $\gamma = 0~\mathrm{dB}$, where we assume an average of $N_\text{ny}$ components to be detected within one Nyquist sample. 
Accordingly, the mean number of false alarms is approximated as $\mu_\text{fa} = N_\text{ny} \s N_\text{s}\, e^{-\gamma^2}$ with $e^{-\gamma^2} = \int _\gamma^\infty f_{\mathrm{fa}}(\zu)\, \mathrm{d}\zu$ denoting the false alarm probability.

\subsection{New \acp{pva}}

Newly detected \acp{pva}, i.e., actual \acp{va} that generate a measurement for the first time, are modeled by a Poisson point process with mean $\mu_\text{n}$ and PDF $f_\text{n}\big(\overline{\V{x}}^{(j)}_{m,n}|\tilde{\V{x}}_n\big)$. Following \cite{LeitMeyHlaWitTufWin:TWC2019,MeyerProc2018}, newly detected \acp{va} are represented by new \ac{pva} states $\overline{\RV{y}}^{(j)}_{m,n}$, $m \in \{1,\dots,M_n^{(j)}\}$, where each new \ac{pva} state corresponds to a measurement $\RV{z}_{m,n}^{(j)}$; $\overline{r}_{m,n}^{(j)} \rreq 1$ implies that measurement  $\RV{z}_{m,n}^{(j)}$ was generated by a newly detected \ac{va}. 
Since newly detected \acp{va} can potentially produce more than one measurement, we use the multiple-measurement-to-feature probabilistic data association and define this mapping as introduced in \cite{MeyWilJ21,MeyerICASSP2020}. 
We also introduce the joint states $\overline{\RV{y}}_n^{(j)} \!\triangleq \big[\overline{\RV{y}}_{1,n}^{(j)\text{T}} \rmv\cdots\ist \overline{\RV{y}}_{M_{n}^{(j)}\rmv,n}^{(j)\text{T}} \big]^\T\rmv$ and $\overline{\RV{y}}_n \!\triangleq \big[\overline{\RV{y}}_n^{(1)\text{T}} \rmv\cdots\ist \overline{\RV{y}}_n^{(J)\text{T}} \big]^\T\rmv$. 
The vector of all \acp{pva} at time $n$ is given by $\RV{y}_{n} \!\triangleq \big[\underline{\RV{y}}_n^{\text{T}} \,\ist \overline{\RV{y}}_n^{\text{T}}\big]^\T\rmv$. 
Note that the total number of \acp{pva} per PA is given by $K_{n}^{(j)} \rmv= K_{n-1}^{(j)} + M_{n}^{(j)}$. 

Since new \acp{pva} are introduced as new measurements are available at each time, the number of \acp{pva} grows indefinitely. 
Thus, for feasible methods a suboptimal pruning step is employed that removes unlikely \acp{pva} (see Section~\ref{sec:problem}).

\subsection{Association Vectors}
\label{sec:assoc_vec_description}

For each \ac{pa}, measurements $\V{z}_{m,n}^{(j)}$ are subject to a data association uncertainty. 
It is not known which measurement $\V{z}_{m,n}^{(j)}$ is associated with which \ac{pva} $k$, or if a measurement $\V{z}_{m,n}^{(j)}$ did not originate from any \ac{pva} (\emph{false alarm}) or if a \ac{pva} did not give rise to any measurement (\emph{missed detection}). 
The associations between measurements $\V{z}_{m,n}^{(j)}$ and the \acp{pva} at time $n$ is described by the binary \ac{pva}-orientated association variables with entries \cite{MeyerICASSP2020,MeyWilJ21}
\begin{equation}
	\rv{a}_{km,n}^{(j)} \triangleq\rmv \begin{cases} 
		1, &\!\begin{minipage}[t]{65mm}if measurement $m$ was generated by \ac{pva} $k$\end{minipage}\\
		0 \ist, &\!\begin{minipage}[t]{65mm} otherwise\ist.\end{minipage}
	\end{cases}\nn
\end{equation}
We distinguish between legacy and new \ac{pva}-associated variable vectors given, respectively,\vspace*{0.2mm} as $\underline{\RV{a}}_{k,n}^{(j)} \rmv\rmv\triangleq\rmv\rmv [ \underline{\rv{a}}_{k1,n}^{(j)}\iist\cdots\iist \underline{\rv{a}}_{kM_n^{(j)},n}^{(j)}]^\T$ with $k \in \mathcal{K}_{n-1}^{(j)}$ and $\overline{\RV{a}}_{k,n}^{(j)} \rmv\rmv\triangleq\rmv\rmv [ \overline{\rv{a}}_{k1,n}^{(j)}$ $\cdots\iist\overline{\rv{a}}_{kk,n}^{(j)}]^\T$ with $k \rmv\rmv\in\rmv\rmv \mathcal{M}_n^{(j)}$ and $\RV{a}_{k,n}^{(j)} \rmv\rmv\triangleq\rmv\rmv [\underline{\RV{a}}_{k,n}^{(j)\T}\iist\cdots\iist\overline{\RV{a}}_{k,n}^{(j)\T}  ]^\T $ \cite{MeyerICASSP2020}. We also define $\RV{a}_{n}^{(j)} \rmv\rmv\triangleq\rmv\rmv [\RV{a}_{1,n}^{(j)\T}\iist\cdots\iist \RV{a}_{K_n^{(j)},n}^{(j)\T}  ]^\T$ and $\RV{a}_{n} \rmv\rmv\triangleq\rmv\rmv [\RV{a}_{n}^{(1)\T}\iist\cdots\iist \RV{a}_{n}^{(J)\T}  ]^\T \vspace*{0.2mm}$. 
To reduce computational complexity, following \cite{WilliamsLauTAE2014,MeyerProc2018,LeitMeyHlaWitTufWin:TWC2019}, we use the redundant description of association variables, i.e., we introduce measurement-orientated association variable 
\begin{align}
	\rv{b}_{m,n}^{(j)} \triangleq\rmv \begin{cases} 
		k \in \{1,\dots,K_n^{(j)} \}, &\!\begin{minipage}[t]{45mm}if measurement $m$ was \\generated by \ac{pva} $k$\end{minipage}\\
		0 \ist, &\!\begin{minipage}[t]{45mm} otherwise \end{minipage}
	\end{cases}\nn\\[-7mm]\nn
\end{align} 
and define the measurement-oriented association vector $\RV{b}^{(j)}_n = [\rv{b}_{1,n}^{(j)}\iist\cdots\iist\rv{b}_{M_n^{(j)},n}^{(j)}]$. 
We also define $\RV{b}_{n} \triangleq [\RV{b}_{n}^{(1)\T}\iist\cdots\iist \RV{b}_{n}^{(J)\T}  ]^\T$. 
Note that any data association event that can be expressed by both random vectors $\RV{a}_n$ and $\RV{b}_n$ is a valid event, i.e., any measurement can be generated by at most one \ac{pva}. 
This redundant representation of events makes it possible to develop scalable \acp{spa} \cite{WilliamsLauTAE2014, MeyerProc2018, LeitMeyHlaWitTufWin:TWC2019, LiLeiVenTuf:TWC2022}.

\subsection{Joint Posterior \ac{pdf}}
\label{sec:derivationFactorGraph}

By using common assumptions \cite{MeyerProc2018,LeitMeyHlaWitTufWin:TWC2019,LiLeiVenTuf:TWC2022}, and for fixed and thus observed measurements $\V{z}_{1:n}$, it can be shown that the joint posterior PDF of $\tilde{\RV{x}}_{1:n}$ ($\tilde{\RV{x}}_{1:n} \triangleq [\tilde{\RV{x}}_1^\T \cdots \tilde{\RV{x}}_n^\T]^\T$), $\RV{y}_{1:n}$, $ \RV{a}_{1:n}$, and $\RV{b}_{1:n}$, conditioned on $\V{z}_{1:n}$ for all time steps $n' \in \{1,\ist\dots\ist, n \}$ is given by
\begin{align}
&\hspace{-2mm}f( \tilde{\V{x}}_{1:n}, \V{y}_{1:n}, \V{a}_{1:n}, \V{b}_{1:n}| \V{z}_{1:n} ) \nn\\[-1.1mm]
&\hspace{-2mm}\propto  f(\V{x}_{1})f(\V{\psi}_{1}) \Bigg(\prod^{J}_{j'=1} \rmv 
  \prod^{K^{(j')}_{1}}_{k'=1} \! f\big( \underline{\V{y}}^{(j')}_{k'\!,1}\big) \Bigg)  \nn\\[-1.3mm]
&\hspace{-1mm}\times\rmv\ \prod^{n}_{n'=2}  \! f(\V{x}_{n'}|\V{x}_{n'-1}) f(\V{\psi}_{n'}|\V{\psi}_{n'-1})  \nn\\[-1.2mm]
&\hspace{-2mm}\times\rmv\ \prod^{J}_{j=1}  \Bigg(\prod^{K^{(j)}_{n'-1}}_{k=1} g\big(\underline{\V{y}}^{(j)}_{k,n'} \big| \V{y}^{(j)}_{k,n'-1},\tilde{\V{x}}_{n'-1}\big)\nn\\[-1.3mm]
&\hspace{-2mm}\times \prod^{M^{(j)}_{n'}}_{m'=1} \rmv q\big( \tilde{\V{x}}_{n'}, \underline{\V{y}}^{(j)}_{k,n'}, \underline{a}^{(j)}_{km',n'}; \V{z}^{(j)}_{m',n'} \big) \underline{\Psi}\big(\underline{a}^{(j)}_{km',n'} \rmv,b^{(j)}_{m',n'}\big) \Bigg) \nn\\[-1mm]
&\hspace{-2mm}\times \Bigg( \prod^{M^{(j)}_{n'}}_{m=1}  v\big( \tilde{\V{x}}_{n'}, \overline{\V{y}}^{(j)}_{m,n'}, \overline{a}^{(j)}_{mm,n'}; \V{z}^{(j)}_{m,n'} \big) \nn \\[-1mm]
& \hspace{-2mm} \times\hspace{-1mm} \prod^{m-1}_{h=1} u\big( \tilde{\V{x}}_{n'}, \overline{\V{y}}^{(j)}_{m,n'}, \overline{a}^{(j)}_{mh,n'}; \V{z}^{(j)}_{h,n'} \big) \overline{\Psi}(\overline{a}^{(j)}_{mh,n'},b^{(j)}_{h,n'}) \Bigg) 
\label{eq:factorization_post} 
\end{align}
where $g(\underline{\V{y}}^{(j)}_{k,n}|\underline{\V{y}}^{(j)}_{k,n-1},\xn{n-1})$, $q\big( \tilde{\V{x}}_{n}, \underline{\V{y}}^{(j)}_{k,n}, \underline{a}^{(j)}_{km,n}; \V{z}^{(j)}_{m,n} \big)$, $\Psi(a^{(j)}_{km,n},b^{(j)}_{m,n})$, $u\big( \tilde{\V{x}}_{n}, \overline{\V{y}}^{(j)}_{k,n}, \overline{a}^{(j)}_{mh,n}; \V{z}^{(j)}_{h,n} \big)$ and $ v\big(  \tilde{\V{x}}_{n}, \overline{\V{y}}^{(j)}_{m,n}, \overline{a}^{(j)}_{mm,n}; \V{z}^{(j)}_{m,n} \big)$ are explained in what follows. 
The \emph{pseudo state-transition function} is given by
\begin{align}
  &\hspace*{-4mm}g(\underline{\V{y}}^{(j)}_{k,n}|\underline{\V{y}}^{(j)}_{k,n-1},\xn{n-1}) \nn\\
  &\hspace*{-2mm}\triangleq\rmv\rmv 
    \begin{cases}
      e^{-\mumL{k}{n-1}}  f(\underline{\V{x}}^{(j)}_{k,n},1 |\underline{\V{x}}^{(j)}_{k,n-1},\underline{r}^{(j)}_{k,n-1}), &\hspace*{-2mm} \underline{r}^{(j)}_{k,n} \rreq 1 \\[1mm]
      f(\underline{\V{x}}^{(j)}_{k,n},0 |\underline{\V{x}}^{(j)}_{k,n-1},\underline{r}^{(j)}_{k,n-1}), &\hspace*{-2mm} \underline{r}^{(j)}_{k,n} \rreq 0
    \end{cases} \label{eq:g}      
\end{align}
and the \emph{pseudo prior distribution} as
\vspace*{-1mm}
\begin{align}
\hspace*{-2.5mm}f(\overline{\V{y}}^{(j)}_{k,n}|\xn{n}) \rmv\rmv\triangleq \rmv\rmv
    \begin{cases}
      \mu_n f_n\big(\overline{\V{x}}^{(j)}_{k,n}|\tilde{\V{x}}_n\big) e^{-\mumN{k}{n}}, &\hspace*{-2mm}\overline{r}^{(j)}_{k,n} \rreq 1 \\[1mm]
      f_d\big(\overline{\V{x}}^{(j)}_{k,n}\big), &\hspace*{-2mm}\overline{r}^{(j)}_{k,n} \rreq 0 \ist.
    \end{cases} \label{eq:fy}     
\end{align}

The \emph{pseudo likelihood functions} related to legacy \acp{pva} for $k \in \Set{K}_{n-1}^{(j)}$ $q\big( \tilde{\V{x}}_n, \underline{\V{y}}^{(j)}_{k,n}, \underline{a}^{(j)}_{km,n}; \V{z}^{(j)}_{m,n} \big) = q\big( \tilde{\V{x}}_n, \underline{\V{x}}^{(j)}_{k,n}, \underline{r}^{(j)}_{k}, \underline{a}^{(j)}_{km,n}; \V{z}^{(j)}_{m,n} \big)$ is given by
\vspace*{-1mm}
\begin{align}
&q\big( \tilde{\V{x}}_n, \underline{\V{x}}^{(j)}_{k,n}, 1, \underline{a}^{(j)}_{km,n}; \V{z}^{(j)}_{m,n} \big) \nn\\
&\hspace*{5mm}\triangleq 
    \begin{cases}
      \frac{\mumL{k}{n} f (\V{z}^{(j)}_{m,n}| \V{p}_{n}, \V{\psi}_{n}, \underline{\V{x}}_{k,n}^{(j)})}{\mu_\text{fa} f_\text{fa}(\V{z}^{(j)}_{m,n})}, &  \underline{a}^{(j)}_{km,n} \rreq  1\\
      1, &  \underline{a}^{(j)}_{km,n} \rreq 0\\
    \end{cases}  \label{eq:q} 
\\[-7mm]\nn     
\end{align}
and $q\big( \tilde{\V{x}}_n, \underline{\V{x}}^{(j)}_{k,n}, 0, \underline{a}^{(j)}_{km,n}; \V{z}^{(j)}_{m,n} \big) \triangleq \delta_{\underline{a}^{(j)}_{km,n}}$. 
The \emph{pseudo likelihood functions} related to a new \ac{pva} (with $k \in \Set{M}_{n}^{(j)} \backslash m$) is given as
$u\big( \tilde{\V{x}}_n, \overline{\V{y}}^{(j)}_{k,n}, \overline{a}^{(j)}_{km,n}; \V{z}^{(j)}_{m,n} \big) = u\big( \tilde{\V{x}}_n, \overline{\V{x}}^{(j)}_{k,n}, \overline{r}^{(j)}_{k}, \overline{a}^{(j)}_{km,n}; \V{z}^{(j)}_{m,n} \big)$ is given by
\vspace*{-1mm}
\begin{align}
&u\big( \tilde{\V{x}}_n, \overline{\V{x}}^{(j)}_{k,n}, 1, \overline{a}^{(j)}_{km,n}; \V{z}^{(j)}_{m,n} \big) \nn\\
&\hspace*{2mm}\triangleq 
    \begin{cases}
       \frac{f(\overline{\V{y}}^{(j)}_{k,n}|\xn{n}) \mumN{k}{n} f(\V{z}^{(j)}_{m,n}| \V{p}_{n}, \V{\psi}_{n}, \overline{\V{x}}_{k,n}^{(j)})}{\mu_\text{fa} f_\text{fa}(\V{z}^{(j)}_{m,n})}, &  \overline{a}^{(j)}_{km,n} \rreq  1\\
     1, &  \overline{a}^{(j)}_{km,n} \rreq 0\\
    \end{cases}  \label{eq:u}  
\\[-7mm]\nn     
\end{align}
and $u\big( \tilde{\V{x}}_n, \overline{\V{x}}^{(j)}_{k,n}, 0, \overline{a}^{(j)}_{km,n}; \V{z}^{(j)}_{m,n} \big) \triangleq \delta_{\overline{a}^{(j)}_{km,n}}$, whereas for $k=m$ as $v\big(\tilde{\V{x}}_n, \overline{\V{y}}^{(j)}_{m}, \overline{a}^{(j)}_{mm,n}; \V{z}^{(j)}_{m,n} \big)  = v\big(\tilde{\V{x}}_n, \overline{\V{x}}^{(j)}_{m,n},\overline{r}^{(j)}_{m,n}, \overline{a}^{(j)}_{mm,n}; \V{z}^{(j)}_{m,n} \big)$ is given by
\vspace*{-1.5mm}
\begin{align}
  &v\big(\tilde{\V{x}}_n, \overline{\V{x}}^{(j)}_{m,n},1 , \overline{a}^{(j)}_{mm,n}; \V{z}^{(j)}_{m,n} \big) \nn\\
  &\hspace*{1mm}\triangleq 
    \begin{cases}
      \frac{f(\overline{\V{y}}^{(j)}_{m,n}|\xn{n}) \mumN{m}{n} f\big(\V{z}^{(j)}_{m,n}|\V{p}_{n},\V{\psi}_{n},\overline{\V{x}}_{m,n}^{(j)}\big)}{\mu_\text{fa} f_\text{fa}(\V{z}^{(j)}_{m,n})}, & \rrmv \rrmv \overline{a}^{(j)}_{mm,n} \rreq 1\\
      0,  & \rrmv \rrmv \overline{a}^{(j)}_{mm,n} \rreq 0\\
    \end{cases} \label{eq:v}  \\[-7.5mm]\nn
\end{align}
and $v\big(\tilde{\V{x}}_n, \overline{\V{x}}^{(j)}_{m,n},0 , \overline{a}^{(j)}_{mm,n}; \V{z}^{(j)}_{m,n} \big) \triangleq \delta_{\overline{a}^{(j)}_{mm,n}}$.

Finally, the binary \textit{indicator functions} that check consistency for any pair $(a^{(j)}_{km,n},b^{(j)}_{m,n})$ of \ac{pva}-oriented and measurement-oriented association variable at time $n$ are, respectively, given by
\vspace*{-1.5mm}
\begin{align}
  &\underline{\Psi}(\underline{a}^{(j)}_{km,n},b^{(j)}_{m,n}) \nn\\
  &\hspace*{5mm}\triangleq 
    \begin{cases}
      0, & \hspace{-2mm} \underline{a}^{(j)}_{km,n} \rreq 1, \ b^{(j)}_{m,n} \rrmv \neq \rrmv k \ \text{or } \underline{a}^{(j)}_{km,n} \rreq 0, \ b^{(j)}_{m,n} \rreq k\\
     1, & \hspace{-2mm}  \text{else } 
    \end{cases} \label{eq:psi_legacy}\\[-7.5mm]\nn
\end{align}
for $k \in \Set{K}^{(j)}_{n-1}$ and
\vspace*{-1.5mm}
\begin{align}
  &\hspace*{-2mm}\overline{\Psi}(\overline{a}^{(j)}_{km,n},b^{(j)}_{m,n})  \rmv\rmv\triangleq\rmv\rmv 
    \begin{cases}
      0, & \begin{minipage}[t]{45mm}$\overline{a}^{(j)}_{km,n} \rreq 1$,  $b^{(j)}_{m,n}  \neq  K_{n-1}^{(j)} \rmv+\rmv k$ \\
      	 or $\overline{a}^{(j)}_{km,n} \rreq 0$, $b^{(j)}_{m,n}  \rreq K_{n-1}^{(j)} \rmv + \rmv k$ \end{minipage}\\
     1, & \text{else}\ist. 
    \end{cases} \label{eq:psi_new}\\[-7.5mm]\nn   
\end{align}
for $k \in \Set{M}^{(j)}_{n}$. 
The factor graph representing the factorization \eqref{eq:factorization_post} is shown in Fig.~\ref{fig:FG}.

\subsection{Detection of \acp{pva} and State Estimation}\label{sec:problem}

\begin{figure}[!t]
\ifthenelse{\equal{\arxiv}{false}}
		{  
	\tikzsetnextfilename{FG_EOT_SLAM}
	\setlength{\abovecaptionskip}{-0.5mm}
	\setlength{\belowcaptionskip}{0pt}
	\hspace*{-3.2cm}\scalebox{0.97}{\input{plots/FG_EOT_SLAMv3.tex}} }
	{
	\includegraphics[scale=1]{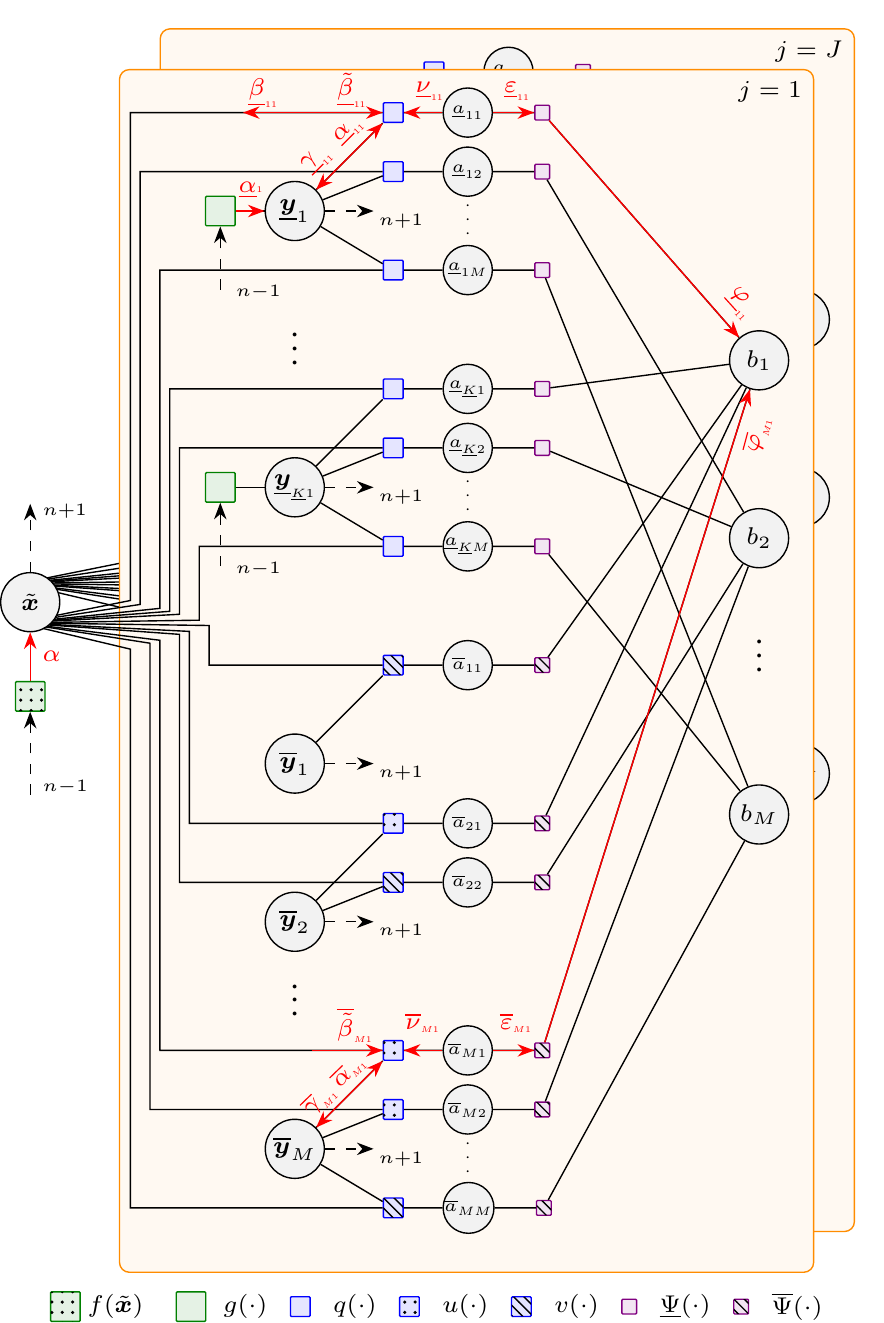} }
	\vspace*{-2.2mm}
	\caption{Factor graph for proposed algorithm. At MP iteration $p$, we use the following short hand notation: $f(\tilde{\V{x}}) \triangleq f(\tilde{\V{x}}_n|\tilde{\V{x}}_{n-1})$, $g(\cdot)$, $q(\cdot)$, $u(\cdot)$, $v(\cdot)$, $\underline{\Psi}(\cdot)$ and  $\overline{\Psi}(\cdot)$  corresponds to \eqref{eq:g}, \eqref{eq:q}, \eqref{eq:u}, \eqref{eq:v}, \eqref{eq:psi_legacy} and \eqref{eq:psi_new}, respectively. Furthermore, we define $\alpha \triangleq \alpha(\tilde{\V{x}}_n)$,  $\underline{\alpha}_k \triangleq \alpha(\underline{\V{x}}^{(j)}_{k,n},\underline{r}^{(j)}_{k,n})$, $\underline{\alpha}_{kl} \triangleq \alpha_l(\underline{\V{x}}^{(j)}_{k,n},\underline{r}^{(j)}_{k,n})$, $\overline{\alpha}_{kl} \triangleq \alpha_l(\overline{\V{x}}^{(j)}_{k,n},\overline{r}^{(j)}_{k,n})$, $\underline{\varepsilon}_{kl} \triangleq \varepsilon(\underline{a}^{(j)}_{kl,n})$, $\overline{\varepsilon}_{kl} \triangleq \varepsilon(\overline{a}^{(j)}_{kl,n})$, $\underline{\gamma}_{kl} \triangleq \gamma_l(\underline{\V{x}}^{(j)}_{k,n},\underline{r}^{(j)}_{k,n})$, $\overline{\gamma}_{kl} \triangleq \gamma_l(\overline{\V{x}}^{(j)}_{k,n},\overline{r}^{(j)}_{k,n})$, $\underline{\nu}_{kl} \triangleq \underline{\nu}_{kl}(\underline{a}^{(j)}_{kl,n})$, $\overline{\nu}_{kl} \triangleq \overline{\nu}_{kl}(\overline{a}^{(j)}_{kl,n})$, $\underline{\varphi}_{kl} \triangleq \underline{\varphi}_{kl}(b_{l,n})$ and $\overline{\varphi}_{kl} \triangleq \overline{\varphi}_{kl}(b_{l,n})$. Due to our proposed scheduling, both $\tilde{\beta}_{kl}$ and $\overline{\tilde{\beta}}_{ml}$ are defined to be $\alpha(\tilde{\V{x}}_n)$ according to \eqref{eq:betaTilde}. Furthermore, $\overline{\beta}_{ml} \triangleq 1$ and $\underline{\beta}_{kl} \triangleq \beta^{(j)}_{kl}(\tilde{\V{x}}_n)$ since the augmented agent state is only updated with messages from legacy \acp{pva}. The time evolution of the agent state and \acp{va} is indicated with dashed arrows.}
	\label{fig:FG}
\end{figure}

We aim to estimate all states using all available measurements $\V{z}_{1:n}\rmv$ from all \acp{pa} up to time $n$. 
In particular, we calculate estimates of the augmented agent state (containing the dispersion parameters) $\tilde{\RV{x}}_n$ by using the \ac{mmse} estimator \vspace{.5mm} \cite[Ch.~4]{Poo:B94}, i.e., \vspace*{-1.5mm}
\begin{align}
	\tilde{\V{x}}^{\text{MMSE}}_{n} &\triangleq\, \int \rmv \tilde{\V{x}}_n \ist f( \tilde{\V{x}}_n|\V{z}_{1:n}) \ist \mathrm{d} \tilde{\V{x}}_n \label{eq:MMSEagent}\\[-6mm]\nn
\end{align}
where $\tilde{\V{x}}^{\text{MMSE}}_{n} = [\V{x}^{\text{MMSE}\iist\T}_{n} \V{\psi}^{\text{MMSE}\iist\T}_{n}]^{\T}$.
The map of the environment is represented by reflective surfaces described by \acp{pva}. 
Therefore, the state $\V{x}_{k,n}^{(j)}$ of the detected \acp{pva} $k \!\in\! \{ 1,\dots,K^{(j)}_n \}$ must be estimated. 
This relies on the marginal posterior existence probabilities $p(r^{(j)}_{k,n} \!=\! 1|\V{z}_{1:n}) = \int f(\V{x}_{k,n}^{(j)} , r^{(j)}_{k,n} \!=\! 1| \V{z}^{(j)}_{1:n} ) \mathrm{d}\V{x}_{k,n}^{(j)}$ and the marginal posterior \acp{pdf} $f(\V{x}_{k,n}^{(j)}| r^{(j)}_{k,n} \!=\! 1, \V{z}_{1:n} ) \rmv\rmv=\rmv\rmv f(\V{x}_{k,n}^{(j)}, r^{(j)}_{k,n} \!=\! 1| \V{z}_{1:n} )/p(r^{(j)}_{k,n} \!=\! 1|\V{z}_{1:n})$. 
A \ac{pva} $k$ is declared to exist if $p(r^{(j)}_{k,n} \!=\! 1|\V{z}_{1:n}) > p_{\text{cf}}$, where $p_{\text{cf}}$ is a confirmation threshold \cite[Ch.~2]{Poo:B94}. 
To avoid that the number of \ac{pva} states grows indefinitely, \ac{pva} states with $p(r^{(j)}_{k,n} \!=\! 1|\V{z}_{1:n})$ below a threshold $p_{\text{pr}}$ are removed from the state space (``pruned''). 
The number $\hat{K}^{(j)}_n$ of \ac{pva} states that are considered to exist is the estimate of the total number $L_n^{(j)}$ of \acp{va} visible at time $n$. 
For existing \acp{pva}, an estimate of its state $\RV{x}_{k,n}^{(j)}$ can again be calculated by the \ac{mmse}
\vspace*{-1.5mm}
\begin{align}
	\V{x}_{k,n}^{(j)\ist\text{MMSE}}  \,\triangleq \int \rmv \V{x}_{k,n}^{(j)}  \ist\ist f(\V{x}_{k,n}^{(j)}\ist | \ist r^{(j)}_{k,n} \!=\! 1, \V{z}_{1:n}) \ist\ist \mathrm{d}\V{x}_{k,n}^{(j)} \rmv. \label{eq:MMSEpva}\\[-7.5mm]\nn
\end{align}
The calculation of $f( \tilde{\V{x}}_n|\V{z}_{1:n})$, $p(r_{k,n} \rmv\rmv=\rmv\rmv 1 |\V{z})$, and $f(\V{x}_{k,n}^{(j)} |$ $r^{(j)}_{k,n} \rmv\rmv=\rmv\rmv 1, \V{z}_{1:n})$ from the joint posterior $f( \tilde{\V{x}}_{1:n}, \V{y}_{1:n},\V{a}_{1:n},$ $\V{b}_{1:n}| \V{z}_{1:n} )$ by direct marginalization is not feasible. 
By performing sequential particle-based \ac{MP} using the SPA rules \cite{KscFreLoe:TIT2001, MeyHliHla:TSPIN2016, MeyerTSP2017, LeitMeyHlaWitTufWin:TWC2019, LeiGreWit:ICC2019,VenLeiTerWit:TWC2023} on the factor graph in Fig.~\ref{fig:FG}, approximations (``beliefs'') $b\big(\tilde{\V{x}}_{n} \big)$ and $b\big(\V{y}^{(j)}_{k,n} \big)$ of the marginal posterior \acp{pdf} $f( \tilde{\V{x}}_n|\V{z}_{1:n})$, $p(r^{(j)}_{k,n} \!=\! 1 |\V{z}_{1:n})$, and $f(\V{x}_{k,n}^{(j)}|$ $r^{(j)}_{k,n} \rmv=\rmv 1, \V{z}_{1:n})$ can be obtained in an efficient way for the agent state as well as all legacy and new \ac{pva} states. 

%% file: inputFiles/SPA.tex

\section{Proposed Sum-Product Algorithm}
\label{sec:SPA}

The factor graph in Fig.~\ref{fig:FG} has cycles, therefore we have to decide on a specific order of message computation \cite{KscFreLoe:TIT2001,Loe:SMP2004_FG}. We use iterative \ac{MP} with \ac{MP} iteration $p \in \{1, \dots, P\}$ where $P$ is the maximum number of \ac{MP} iterations. 
We choose the order according to the following rules: (i) messages are only sent forward in time; (ii) for each PA, messages are updated in parallel; (iii) along an edge connecting the augmented agent state variable node and a new \ac{pva}, messages are only sent from the former to the latter; (iv) the augmented agent state variable node is only updated at \ac{MP} iteration $P$. The corresponding messages are shown in Fig.~\ref{fig:FG}. Note, that this scheduling is suboptimal since the extrinsic messages of the augmented agent state are neglected. This calculation order is solely chosen to reduce the computational demand. With these rules, the message passing equations of the \ac{spa} \cite{KscFreLoe:TIT2001} yield the following operations at each time step.

\subsection{Prediction Step}
A prediction step is performed for the augmented agent state and all legacy \acp{va} $k \in \Set{K}_{n-1}^{(j)}$. 
It has the form of
\vspace*{-1mm}
\begin{align}
\alpha(\tilde{\V{x}}_n) &=  \int f(\tilde{\V{x}}_n|\tilde{\V{x}}_{n-1}) b(\tilde{\V{x}}_{n-1}) \mathrm{d}\tilde{\V{x}}_{n-1} \label{eq:mp_agentPredic}\\
\alpha(\underline{\V{x}}^{(j)}_{k,n},\underline{r}^{(j)}_{k,n}) &=  \hspace{-2mm} \hspace{-4mm}\sum_{r^{(j)}_{k,n-1} \rmv \in \{0,1\}} \hspace{-5mm} \iint g(\underline{\V{x}}^{(j)}_{k,n},\underline{r}^{(j)}_{k,n}|\V{x}^{(j)}_{k,n-1},r^{(j)}_{k,n-1},\xn{n-1}) \nn \\
	& \hspace*{-5mm}\times b(\V{x}^{(j)}_{k,n-1},r^{(j)}_{k,n-1})  b(\tilde{\V{x}}_{n-1}) \mathrm{d}\V{x}^{(j)}_{k,n-1} \mathrm{d}\tilde{\V{x}}_{n-1} \label{eq:mp_VaPredic}
\end{align}
with $b(\tilde{\V{x}}_{n-1})$ and $b(\V{x}^{(j)}_{k,n-1},r^{(j)}_{k,n-1})$ denoting the beliefs of the augmented agent state and the legacy \ac{va} $k$ calculated at the previous time step, respectively. 
The summation in \eqref{eq:mp_VaPredic}, can be further written as
\begin{align}
\alpha(\underline{\V{x}}^{(j)}_{k,n},\underline{r}^{(j)}_{k,n} \rreq 1) &= p_\text{s} \rmv\rmv\rmv \iint \rmv\rmv\rmv e^{-\mumL{k}{n-1}}\rmv\rmv f(\underline{\V{x}}^{(j)}_{k,n},1|\V{x}^{(j)}_{k,n-1},1) \nn \\
	& \times b(\V{x}^{(j)}_{k,n-1},1) b(\xn{n-1}) \mathrm{d}\V{x}^{(j)}_{k,n-1} \mathrm{d}\xn{n-1}
\end{align} 
and $\alpha(\underline{\V{x}}^{(j)}_{k,n},\underline{r}^{(j)}_{k,n} \rreq 0) = \underline{\alpha}^{\text{n},(j)}_k f_\text{d}(\underline{\V{x}}^{(j)}_{k,n})$ with 
\vspace*{-1mm}
\begin{align}
	\underline{\alpha}^{\text{n},(j)}_k & \triangleq \tilde{b}_{k,n-1} + (1-p_\text{s}) \int b(\Lxn{n-1},1) \mathrm{d}\Lxn{n-1} \nn \\
	& = \tilde{b}_{k,n-1} + (1-p_\text{s})(1-\tilde{b}_{k,n-1})
\end{align}
where $\tilde{b}_{k,n-1} = \int b(\Lxn{n-1},0) \mathrm{d}\Lxn{n-1}$ approximates the probability of non-existence of legacy \ac{va} $k$. 

\subsection{Measurement Evaluation}
The messages $\varepsilon^{[p]}(\al)$ sent from factor nodes $q(\x,\underline{\V{y}}^{(j)}_{k,n},\al,\V{z}^{(j)}_{l,n})$ to variable nodes $\al$ at \ac{MP} iteration $p$ with $k \in \{1,\dots,K_{n-1}^{(j)}\}$ and $l \in \{1,\dots,M_n^{(j)}\}$ are defined as
\begin{align}
	\varepsilon^{[p]}(\al) =&   \iint \tilde{\beta}^{[p]}_{kl}(\tilde{\V{x}}_n) \alpha^{[p]}_l(\underline{\V{y}}^{(j)}_{k,n}) \nn \\
	& \times q(\x,\underline{\V{y}}^{(j)}_{k,n},\al,\V{z}^{(j)}_{l,n}) \mathrm{d}\tilde{\V{x}}_n \mathrm{d}\underline{\V{y}}^{(j)}_{k,n}.
	\label{eq:message_epsilon1}
\end{align}
The messages from factor nodes $u(\x,\overline{\V{y}}^{(j)}_{k,n},\an{k}{l},\V{z}^{(j)}_{l,n})$ to variable nodes $\an{k}{l}$ where $k \in\{1,\dots,M_n^{(j)}\}$ and $l \in \{1,\dots,M_n^{(j)} \}\backslash k $, are given as
\begin{align}
	\varepsilon^{[p]}(\an{k}{l}) =&   \iint \tilde{\beta}^{[p]}_{kl}(\tilde{\V{x}}_n) \alpha^{[p]}_l(\overline{\V{y}}^{(j)}_{k,n}) \nn \\
	& \times u(\tilde{\V{x}}_n,\overline{\V{y}}^{(j)}_{k,n},\an{k}{l},\V{z}^{(j)}_{l,n}) \mathrm{d}\tilde{\V{x}}_n \mathrm{d}\overline{\V{y}}^{(j)}_{k,n}
	\label{eq:message_epsilon2}
\end{align}
and the messages from factor nodes $v(\x,\overline{\V{y}}^{(j)}_{m,n},\overline{a}^{(j)}_{mm,n},\V{z}^{(j)}_{m,n})$ to variable nodes $\overline{a}^{(j)}_{mm,n}$, $m \in\{1,\dots,M_n^{(j)}\}$, are given as
\begin{align}
	\varepsilon^{[p]}(\overline{a}^{(j)}_{mm,n}) =&   \iint \tilde{\beta}^{[p]}_{mm}(\tilde{\V{x}}_n) \alpha^{[p]}_m(\overline{\V{y}}^{(j)}_{m,n}) \nn \\
	& \times v(\tilde{\V{x}}_n,\overline{\V{y}}^{(j)}_{m,n},\overline{a}^{(j)}_{mm,n},\V{z}^{(j)}_{m,n}) \mathrm{d}\tilde{\V{x}}_n \mathrm{d}\overline{\V{y}}^{(j)}_{m,n}.
	\label{eq:message_epsilon3}
\end{align}
Note that $\alpha^{[p=1]}_l(\underline{\V{y}}^{(j)}_{k,n}) \triangleq \alpha(\underline{\V{x}}^{(j)}_{k,n},\underline{r}^{(j)}_{k,n})$ and $\alpha^{[p=1]}_l(\overline{\V{y}}^{(j)}_{k,n}) \triangleq 1$. 
For $p>1$, $\alpha^{[p]}_l({\V{y}}^{(j)}_{k,n})$ is calculated according to Section~\ref{sec:ExtrInfo}. 
The message $\tilde{\beta}^{[p]}_{kl}(\tilde{\V{x}}_n)$ will be defined in Section~\ref{sec:MP_updateAgent}.
Using \eqref{eq:message_epsilon1}, $\varepsilon^{[p]}(\al)$ is further investigated. 
For the messages containing information about legacy \acp{va}, it results in
\begin{align}
	\varepsilon^{[p]}(\underline{a}^{(j)}_{kl,n} \rreq 1) =& \iint \tilde{\beta}^{[p]}_{kl}(\tilde{\V{x}}_n) \alpha^{[p]}_l(\underline{\V{x}}^{(j)}_{k,n},\underline{r}^{(j)}_{k,n} \rreq 1)  \nn \\
	& \times \frac{\mu_\text{m}\big(\tilde{\V{x}}_n,\underline{\V{x}}^{(j)}_{k,n} \big) f(\V{z}^{(j)}_{l,n}|\tilde{\V{x}}_n,\underline{\V{x}}^{(j)}_{k,n})}{\mu_\text{fa} f_\text{fa}(\V{z}^{(j)}_{l,n})} \mathrm{d}\underline{\V{x}}^{(j)}_{k,n} \mathrm{d}\tilde{\V{x}}_n \nonumber \\
	\varepsilon^{[p]}(\underline{a}^{(j)}_{kl,n} \rreq 0) =& \iint  \tilde{\beta}^{[p]}_{kl}(\tilde{\V{x}}_n) \Big( \alpha^{[p]}_l(\underline{\V{x}}^{(j)}_{k,n},\underline{r}^{(j)}_{k,n} \rreq 1)  \nn \\
	&   \hspace*{4mm} + \alpha^{[p]}_l(\underline{\V{x}}^{(j)}_{k,n},\underline{r}^{(j)}_{k,n} \rreq 0)\Big) \mathrm{d}\underline{\V{x}}^{(j)}_{k,n} \mathrm{d}\tilde{\V{x}}_n.
	\label{eq:epsilon_legacy}
\end{align}
This can be further simplify by dividing both messages by $\varepsilon^{[p]}(\underline{a}^{(j)}_{kl,n} = 0)$. 
With an abuse of notation, it results in $\varepsilon^{[p]}(\underline{a}^{(j)}_{kl,n} \rreq 0) = 1$. 

The messages $\varepsilon^{[p]}(\overline{a}^{(j)}_{kl,n})$ can be obtained similarly by using \eqref{eq:message_epsilon2} and \eqref{eq:message_epsilon3}, yielding
\begin{align}
&\varepsilon^{[p]}(\overline{a}^{(j)}_{kl,n} \rreq 1) = \iint \tilde{\beta}^{[p]}_{kl}(\tilde{\V{x}}_n) \alpha^{[p]}_l(\overline{\V{x}}^{(j)}_{k,n},\overline{r}^{(j)}_{k,n} \rreq 1) \nn \\
	& \hspace{2mm}  \times \frac{f(\Nx{k}|\x) \mumN{k}{n} f(\V{z}^{(j)}_{l,n}|\tilde{\V{x}}_n,\overline{\V{x}}^{(j)}_{k,n})}{\mu_\text{fa} f_\text{fa}(\V{z}^{(j)}_{l,n})} \mathrm{d}\overline{\V{x}}^{(j)}_{k,n} \mathrm{d}\tilde{\V{x}}_n \\
&\varepsilon^{[p]}(\overline{a}^{(j)}_{kl,n} \rreq 0) = \iint \tilde{\beta}^{[p]}_{kl}(\tilde{\V{x}}_n) \Big( \alpha^{[p]}_l(\overline{\V{x}}^{(j)}_{k,n},\overline{r}^{(j)}_{k,n} \rreq 1)  \nn \\
	& \hspace{23mm}  + \alpha^{[p]}_l(\overline{\V{x}}^{(j)}_{k,n},\overline{r}^{(j)}_{k,n} \rreq 0)  \Big) \mathrm{d}\overline{\V{x}}^{(j)}_{k,n} \mathrm{d}\tilde{\V{x}}_n
\end{align}
\begin{align}
&\varepsilon^{[p]}(\overline{a}^{(j)}_{mm,n} \rreq 1)  =  \iint \tilde{\beta}^{[p]}_{mm}(\tilde{\V{x}}_n) \alpha^{[p]}_m(\overline{\V{x}}^{(j)}_{m,n},\overline{r}^{(j)}_{m,n} \rreq 1)  \nn \\
 &\hspace{0mm}\times \frac{f(\Nx{m}|\x) \mu_\text{m}\big(\tilde{\V{x}}_n,\overline{\V{x}}^{(j)}_{m,n} \big) f(\V{z}^{(j)}_{m,n}|\tilde{\V{x}}_n,\overline{\V{x}}^{(j)}_{m,n})}{\mu_\text{fa} f_\text{fa}(\V{z}^{(j)}_{m,n})} \mathrm{d}\overline{\V{x}}^{(j)}_{m,n} \mathrm{d}\tilde{\V{x}}_n\\
& \varepsilon^{[p]}(\overline{a}^{(j)}_{mm,n} \rreq 0) \rmv\rmv=\rmv\rmv\rmv\rmv\rmv  \nn \\
&\hspace{10mm} \iint \rmv\rmv \tilde{\beta}^{[p]}_{mm}(\tilde{\V{x}}_n) \alpha^{[p]}_m(\overline{\V{x}}^{(j)}_{m,n},\overline{r}^{(j)}_{m,n} \rreq 0)  \mathrm{d}\overline{\V{x}}^{(j)}_{m,n} \mathrm{d}\tilde{\V{x}}_n
\end{align}
The expressions can be simplified by dividing all messages by $\varepsilon(\overline{a}^{(j)}_{kl,n} = 0)$. With an abuse of notation, it results in $\varepsilon(\overline{a}^{(j)}_{kl,n} = 0) = 1$ and
\begin{align}
&\varepsilon^{[p]}(\overline{a}^{(j)}_{mm,n} \rreq 0)   \nn \\
&\hspace*{0mm}	=\frac{\iint  \tilde{\beta}^{[p]}_{mm}(\tilde{\V{x}}_n) \alpha^{[p]}_m(\overline{\V{x}}^{(j)}_{m,n}, 0) \mathrm{d}\overline{\V{x}}^{(j)}_{m,n} \mathrm{d}\tilde{\V{x}}_n}{\iint  \tilde{\beta}^{[p]}_{mm}(\tilde{\V{x}}_n) \Big(\alpha^{[p]}_m(\overline{\V{x}}^{(j)}_{m,n},1) + \alpha^{[p]}_m(\overline{\V{x}}^{(j)}_{m,n}, 0)\Big)\mathrm{d}\overline{\V{x}}^{(j)}_{m,n} \mathrm{d}\tilde{\V{x}}_n}
\end{align} 

\subsection{Data Association}

The messages $\varphi^{[p]}_{kl}(b^{(j)}_{l,n})$ sent from factor node $\Psi({a}^{(j)}_{kl},b^{(j)}_l)$ to variable node $b^{(j)}_{l,n}$ and the message $\nu^{[p]}_{kl}(a^{(j)}_{kl,n})$ sent from factor node  $\Psi({a}^{(j)}_{kl},b^{(j)}_l)$ to variable node $a^{(j)}_{kl,n}$ are calculated using the measurement evaluation messages in  \eqref{eq:message_epsilon1}, \eqref{eq:message_epsilon2} and \eqref{eq:message_epsilon3}. Details can be found in Appendix~\ref{sec:DA}.

\subsection{Measurement update for \acp{pva}}
Next, we determine the messages sent from factor node $q(\tilde{\V{x}}_n,\underline{\V{y}}^{(j)}_{k,n},\underline{a}^{(j)}_{kl},\V{z}^{(j)}_{l,n})$ to variable node $\underline{\V{y}}^{(j)}_{k,n}$ as
\vspace*{-1mm}
\begin{align}
\gamma^{[p]}_l(\underline{\V{y}}^{(j)}_{k,n}) &= \hspace{-3mm} \sum_{\underline{a}^{(j)}_{kl,n} \in \{0,1\}} \hspace{-1mm}  \int  q(\tilde{\V{x}}_n,\underline{\V{x}}^{(j)}_{k,n},\underline{r}^{(j)}_{k,n},\underline{a}^{(j)}_{kl,n},\V{z}^{(j)}_{l,n})\nn\\
&\hspace*{4mm}\times \underline{\nu}^{[p]}_{kl}(\underline{a}^{(j)}_{kl,n}) \mathrm{d}\tilde{\V{x}}_n\\[-7mm]\nn
\end{align}
which results after marginalizing $\underline{a}^{(j)}_{kl,n}$ in 
\begin{align}
\gamma^{[p]}_l(\underline{\V{x}}^{(j)}_{k,n},\underline{r}^{(j)}_k \rreq 1) &= \int q(\tilde{\V{x}}_n,\underline{\V{x}}^{(j)}_{k,n},1,1,\V{z}^{(j)}_{l,n})\underline{\nu}^{[p]}_{kl}(1) \mathrm{d}\tilde{\V{x}}_n \nn \\
&\hspace*{4mm} + \underline{\nu}^{[p]}_{kl}(0) \\
\gamma^{[p]}_l(\underline{\V{x}}^{(j)}_{k,n},\underline{r}^{(j)}_k \rreq 0) &=  \underline{\nu}^{[p]}_{kl}(0)\ist.
\end{align}
The messages from factor node $u(\tilde{\V{x}}_n,\overline{\V{y}}^{(j)}_{k,n},\overline{a}^{(j)}_{kl},\V{z}^{(j)}_{l,n})$ to variable node $\overline{\V{y}}^{(j)}_{k,n}$ are given as
\begin{align}
\gamma^{[p]}_l(\overline{\V{y}}^{(j)}_{k,n}) &= \hspace{-3mm} \sum_{\overline{a}^{(j)}_{kl,n} \in \{0,1\}} \hspace{-1mm}  \int  u(\tilde{\V{x}}_n,\overline{\V{x}}^{(j)}_{k,n},\overline{r}^{(j)}_{k,n},\overline{a}^{(j)}_{kl,n},\V{z}^{(j)}_{l,n})\nn\\
&\hspace*{4mm}\times \overline{\nu}^{[p]}_{kl}(\overline{a}^{(j)}_{kl,n}) \mathrm{d}\tilde{\V{x}}_n\\[-7mm]\nn
\end{align}
which results after marginalizing $\overline{a}^{(j)}_{kl,n}$ in 
\begin{align}
\gamma^{[p]}_l(\overline{\V{x}}^{(j)}_{k,n},\overline{r}^{(j)}_k \rreq 1) &= \int u(\tilde{\V{x}}_n,\overline{\V{x}}^{(j)}_{k,n},1,1,\V{z}^{(j)}_{l,n})\overline{\nu}^{[p]}_{kl}(1) \mathrm{d}\tilde{\V{x}}_n \nn \\
&\hspace*{4mm} + \overline{\nu}^{[p]}_{kl}(0) \\
\gamma^{[p]}_l(\overline{\V{x}}^{(j)}_{k,n},\overline{r}^{(j)}_k \rreq 0) &=  \overline{\nu}^{[p]}_{kl}(0)\ist.
\end{align}
The message from factor node $v(\tilde{\V{x}}_n,\overline{\V{y}}^{(j)}_{m,n},\overline{a}^{(j)}_{mm},\V{z}^{(j)}_{m,n})$ to variable node $\overline{\V{y}}^{(j)}_{m,n}$ is given by
\begin{align}
\gamma^{[p]}_m(\overline{\V{y}}^{(j)}_{m,n}) &= \sum_{\overline{a}^{(j)}_{mm,n} \in \{0,1\}} \hspace{-1mm}  \int \rrmv v(\tilde{\V{x}}_n,\overline{\V{y}}^{(j)}_{m,n},\overline{a}^{(j)}_{mm,n},\V{z}^{(j)}_{m,n})\nn \\
& \hspace*{10mm}\times \overline{\nu}^{[p]}_{mm}(\overline{a}^{(j)}_{mm,n}) \mathrm{d}\tilde{\V{x}}_n
\end{align}
resulting in 
\begin{align}
\hspace*{-2mm}\gamma^{[p]}_m(\overline{\V{x}}^{(j)}_{m,n}, 1) &= \int v(\tilde{\V{x}}_n,\overline{\V{x}}^{(j)}_{m,n},1,1,\V{z}^{(j)}_{m,n})\overline{\nu}^{[p]}_{mm}(1)\mathrm{d}\tilde{\V{x}}_n \\
\hspace*{-2mm}\gamma^{[p]}_m(\overline{\V{x}}^{(j)}_{m,n}, 0) &=  \overline{\nu}^{[p]}_{mm}(0)\ist.
\end{align}
The messages are initialized with $\gamma^{[p=1]}_\ell(\V{y}^{(j)}_{k,n}) = 1$.

\subsection{Extrinsic Information}
\label{sec:ExtrInfo}
For each legacy \ac{va}, the messages sent from variable node $\underline{\V{y}}^{(j)}_{k,n}$ to factor nodes $q( \tilde{\V{x}}_n, \underline{\V{y}}^{(j)}_{k,n}, \underline{a}^{(j)}_{kl,n}; \V{z}^{(j)}_{l,n} )$ with $k \in \Set{K}^{(j)}_{n-1}$, $l \in \Set{M}^{(j)}_{n}$ at \ac{MP} iteration $p+1$ are defined as
\vspace*{-1.5mm}
\begin{align}
	\alpha^{[p+1]}_l(\underline{\V{y}}^{(j)}_{k,n}) & = \alpha(\underline{\V{y}}^{(j)}_{k,n}) \prod_{\substack{\ell = 1 \\ \ell \neq l}}^{M_n^{(j)}} \gamma^{[p]}_\ell(\underline{\V{y}}^{(j)}_{k,n})\ist.\\[-7.5mm]\nn
\end{align}
For new \acp{va}, a similar expression can be obtained for the messages from variable node  $\overline{\V{y}}^{(j)}_{m,n}$ to factor nodes $u( \tilde{\V{x}}_n, \overline{\V{y}}^{(j)}_{m,n}, \overline{a}^{(j)}_{ml,n}; \V{z}^{(j)}_{l,n} )$ and factor node $v(\tilde{\V{x}}_n, \overline{\V{y}}^{(j)}_{m,n}, \overline{a}^{(j)}_{mm,n}; \V{z}^{(j)}_{m,n} )$, i.e.,
\vspace*{-1mm}
\begin{align}
\alpha^{[p+1]}_l(\overline{\V{y}}^{(j)}_{m,n}) & = \alpha(\overline{\V{y}}^{(j)}_{m,n}) \prod_{\substack{\ell = 1 \\ \ell \neq l}}^m \gamma^{[p]}_\ell(\overline{\V{y}}^{(j)}_{m,n})\ist.\\[-7.5mm]\nn
\end{align}

\subsection{Measurement update for augmented agent state}
\label{sec:MP_updateAgent}
Due to the proposed scheduling, the augmented agent state is only updated by messages of legacy PVAs and only at the end of the iterative message passing. 
This results in 
\begin{align}
\tilde{\beta}^{[p]}_{kl}(\tilde{\V{x}}_n) &= \alpha(\tilde{\V{x}}_n) \label{eq:betaTilde}\\
\beta^{[p](j)}_{kl}(\tilde{\V{x}}_n) &= \sum_{\underline{a}^{(j)}_{kl,n} \in \{0,1\}} \sum_{\underline{r}^{(j)}_{k,n} \in \{0,1\}} \int \alpha_l^{[p]}(\underline{\V{x}}^{(j)}_{k,n},\underline{r}^{(j)}_{k,n}) \nn \\
& \hspace{-7mm} \times q(\tilde{\V{x}}_n,\underline{\V{x}}^{(j)}_{k,n},\underline{r}^{(j)}_{k,n},\underline{a}^{(j)}_{kl,n},\V{z}^{(j)}_{l,n})\underline{\nu}^{[p]}_{kl}(\underline{a}^{(j)}_{kl,n}) \mathrm{d}\underline{\V{x}}^{(j)}_{k,n}
\end{align}
which can be further simplified to 
\begin{align}
\beta^{[p](j)}_{kl}(\tilde{\V{x}}_n) &= \int \alpha_l^{[p]}(\V{x}^{(j)}_{k,n},1) \Big( q(\tilde{\V{x}}_n,\underline{\V{x}}^{(j)}_{k,n},1,1,\V{z}^{(j)}_{l,n})\underline{\nu}^{[p]}_{kl}(1)   \nn \\
&\hspace*{3mm} + \underline{\nu}^{[p]}_{kl}(0) \Big)\mathrm{d}\underline{\V{x}}^{(j)}_{k,n} +\underline{\alpha}_k^{\text{n},(j)} \underline{\nu}^{[p]}_{kl}(0).
\end{align}

\subsection{Belief calculation}
Once all messages are available and $p=P$, the beliefs approximating the desired marginal posterior \ac{pdf}s are obtained. 
The belief for the augmented agent state is given, up to a normalization factor, by 
\begin{align}
b(\tilde{\V{x}}_n) \propto  \alpha(\tilde{\V{x}}_n)  \prod_{j = 1}^J \ \prod_{k = 1}^{K_{n-1}^{(j)}} \ \prod_{m = 1}^{M_{n}^{(j)}} \beta^{[P](j)}_{km}(\tilde{\V{x}}_n)
\end{align}
where we only use messages from legacy \acp{va}.
This belief (after normalization) provides an approximation of the marginal posterior \ac{pdf} $f(\tilde{\V{x}}_n |\V{z}_{1:n})$, and it is used instead of $f(\tilde{\V{x}}_n |\V{z}_{1:n})$ in \eqref{eq:MMSEagent}. Furthermore, the beliefs of the legacy \acp{va} $b(\underline{\V{y}}^{(j)}_k)$ and new \acp{va} $b(\overline{\V{y}}^{(j)}_k)$ are given as
\vspace*{-2mm}
\begin{align}
b(\underline{\V{y}}^{(j)}_{k,n}) &\propto \alpha(\underline{\V{y}}^{(j)}_{k,n}) \prod_{l = 1}^{M_{n}^{(j)}} \gamma^{[P]}_l(\underline{\V{y}}^{(j)}_{k,n}) \\
b(\overline{\V{y}}^{(j)}_{m,n}) &\propto \alpha(\overline{\V{y}}^{(j)}_{m,n}) \prod_{l = 1}^m \gamma^{[P]}_l(\overline{\V{y}}^{(j)}_{m,n})\\[-7mm]\nn
\end{align}
A computationally feasible approximate calculation of the various messages and beliefs can be based on the sequential Monte Carlo (particle-based) implementation approach introduced in \cite{MeyHliHla:TSPIN2016,MeyerProc2018, LiLeiVenTuf:TWC2022}.

%% file: inputFiles/simulation_results.tex

\begin{figure}[!t]
\ifthenelse{\equal{\arxiv}{false}}
		{  
	\tikzsetnextfilename{floorplan}
	\centering
	\setlength{\abovecaptionskip}{-0.5mm}
	\setlength{\belowcaptionskip}{0pt}
	\input{plots/floorplan.tikz}}
	{
	\includegraphics[scale=1]{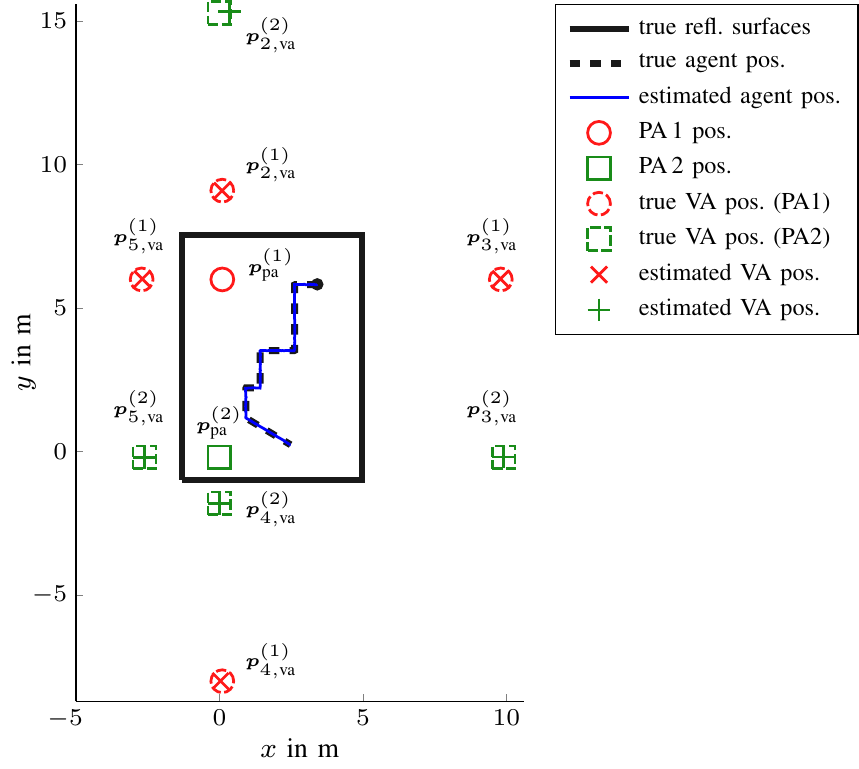}}
	\caption{Considered scenario for performance evaluation in a rectangular room with two \acp{pa}, four reflective surfaces and the corresponding \acp{va}. The estimated agent track for a single realization is shown in blue.}
	\label{fig:floorplan}
\end{figure}

\def\figHpa{0.2\columnwidth}
\def\figWpa{0.8\columnwidth}
\begin{figure}[!t]
\ifthenelse{\equal{\arxiv}{false}}
		{ 
	\setlength{\abovecaptionskip}{-0.5mm}
	\setlength{\belowcaptionskip}{0pt}
	\captionsetup[subfigure]{captionskip=0pt,labelformat=empty}
	\centering
	\subfloat[]{
		\tikzsetnextfilename{Final_PA1measDistVsTime}
		\input{plots/Final_PA1measDistVsTime.tex}\label{fig:PA1_meas}}\\
	\vspace*{-10mm}
	\subfloat[]{
		\tikzsetnextfilename{Final_PA2measDistVsTime}
		\input{plots/Final_PA2measDistVsTime.tex}\label{fig:PA2_meas}} \\
	\vspace*{-6mm}
	\subfloat{\tikzsetnextfilename{legend_PA} \pgfref{s_PAlegend}\hfill}\\
	\legendref{s_PAlegend}\hfill \\}
	{
	\captionsetup[subfigure]{captionskip=0pt,labelformat=empty}
	\centering
	\subfloat[]{
		\includegraphics{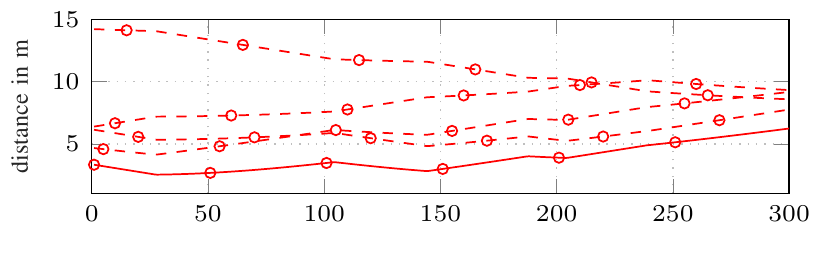}\label{fig:PA1_meas}}\\
	\vspace*{-10mm}
	\subfloat[]{
		\includegraphics{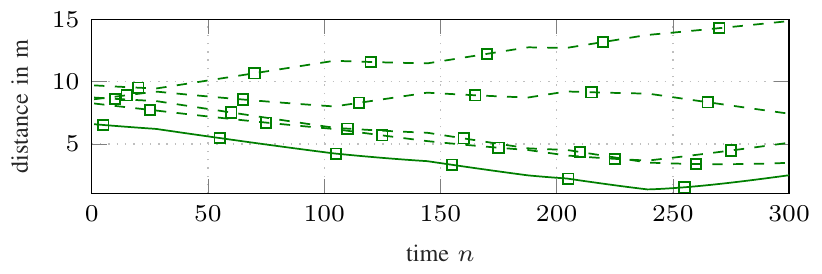}\label{fig:PA2_meas}} \\
	\vspace*{-6mm}
	\subfloat{ \includegraphics{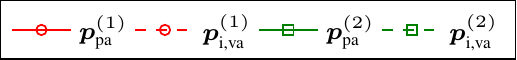}\hfill}\\
}
	\caption{Distances of main components (between the \ac{pa} positions as well as their corresponding \ac{va} positions and the agent positions) versus time $n$.}
	\label{fig:PA_meas}
\end{figure}

\def\figHresults{0.17\columnwidth}
\def\figWresults{0.55\columnwidth}
\begin{figure*}[!t]
	\setlength{\abovecaptionskip}{-0.5mm}
	\setlength{\belowcaptionskip}{0pt}
	\captionsetup[subfigure]{captionskip=0pt}
	\centering
	\ifthenelse{\equal{\arxiv}{false}}
		{ 
	\subfloat[]{\tikzsetnextfilename{RMSEoverTime}\hspace{8mm}
		\input{plots/RMSEoverTime.tex}\label{fig:rmseVsTime}}
	\subfloat[]{\tikzsetnextfilename{Extent_ed}\hspace{8mm}
		\input{plots/Extent_ed.tex}\label{fig:extend_d}}
	\subfloat[]{\tikzsetnextfilename{Extent_eu}\hspace{8mm}
		\input{plots/Extent_eu.tex}\label{fig:extend_u}} \\\vspace{-6mm}
	\subfloat[]{\tikzsetnextfilename{MOSPA_PA1}\hspace{8mm}
		\input{plots/MOSPA_PA1.tex}\label{fig:mospaPA1}}
	\subfloat[]{\tikzsetnextfilename{FeatureErrorPA1}\hspace{8mm}
		\input{plots/FeatureErrorPA1.tex}\label{fig:featureRMSE_PA1}} 
	\subfloat[]{\tikzsetnextfilename{CardError_PA1}\hspace{8mm}
		\input{plots/CardError_PA1.tex}\label{fig:cardErrorPA1}} \\\vspace{-6mm}
	\subfloat[]{\tikzsetnextfilename{MOSPA_PA2}\hspace{9mm}
		\input{plots/MOSPA_PA2.tex}\label{fig:mospaPA2}} 
	\subfloat[]{\tikzsetnextfilename{FeatureErrorPA2}\hspace{8mm}
		\input{plots/FeatureErrorPA2.tex}\label{fig:featureRMSE_PA2}} 
	\subfloat[]{\tikzsetnextfilename{CardError_PA2}\hspace{8mm}
		\input{plots/CardError_PA2.tex}\label{fig:cardErrorPA2}}\\[1mm]
	\hspace*{0cm}
	\subfloat{\tikzsetnextfilename{legend_Ex1} \pgfref{nameS}\hfill}\\
	\legendref{nameS}\hfill \\}
	{
	\subfloat[]{\hspace{3mm}
		\includegraphics{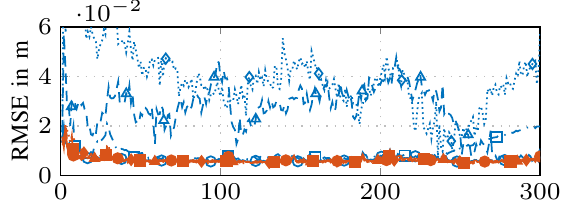}\label{fig:rmseVsTime}}
	\subfloat[]{\hspace{-3mm}
		\includegraphics{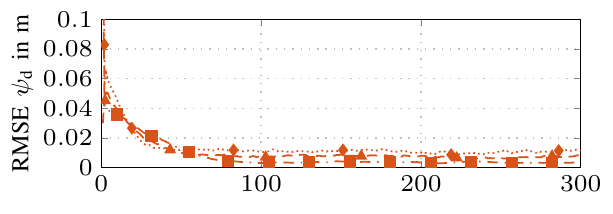}\label{fig:extend_d}}
	\subfloat[]{\hspace{-3mm}
		\includegraphics{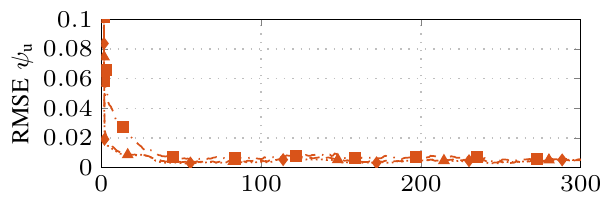}\label{fig:extend_u}} \\\vspace{-6mm}
	\subfloat[]{\hspace{3mm}
		\includegraphics{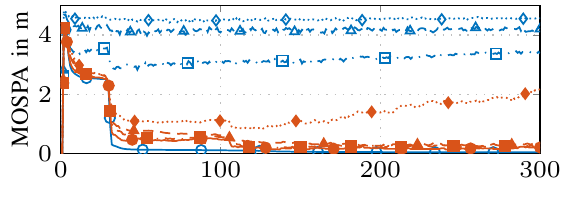}\label{fig:mospaPA1}}
	\subfloat[]{\hspace{1.5mm}
		\includegraphics{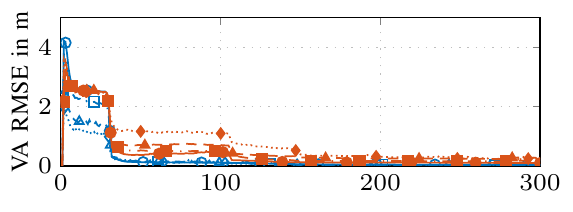}\label{fig:featureRMSE_PA1}} 
	\subfloat[]{\hspace{-1.5mm}
		\includegraphics{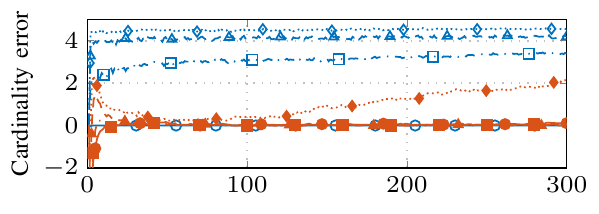}\label{fig:cardErrorPA1}} \\\vspace{-6mm}
	\subfloat[]{\hspace{4.3mm}
		\includegraphics{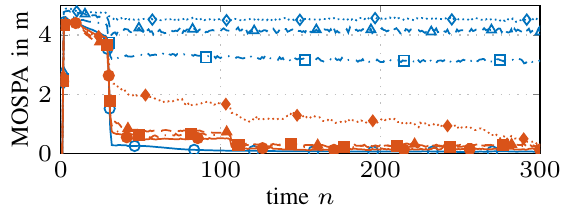}\label{fig:mospaPA2}} 
	\subfloat[]{\hspace{1.5mm}
		\includegraphics{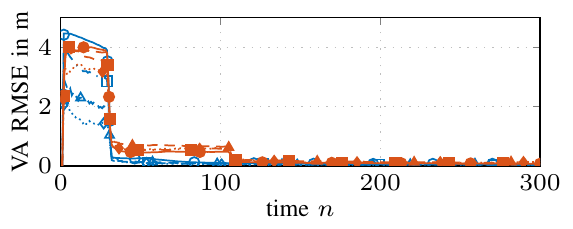}\label{fig:featureRMSE_PA2}} 
	\subfloat[]{\hspace{-1.5mm}
		\includegraphics{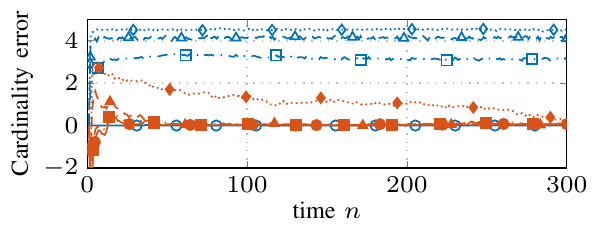}\label{fig:cardErrorPA2}}\\[1mm]
	\hspace*{0cm}
	\subfloat{\includegraphics{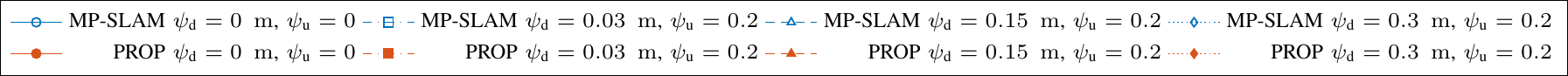}\hfill}\\
	}
	\vspace*{1mm}
	\caption{Experiment 1: Results for converged simulation runs. (a) shows the \ac{rmse} of the agent position over the whole track. (b) and (c) present the \ac{rmse} of the dispersion parameters. (d) and (g) present the map error in terms of the \ac{mospa} for \ac{pa}~$1$ and \ac{pa}~$2$, respectively. (e) and (h) show the \ac{rmse} of the estimated VA positions for \ac{pa}~$1$ and \ac{pa}~$2$, respectively. (f) and (i) show the cardinality error of the estimated \acp{va} for \ac{pa}~$1$ and \ac{pa}~$2$, respectively.}
	\label{fig:results}
	\vspace*{-2mm}
\end{figure*}

\begin{figure}[!t]
	\setlength{\abovecaptionskip}{-0.5mm}
	\setlength{\belowcaptionskip}{0pt}
	\centering
	\ifthenelse{\equal{\arxiv}{false}}
		{ 
	\tikzsetnextfilename{CDF_all}
	\input{plots/CDF_all.tex}}
	{
	\includegraphics{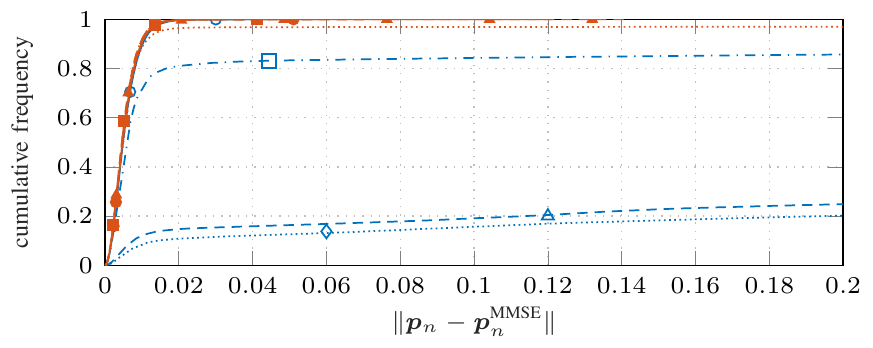}}
	\caption{Experiment 1: Cumulative frequency of the deviation of the \ac{mmse} estimate of the agent position from the true agent position for all simulation runs and time instances. The legend is given in Fig.~\ref{fig:results}.}
	\label{fig:cdf}
	\vspace*{-2mm}
\end{figure} 

\section{Numerical Results}
\label{sec:results}

The performance of the proposed algorithm (PROP) is validated and compared with the \ac{Mpslam} from \cite{LeitMeyHlaWitTufWin:TWC2019,LeiGreWit:ICC2019}, which assumes that each \ac{va} generates at most one measurement and that a measurement originates from at most one \ac{va}.
The validation of the algorithms is based on synthetic measurements in two settings.
\begin{enumerate}[leftmargin=6mm]
	\item Experiment 1 in Section~\ref{sec:ResultsModel} is based on measurements directly generated from the measurement model introduced in Section~\ref{sec:system_model}.
	\item Experiment 2 in Section~\ref{sec:ResultsCEDA} is based on measurements provided by a \ac{ceda} applied to radio signals that are generated with parameters according to the measurement model introduced in Section~\ref{sec:system_model}.
\end{enumerate}

\subsection{Simulation Scenario and Common Simulation Parameters}\label{sec:commonSimP}

\begin{figure*}[h!]
	\setlength{\abovecaptionskip}{-0.5mm}
	\setlength{\belowcaptionskip}{0pt}
	\captionsetup[subfigure]{captionskip=0pt}
	\centering
	\setlength{\figurewidth}{0.44\columnwidth}
	\setlength{\figureheight}{0.175\columnwidth}  
	\ifthenelse{\equal{\arxiv}{false}}
		{   
	\subfloat{\centering
		\def\datapath{./plots/measurements/PA1_step2_data}
		\tikzsetnextfilename{PA1_step2_data}
		\input{\datapath/PA1_step2_data.tikz}\hspace{-6mm}
		\def\datapath{./plots/measurements/PA1_step20_data}
		\tikzsetnextfilename{PA1_step20_data}
		\input{\datapath/PA1_step20_data.tikz}\hspace{-6mm}
		\def\datapath{./plots/measurements/PA1_step40_data}
		\tikzsetnextfilename{PA1_step40_data}
		\input{\datapath/PA1_step40_data.tikz}\hspace{-6mm}
		\def\datapath{./plots/measurements/PA1_step100_data}
		\tikzsetnextfilename{PA1_step100_data}
		\input{\datapath/PA1_step100_data.tikz}}}
		{
		\subfloat{\centering
		\includegraphics{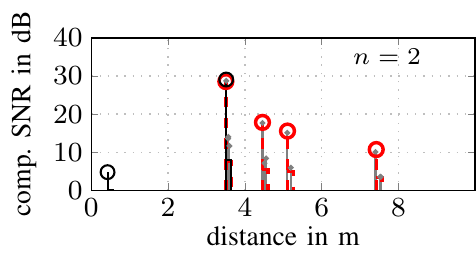}\hspace{0mm}
		\includegraphics{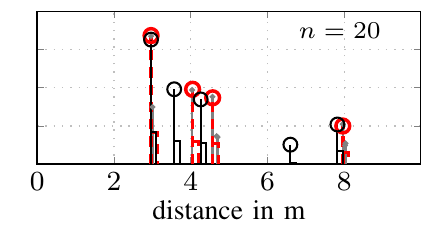}\hspace{0mm}
		\includegraphics{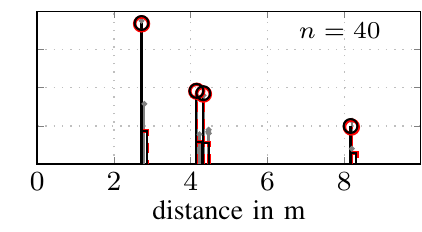}\hspace{0mm}
		\includegraphics{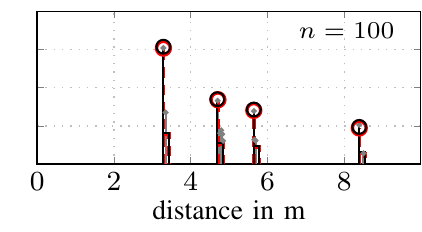}}}
	\caption{Estimated distances and dispersion parameters for \ac{pa}~$1$ for a single simulation run, represented by dot markers and boxes, respectively. The true components and respective dispersion parameters are indicated in red. All measurements are indicated in gray. Estimated components and respective dispersion parameters are indicated in black.}
	\label{fig:evolutionOfComponents}
	\vspace*{-2mm}
\end{figure*}

We consider an indoor scenario shown in Fig.~\ref{fig:floorplan}. 
The scenario consists of two \acp{pa} at positions $\V{p}_{\mathrm{pa}}^{(1)} = [0.1 \iist 6]^{\T}$, and $\V{p}_{\mathrm{pa}}^{(2)} = [0 \iist -\rmv\rmv0.2]^{\T}$ and four reflective surfaces, i.e., $4$ \acp{va} per \ac{pa}. 
The agent moves along a track which is observed for $300$ time instances $n$ with observation period $\Delta T = 1\,$s. 
For simplicity, we restrict the simulations to single-bounce reflections. 
The distances of the main components are calculated based on the \ac{pa} and the corresponding \acp{va} positions as well as agent positions (see Section~\ref{sec:signal_model}). 
Fig.~\ref{fig:PA_meas} shows the distances of the main components versus time $n$. 
The signal \ac{snr} is set to {$30\,\mathrm{dB}$} at an \ac{los} distance of $1\,$m. 
The amplitudes of the main components (\ac{los} component and the \acp{mpc}) are calculated using a free-space path loss model and an additional attenuation of {$1\,\mathrm{dB}$} for each reflection at a flat surface.
We use $20000$ particles. The particles for the initial agent state are drawn from a 4-D uniform distribution with center $\V{x}_0 = [\V{p}_{0}^{\T}\;0\;\, 0]^{\T}\rmv$, where $\V{p}_{0}$ is the starting position of the actual agent track, and the support of each position component about the respective center is given by $[-0.1\,\mathrm{m}, 0.1\,\mathrm{m}]$ and of each velocity component is given by $[-0.01\,\mathrm{m/s}, 0.01\,\mathrm{m/s}]$. 
At time $n \rmv=\rmv 0$, the number of \acp{va} is $0$, i.e., no prior map information is available. 
The prior distribution for new \ac{pva} states $f_\text{n}\big(\overline{\V{x}}^{(j)}_{m,n}|\tilde{\V{x}}_n\big)$ is uniform on the square region given by $[-\text{15 m}, \text{15 m}] \times [-\text{15 m}, \text{15 m}]$ around the center of the floor plan shown in Fig.~\ref{fig:floorplan} and the mean number of new \acp{pva} at time $n$ is  $\mu_\text{n} = 0.01$. 
The probability of survival is $p_{\mathrm{s}} = 0.999$. The confirmation threshold as well as the pruning threshold are given as $p_{\mathrm{cf}} = 0.5$ and $p_{\mathrm{pr}} = 10^{-3}$, respectively. For the sake of numerical stability, we introduce a small amount of regularization noise to the \ac{va} state $\V{p}_{k,\mathrm{va}}$ at each time step $n$, i.e., $\underline{\V{p}}^{(j)}_{k,\mathrm{va}} \rmv\rmv=\rmv\rmv \V{p}^{(j)}_{k,\mathrm{va}} \rmv+\rmv \V{\omega}_{k}$, where $\V{\omega}_{k}$ is \ac{iid} across $k$, zero-mean, and Gaussian with covariance matrix $\sigma_a^2\, \bold{I}_2$ and $\sigma_a = 10^{-3}\,\text{m}$. 
The state transition variances are set as $\sigma_\text{w} = 10^{-3}\,\mathrm{m/s^2}$, $q_{\tau} = q_{\text{u}} = 10^{4}$ \cite{Koc:TAES2008_EOT,MeyWilJ21}, and  $\sigma_{\text{u},k} = 0.05 \, u_{k,n\rmv - \rmv 1}^{(j)\s \text{MMSE}}$. 
Note that for the normalized amplitude state, we use a value proportional to the \ac{mmse} estimate of the previous time step $n \minus 1$ as a heuristic. 
The dispersion parameters are set to fixed values over time $n$, i.e., ${\psi}_{\tau,n} = \psi_\tau = \psi_\text{d}/c$ and ${\psi}_{\text{u},n} = {\psi}_{\text{u}}$.\footnote{For better readability, we introduce $\psi_\text{d}$ as a scaled version of $\psi_\tau $.}
The performance of the different methods discussed is measured in terms of the \ac{rmse} of the agent position and the dispersion parameters as well as the \acf{ospa} error \cite{Schuhmacher2008} of all \acp{va} with cutoff parameter and order set to 5~m and 2, respectively. 
The \ac{mospa} errors and \acp{rmse} of each unknown variable are obtained by averaging over all converged simulation runs. 
We declare a simulation run to be converged if $\{\forall n: \|\V{p}_n - \V{p}^{\text{MMSE}}_n \| < d_\text{cv}\,\text{m}\}$, where $d_\text{cv}$ is the convergence threshold.

\subsection{Experiment 1: Measurement Model}
\label{sec:ResultsModel}

We investigate PROP with four different dispersion parameter settings, given as $\psi_\text{d}$, which takes values of $0\,$m, $0.03\,$m, $0.15\,$m and $0.3\,$m, and ${\psi}_{\text{u}}$, which is either set to $0$ for $\psi_\text{d} = 0\,$m or $0.2$ otherwise. Furthermore, we set $N_\text{ny}=4$. We performed $100$ simulation runs. In each simulation run, we generated noisy measurements $\V{z}_{m,n}^{(j)}$ according to the measurement model proposed in Section~\ref{sec:measurementModel} using the main components calculated as described in Section~\ref{sec:commonSimP}. In the case $\psi_\text{d} = 0\,\mathrm{m}$ only main-component measurements are generated, which is equivalent to the system model in \cite{LeiGreWit:ICC2019}.  The detection threshold is given by $\gamma = 2.5$. For numerical stability, we reduced the root mean squared bandwidth $\beta_\text{bw}$ for \acp{va} by a factor of $4$. The convergence threshold is set to $d_\text{cv} = 0.2$.

\begin{table}[!h]
	\caption{Experiment 1: Convergence rate and mean number of estimated \acp{va} for different algorithms and dispersion settings.}
	\begin{center}
		\begin{tabular}{ r r | c| c}
			&setting & convergence & $\hat{K}$  \\
			\hline \hline
			\multirow{4}{1.5cm}{MP-SLAM} & $\psi_\text{d} = 0.00\,\mathrm{m}$ & 100 \% & 4 \\
			& $\psi_\text{d} = 0.03\,\mathrm{m}$ & 82 \% & 9 \\
			& $\psi_\text{d} = 0.15\,\mathrm{m}$ & 15 \% & 16 \\
			& $\psi_\text{d} = 0.30\,\mathrm{m}$ & 11 \% & 30 \\\hline
			\multirow{4}{1.5cm}{PROP} & $\psi_\text{d} = 0.00\,\mathrm{m}$  & 100 \% & 4 \\
			& $\psi_\text{d} = 0.03\,\mathrm{m}$  & 100 \% & 4\\ 
			& $\psi_\text{d} = 0.15\,\mathrm{m}$  & 100 \% & 4\\ 
			& $\psi_\text{d} = 0.30\,\mathrm{m}$  & 96 \% & 5\\ \hline \hline
		\end{tabular}
	\end{center}
	\label{tb:convergence}
\end{table} 

Table~\ref{tb:convergence} summarizes the number of converged runs (in percentage) as well as the mean number of detected \acp{va} $\hat{K}$ (averaged over all simulation runs and time steps) for all investigated dispersion parameter settings. 
The results are summarized in Fig.~\ref{fig:results}. 
In particular, Fig.~\ref{fig:rmseVsTime} shows the \ac{rmse} of the agent positions, Fig.~\ref{fig:extend_d} and \ref{fig:extend_u} show the \ac{rmse} of the dispersion parameters, Fig.~\ref{fig:mospaPA1}~-~Fig.~\ref{fig:cardErrorPA1} as well as  Fig.~\ref{fig:mospaPA2}~-~Fig.~\ref{fig:cardErrorPA2} show the \ac{mospa} error and its \ac{va} position error and mean cardinality error contributions for \ac{pa}~$1$ and \ac{pa}~$2$, respectively. The results in all figures are presented versus time $n$ (and for all investigated dispersion parameter settings). 
Fig.~\ref{fig:rmseVsTime} shows that the \ac{rmse} of the agent position of PROP is similar for all dispersion parameter settings. While PROP significantly outperforms MP-SLAM in terms of converged runs for dispersion parameter settings $\psi_\text{d} > 0\,\mathrm{m}$, it shows slightly reduced performance for $\psi_\text{d} = 0\,\mathrm{m}$. Additionally, Fig.~\ref{fig:cdf} shows the cumulative frequencies of the individual agent errors, i.e., $\|\V{p}_n - \V{p}^{\text{MMSE}}_n \|$ for all simulation runs and time instances. It can be observed that the \ac{mmse} positions of the agent of PROP show almost no large deviations, while the estimates of MP-SLAM exhibit large errors in many simulation runs.
For dispersion parameter settings $\psi_\text{d} > 0\,\mathrm{m}$, measurements of the sub-components are available. Thus, as Fig.~\ref{fig:extend_d} and \ref{fig:extend_u} show, the dispersion parameters are well estimated indicated by the small \acp{rmse}. For the setting $\psi_\text{d} = 0\,\mathrm{m}$, estimation of the dispersion parameters is not possible because there are no sub-component measurements, i.e., there is only one measurement generated by each \ac{va}. However, as Fig.~\ref{fig:rmseVsTime} shows, this does not affect the accuracy of the agent's position estimation.

The \ac{mospa} errors (and their \ac{va} positions and the mean cardinality error contributions) of PROP, shown in Fig.~\ref{fig:mospaPA1} and \ref{fig:mospaPA2}, are very similar for all dispersion parameter settings. They slightly increase with increased dispersion parameter $\psi_\text{d}$. Only for the setting $\psi_\text{d} = 0.3\,\mathrm{m}$, PROP shows a larger cardinality error. This can be explained by looking at the distances from \ac{pa} $1$ and its corresponding \acp{va} as shown in Fig.~\ref{fig:PA_meas}. At the end of the agent track, many \acp{va} show similar distances to the agent's position making it difficult to resolve the individual components. For larger dispersion parameter $\psi_\text{d}$, this becomes even more challenging leading to increased \ac{mospa} errors. For \ac{pa} $2$ and the corresponding \acp{va}, Fig.~\ref{fig:PA_meas} shows that all components are well separated by their distances at the end of the agent track, which makes it easier for PROP to correctly estimate the number and positions of \acp{va}. Unlike PROP, MP-SLAM completely fails to estimate the correct number of \acp{va} for larger $\psi_\text{d}$ (and ${\psi}_{\text{u}}$), resulting in a large cardinality error. This can be explained by the fact that MP-SLAM does not consider additional sub-components in the measurement and system model. We suspect that this estimation of additional spurious \acp{va} is the reason for the large number of divergent simulation runs. 
As an example, Fig.~\ref{fig:evolutionOfComponents} depicts the time evolution of the estimated distances (using the \ac{pa} position, the estimated \ac{va} positions, and estimated agent positions) with according component \acp{snr} as well as the respective dispersion parameters for \ac{pa} $1$.

\def\figHresults{0.17\columnwidth}
\def\figWresults{0.55\columnwidth}
\begin{figure*}[t]
	\setlength{\abovecaptionskip}{-0.5mm}
	\setlength{\belowcaptionskip}{0pt}
	\captionsetup[subfigure]{captionskip=0pt}
	\centering
	\ifthenelse{\equal{\arxiv}{false}}
		{   
	\subfloat[]{\tikzsetnextfilename{RMSE_CEDA}\hspace{8mm}
		\input{plots/RMSE_CEDA.tex}\label{fig:rmseCEDA}}
	\subfloat[]{\tikzsetnextfilename{PsiError_tau}\hspace{8mm}
		\input{plots/PsiError_tau.tex}\label{fig:tauCEDA}}
	\subfloat[]{\tikzsetnextfilename{PsiError_u}\hspace{8mm}
		\input{plots/PsiError_u.tex}\label{fig:uCEDA}}\\\vspace{-5mm}
	\subfloat[]{\tikzsetnextfilename{MOSPA_PA1_CEDA}\hspace{8mm}
		\input{plots/MOSPA_PA1_CEDA.tex}\label{fig:mospa_pa1_ceda}}
	\subfloat[]{\tikzsetnextfilename{VAerror_PA1_CEDA}\hspace{8mm}
		\input{plots/VAerror_PA1_CEDA.tex}\label{fig:va_pa1_ceda}}  
	\subfloat[]{\tikzsetnextfilename{Cerror_PA1_CEDA}\hspace{8mm}
		\input{plots/Cerror_PA1_CEDA.tex}\label{fig:c_pa1_ceda}} \\\vspace{-5mm}
	\subfloat[]{\tikzsetnextfilename{MOSPA_PA2_CEDA}\hspace{8mm}
		\input{plots/MOSPA_PA2_CEDA.tex}\label{fig:mospa_pa2_ceda}}
	\subfloat[]{\tikzsetnextfilename{VAerror_PA2_CEDA}\hspace{8mm}
		\input{plots/VAerror_PA2_CEDA.tex}\label{fig:va_pa2_ceda}}  
	\subfloat[]{\tikzsetnextfilename{Cerror_PA2_CEDA}\hspace{8mm}
		\input{plots/Cerror_PA2_CEDA.tex}\label{fig:c_pa2_ceda}} \\
	\subfloat{\tikzsetnextfilename{legend_Ex2} \pgfref{nameSe}\hfill}\\
	\legendref{nameSe}\hfill \\}
	{
	\subfloat[]{\hspace{-3mm}
		\includegraphics{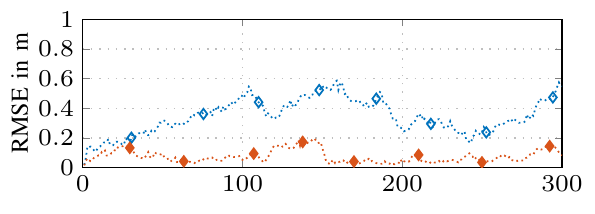}\label{fig:rmseCEDA}}
	\subfloat[]{\hspace{-3mm}
		\includegraphics{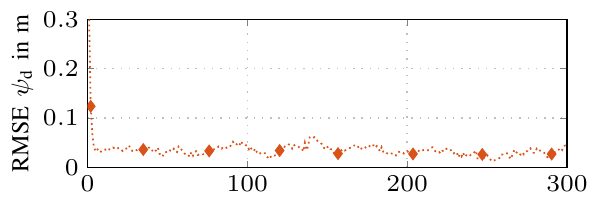}\label{fig:tauCEDA}}
	\subfloat[]{\hspace{-3mm}
		\includegraphics{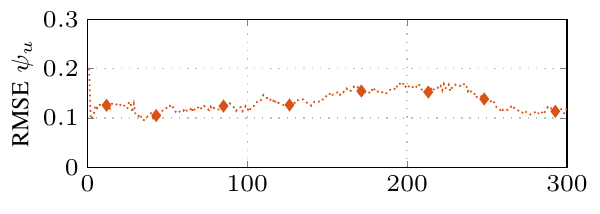}\label{fig:uCEDA}}\\\vspace{-5mm}
	\subfloat[]{\hspace{0mm}
		\includegraphics{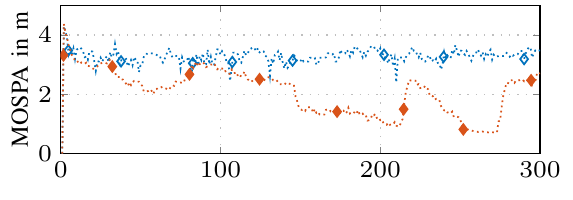}\label{fig:mospa_pa1_ceda}}
	\subfloat[]{\hspace{-1.2mm}
		\includegraphics{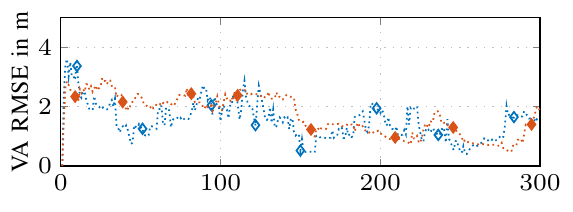}\label{fig:va_pa1_ceda}}  
	\subfloat[]{\hspace{-1mm}
		\includegraphics{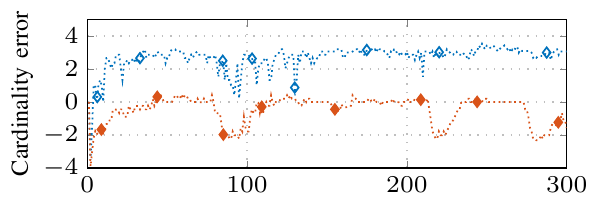}\label{fig:c_pa1_ceda}} \\\vspace{-5mm}
	\subfloat[]{\hspace{0mm}
		\includegraphics{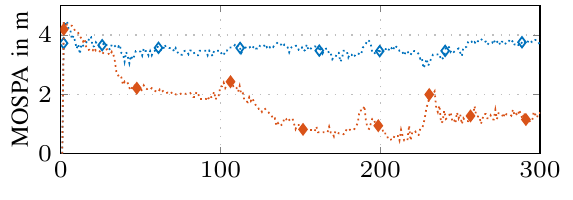}\label{fig:mospa_pa2_ceda}}
	\subfloat[]{\hspace{-1.2mm}
		\includegraphics{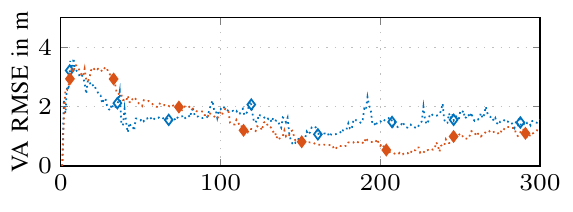}\label{fig:va_pa2_ceda}}  
	\subfloat[]{\hspace{-1mm}
		\includegraphics{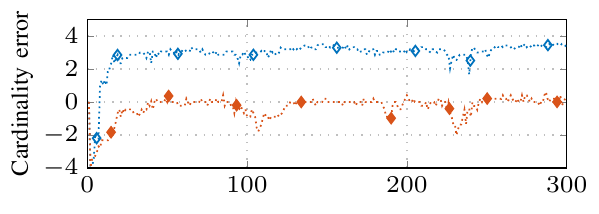}\label{fig:c_pa2_ceda}} \hfill \\
		\vspace*{-2mm}
		\captionsetup[subfigure]{labelformat=empty}
	\subfloat[]{\hspace{8mm}
		\includegraphics{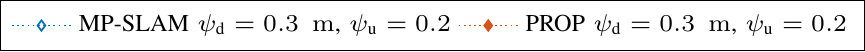}\label{fig:c_pa2_ceda}} \\
	}
	\vspace*{1mm}
	\caption{Experiment 2: Results for converged simulation runs based on estimates from \ac{ceda}. (a) shows the \ac{rmse} of the agent position over the whole track. (b) and (c) present the \ac{rmse} of the dispersion parameters. (d) and (g) present the map error in terms of the \ac{mospa} for \ac{pa}~$1$ and \ac{pa}~$2$, respectively. (e) and (h) show the \ac{rmse} of the estimated \ac{va} positions for \ac{pa}~$1$ and \ac{pa}~$2$, respectively. (f) and (i) show the cardinality error of the estimated \acp{va} for \ac{pa}~$1$ and \ac{pa}~$2$, respectively.}
	\label{fig:resultsCEDA}
\end{figure*}

\subsection{Experiment 2: Radio Signals}\label{sec:ResultsCEDA}

In this section, we use a dispersion parameter setting of $\psi_\text{d} = 0.3\,\mathrm{m}$ and $\psi_\text{u} = 0.2$. The signal spectrum of the transmit pulse  $s(t)$ has a root-raised-cosine shape with a roll-off factor of $0.6$ and a $3\,\mathrm{dB}$ bandwidth of $B = 1\,\mathrm{GHz}$. The signal is critically sampled, i.e., $T_\text{s} = 1/(1.6 B)$, with a total number of $N_\text{s} = 161$ samples resulting in a maximum distance $d_\text{max} = 60\,\mathrm{m}$. For the data generation, we use $N_\text{ny}=2$. We perform $10$ simulation runs. In each simulation run, we generate a received signal vector (see \eqref{eq:signal_model_sampled}) using the main components calculated as described in Section~\ref{sec:commonSimP} and uniformly distributed sub-components (see \eqref{eq:dispersion}). To obtain the measurements, we use the \ac{ceda} in \cite{HanBadFleRao:SAM2014} with a detection threshold of $\gamma = 2$, i.e., corresponding to $6\,\mathrm{dB}$ \cite{GreLeiFleWit:Arxiv2023}. For numerical stability, we reduced the root mean squared bandwidth $\beta_\text{bw}$ for \acp{va} by a factor of $4$ and increased the factor $1/2$ in amplitude scale parameter in \eqref{eq:scaleParam_amplitude} to $4$. The convergence threshold is $d_\text{cv} = 2$.

\begin{table}[!h]
	\caption{Experiment 2: Convergence rate and mean number of estimated \acp{va} for different algorithms.}
	\begin{center}
		\begin{tabular}{ r | r | r}
			setting & convergence & $\hat{K}$ \\
			\hline \hline
			MP-SLAM $\psi_\text{d} = 0.30\,\mathrm{m}$ & 20 \% & 7.5 \\
			PROP $\psi_\text{d} = 0.30\,\mathrm{m}$  & 100 \% & 3.7\\ \hline \hline
		\end{tabular}
	\end{center}
	\label{tb:convergenceCEDA}
	\vspace*{-2mm}
\end{table} 

Table~\ref{tb:convergenceCEDA} again summarizes the number of converged runs and the mean number of detected \acp{va}. For PROP, none of the simulation runs diverged, but $80\,\%$ of the MP-SLAMs simulation runs diverged, showing that PROP significantly outperforms MP-SLAM. The results shown in Fig.~\ref{fig:resultsCEDA} follow a similar trend as the results shown in Fig.~\ref{fig:results}. The only significant difference is observed in the \ac{rmse} of the dispersion parameter $\psi_\text{u}$, which remains relatively large (see Fig.~\ref{fig:uCEDA}). This is because the variance of the estimated normalized amplitudes provided by the \ac{ceda} is very large. This may be explained by two factors: (i) the \ac{ceda} also needs to estimate the noise variance, which is only approximately covered by the amplitude scale parameter given in \eqref{eq:scaleParam_amplitude}, and (ii) the sub-components are very close in the delay domain, resulting in strongly correlated amplitude estimates. 
The steps in Fig.~\ref{fig:mospa_pa1_ceda} and \ref{fig:c_pa1_ceda} are due to crossings where the delays from two or more VAs to the agent are equal. Hence, one of the VAs is discarded, leading to an overall underestimated number of VAs.

%% file: inputFiles/conclusions.tex
We have proposed a new \ac{Mpslam} method that can cope with multiple-measurements being generated by a single environment feature, i.e., a single \ac{va}. It is based on a novel statistical measurement model that is derived from the radio signal introducing dispersion parameters to \acp{mpc}. The resulting likelihood function model allows to capture the measurement spread originating from non-ideal effects such as rough reflective surfaces or non-calibrated antennas. The performance results show that the proposed method is able to cope with multiple measurements being produced per \ac{va} and outperforms classical \ac{Mpslam} in terms of the agent positioning error and the map \ac{mospa} error. We show that multiple measurements get correctly associated to their corresponding VA, resulting in a correctly estimated number of VAs. Furthermore, the results indicate that the proposed algorithm generalizes to the classical multipath-based \ac{slam} for a single measurement per VA. Possible directions of future research include the extension to individual dispersion parameters for each feature as well as incorporating multiple-measurements-to-feature data association into the \ac{mva}-based \ac{slam} method \cite{VenLeiTerWit:TWC2023}.

%% file: inputFiles/appendix_DA.tex

\section{Radio Signal Model}
\label{sec:app_signal_model}

In this section we derive the radio signal model described in Section~\ref{sec:signal_model}. 
Usually, specular reflections of radio signals at flat surfaces are modeled by \acp{va} that are mirror images of the \acp{pa} \cite{LeitingerJSAC2015,WitrisalSPM2016,LeitMeyHlaWitTufWin:TWC2019,MenMeyBauWin:JSTSP2019}. 
We start by defining the typical channel impulse response, given for time $n$ and anchor $j$ as
\vspace*{-2mm}
\begin{align}
	h_{\text{c}, n}^{(j)}(\tau) = \rmv\rmv\rmv  \sum_{l = 1}^{{L}_n^{(j)}} {\alpha}_{l,n}^{(j)} \delta\big(\tau\minus {\tau}_{l,n}^{(j)}\big)
\rmv \ist.\\[-7mm]\nn
\end{align}
The first summand describes the \ac{los} component and the sum of ${L}_n^{(j)} \minus 1$ the specular \acp{mpc} with their corresponding complex amplitudes ${\alpha}_{l,n}^{(j)}$ and delays ${\tau}_{l,n}^{(j)}$, respectively. 
In non-ideal radio channels we observe rays to arrive as clusters \cite{SalVal:JSAC1987,KulmerPIMRC2018,PedersenJTAP2018,WenKulWitWym:TWC2021}. The reason for this observation is manifold. Typical examples are non-calibrated antennas, the scattering from a user-body as well as non-ideal reflective surfaces. Fig.~\ref{fig:overview} visualizes these effects, introducing generic impulse responses $h_{\text{ant}, n}^{(j)}(\tau) $ and $h_{\text{surf}, n}^{(j)}(\tau)$. We propose to model the overall impulse response encompassing all considered dispersion effects 
as
\vspace{-2mm}
\begin{equation}\vspace{-2mm}
	h_{\text{d}, n}^{(j)}(\tau) =	\delta(\tau) + \sum_{i=1}^{S_l^{(j)}} \beta^{(j)}_{l,i,n} 	\delta(\tau -\rmv\rmv\nu^{(j)}_{l,i,n}\big)
\end{equation}
where $\beta^{(j)}_{l,i,n} \in \mathbb{R}$ is a relative dampening variable and $\nu^{(j)}_{l,i,n}$ is the excess delay. The presented model denotes a marked Possion point process \cite{PedersenJTAP2018}. Its statistical properties, i.e, the distribution of $\nu^{(j)}_{l,i,n}$, $\beta^{(j)}_{l,i,n}$, and $S_l^{(j)}$, are discussed in Section~\ref{sec:signal_model} and \ref{sec:system_model} in detail.
We obtain the complex baseband signal received at the $j$th anchor given by the convolution of $h_{\text{d}, n}^{(j)}(\tau)$ and $h_{\text{c}, n}^{(j)}(\tau)$ with the transmitted signal $s(t)$ as \vspace{-2mm}
\begin{align}\label{eq:signal_cont}
\RV{s}_{\text{rx},n}^{(j)} &=   \sum_{l = 1}^{{L}_n^{(j)}} {\alpha}_{l,n}^{(j)} \Big(s(t\minus {\tau}_{l,n}^{(j)} ) \nn\\
&\hspace*{3mm}	+ \sum_{i=1}^{S_l^{(j)}} \beta^{(j)}_{l,i,n} 	s(t \minus {\tau}_{l,n}^{(j)} \minus \rmv\rmv\nu^{(j)}_{l,i,n}) \Big)  +  \rv{\noise{}}_{n}^{(j)}(t)\ist.\\[-7mm]\nn
\end{align}
The second term $\rv{\noise{}}_{n}^{(j)}(t)$ represents an additive white Gaussian noise process with double-sided power spectral density ${N}_{0}^{(j)}/\s 2$.   
%
%
%
%
%
%
\section{Data Association}
\label{sec:DA}
This section contains the detailed derivation of the data association-related messages $\varphi^{[p]}_{kl}(b^{(j)}_{l,n})$ and $\nu^{[p]}_{kl}(a^{(j)}_{kl,n})$. Using the measurement evaluation messages in \eqref{eq:message_epsilon1}, \eqref{eq:message_epsilon2} and \eqref{eq:message_epsilon3}, the messages $\underline{\varphi}^{[p]}_{kl}(b^{(j)}_{l,n})$ and $\overline{\varphi}^{[p]}_{ml}(b^{(j)}_{l,n})$ are calculated by
\begin{align}
	\underline{\varphi}^{[p]}_{kl}(b^{(j)}_{l,n}) &= \sum_{\underline{a}^{(j)}_{kl,n} \in \{0,1\}} \varepsilon^{[p]}(\underline{a}^{(j)}_{kl,n})  \underline{\Psi}(\underline{a}^{(j)}_{kl,n},b^{(j)}_{l,n}) \label{eq:app_eq_1}\\
	\overline{\varphi}^{[p]}_{ml}(b^{(j)}_{l,n}) &= \sum_{\overline{a}^{(j)}_{ml,n} \in \{0,1\}} \varepsilon^{[p]}(\overline{a}^{(j)}_{ml,n})  \overline{\Psi}(\overline{a}^{(j)}_{ml,n},b^{(j)}_{l,n})\label{eq:app_eq_2}
\\[-7mm]\nn
\end{align}
for $k \in \{1,\dots,\underline{K}\}$ with $\underline{K} \triangleq K^{(j)}_{n-1}$ and  $m,l  \in \{1,\dots,M^{(j)}_{n}\}$ and are sent from factor node $\underline{\Psi}(\underline{a}^{(j)}_{kl,n},b^{(j)}_{l,n})$ and $\overline{\Psi}(\overline{a}^{(j)}_{ml,n},b^{(j)}_{l,n})$ to variable node $b^{(j)}_{l,n}$, respectively.
By making use of the indicator functions given in \eqref{eq:psi_legacy} and \eqref{eq:psi_new}, respectively, \eqref{eq:app_eq_1} and \eqref{eq:app_eq_2} are also given as
\vspace*{-1mm}
\begin{align}
	\underline{\varphi}^{[p]}_{kl}(b^{(j)}_{l,n}=k) &= \varepsilon^{[p]}(\underline{a}^{(j)}_{kl,n} = 1)  \label{eq:phi_1}\\
	\underline{\varphi}^{[p]}_{kl}(b^{(j)}_{l,n}\neq k) &= \varepsilon^{[p]}(\underline{a}^{(j)}_{kl,n} = 0)  \\
	\overline{\varphi}^{[p]}_{ml}(b^{(j)}_{l,n}= \underline{K} + m) &= \varepsilon^{[p]}(\overline{a}^{(j)}_{ml,n} = 1)  \\
	\overline{\varphi}^{[p]}_{ml}(b^{(j)}_{l,n}\neq \underline{K} + m) &= \varepsilon^{[p]}(\overline{a}^{(j)}_{ml,n} = 0) \label{eq:phi_4}\\[-7mm]\nn
\end{align}
The messages in \eqref{eq:phi_1} - \eqref{eq:phi_4} can be rewritten in the form of
\begin{align}
	\underline{\varphi}^{[p]}_{kl}(b^{(j)}_{l,n}) &= 
	\begin{cases}
		\frac{\varepsilon^{[p]}(\underline{a}^{(j)}_{kl,n} = 1)}{\varepsilon^{[p]}(\underline{a}^{(j)}_{kl,n} = 0)}, &   b^{(j)}_{l,n} = k \\
		1, &  b^{(j)}_{l,n} \neq k
	\end{cases} \label{eq:phi_legacy}\\
	\overline{\varphi}^{[p]}_{ml}(b^{(j)}_{l,n}) &= 
	\begin{cases}
		\frac{\varepsilon^{[p]}(\overline{a}^{(j)}_{ml,n} = 1)}{\varepsilon^{[p]}(\overline{a}^{(j)}_{ml,n} = 0)}, &   b^{(j)}_{l,n} = \underline{K} + m \\
		1, &  b^{(j)}_{l,n} \neq \underline{K} + m.
	\end{cases} \label{eq:phi_new}  
\end{align}

The messages $\underline{\nu}^{[p]}_{kl}(\underline{a}^{(j)}_{kl,n})$ and $\overline{\nu}^{[p]}_{ml}(\overline{a}^{(j)}_{ml,n})$ represent the messages from variable node $\underline{a}^{(j)}_{kl,n}$ to factor node $q( \tilde{\V{x}}_n, {\underline{\V{y}}}^{(j)}_{k,n}, \underline{a}^{(j)}_{kl,n}; \V{z}^{(j)}_{l,n} )$ and from variable node $\overline{a}^{(j)}_{ml,n}$ to factor node $u( \tilde{\V{x}}_n, {\overline{\V{y}}}^{(j)}_{m,n}, \overline{a}^{(j)}_{ml,n}; \V{z}^{(j)}_{l,n} )$, respectively. $\overline{\nu}^{[p]}_{mm}(\overline{a}^{(j)}_{mm,n})$ represents the messages from variable node $\overline{a}^{(j)}_{mm,n}$ to factor node $v( \tilde{\V{x}}_n, {\overline{\V{y}}}^{(j)}_{m,n}, \overline{a}^{(j)}_{mm,n}; \V{z}^{(j)}_{m,n} )$. They are defined as
\begin{align}
\underline{\nu}^{[p]}_{kl}(\underline{a}^{(j)}_{kl,n}) &= \sum_{b^{(j)}_{l,n} = 0}^{K^{(j)}_{n}}  \prod_{\substack{i=1 \\ i\neq k}}^{\underline{K}} \underline{\varphi}^{[p]}_{il}(b^{(j)}_{l,n}) \prod_{m = l}^{M^{(j)}_n}   \overline{\varphi}^{[p]}_{ml}(b^{(j)}_{l,n}) \label{eq:app_eq_3}\\
	\overline{\nu}^{[p]}_{ml}(\overline{a}^{(j)}_{ml,n}) &= \sum_{b^{(j)}_{l,n} = 0}^{K^{(j)}_{n}}  \prod_{i=1}^{\underline{K}} \underline{\varphi}^{[p]}_{il}(b^{(j)}_{l,n}) \prod_{\substack{h=l \\ h\neq m}}^{M^{(j)}_n}  \overline{\varphi}^{[p]}_{hl}(b^{(j)}_{l,n}).\label{eq:app_eq_4}\\[-7mm]\nn
\end{align}
Using the results from \eqref{eq:phi_legacy} and \eqref{eq:phi_new}, \eqref{eq:app_eq_3} and \eqref{eq:app_eq_4} are, respectively, rewritten as
\begin{align}
\underline{\nu}^{[p]}_{kl}(\underline{a}^{(j)}_{kl,n} \rreq 1) =& \prod_{\substack{i=1 \\ i\neq k}}^{\underline{K}} \underline{\varphi}^{[p]}_{il}(b^{(j)}_{l,n} \rreq k) \prod_{m = l}^{M^{(j)}_n}  \overline{\varphi}^{[p]}_{ml}(b^{(j)}_{l,n} \rreq \underline{K} \rrmv + \rrmv k) \nn\\[-4mm]
\text{}\label{eq:nu_final1}\\
\underline{\nu}^{[p]}_{kl}(\underline{a}^{(j)}_{kl,n} = 0) =& \hspace{-5mm} \sum_{\substack{b^{(j)}_{l,n}=0 \\ b^{(j)}_{l,n} \notin \{k,\underline{K} + k\}}}^{K^{(j)}_{n}} \hspace{-2mm} \prod_{\substack{i=1 \\ i\neq k}}^{\underline{K}} \underline{\varphi}^{[p]}_{il}(b^{(j)}_{l,n}) \prod_{m = l}^{M^{(j)}_n}  \overline{\varphi}^{[p]}_{ml}(b^{(j)}_{l,n})\nn \\[-5mm]
\text{}
\end{align}
and 
\begin{align}
\overline{\nu}^{[p]}_{ml}(\overline{a}^{(j)}_{ml,n}  \rreq 1) =& \prod_{i=1}^{\underline{K}} \underline{\varphi}^{[p]}_{il}(b^{(j)}_{l,n}  \rreq  m) \rrmv \rrmv \prod_{\substack{h=l \\ h\neq m}}^{M^{(j)}_n} \rrmv \overline{\varphi}^{[p]}_{hl}(b^{(j)}_{l,n}  \rreq \underline{K}  \rrmv + \rrmv m) \nn\\[-2mm]
\text{}\label{eq:nu_final2}\\
\overline{\nu}^{[p]}_{ml}(\overline{a}^{(j)}_{ml,n} = 0) =&\hspace{-5mm} \sum_{\substack{b^{(j)}_{l,n}=0 \\ b^{(j)}_{l,n} \notin \{m,\underline{K} + m\}}}^{K^{(j)}_{n}} \hspace{-2mm} \prod_{i=1}^{\underline{K}} \underline{\varphi}^{[p]}_{il}(b^{(j)}_{l,n}) \prod_{\substack{h=l \\ h\neq m}}^{M^{(j)}_n}  \overline{\varphi}^{[p]}_{hl}(b^{(j)}_{l,n})\nn \\[-5mm]
\text{}
\end{align}
Note that $\varphi^{[p]}_{kl}(b^{(j)}_{l,n} \rreq 0) = 1$. By normalizing \eqref{eq:nu_final1} by $\underline{\nu}^{[p]}_{kl}(\underline{a}^{(j)}_{kl,n} = 0)$ and \eqref{eq:nu_final2} by $\overline{\nu}^{[p]}_{ml}(\overline{a}^{(j)}_{ml,n} = 0)$, equivalent expressions for \eqref{eq:app_eq_3} and \eqref{eq:app_eq_4} are given as
\begin{align}
&\underline{\nu}^{[p]}_{kl}(\underline{a}^{(j)}_{kl,n}) \nn \\
&= \begin{cases}
		\frac{\prod_{\substack{\scalebox{0.6}{$i=1$}\\ \scalebox{0.6}{$i\neq k$}}}^{\underline{K}} \underline{\varphi}^{[p]}_{il}(b^{(j)}_{l,n} = k) \prod_{m = l}^M  \overline{\varphi}^{[p]}_{ml}(b^{(j)}_{l,n} = \underline{K} + k)}{\sum_{\substack{\hspace{-8mm}\scalebox{0.6}{$b^{(j)}_{l,n}=0$} \\ \scalebox{0.6}{$b^{(j)}_{l,n} \notin \{k,\underline{K} + k\}$}}}^{K^{(j)}_{n}}  \hspace{-4mm} \prod_{\substack{\scalebox{0.6}{$i=1$}\\ \scalebox{0.6}{$i\neq k$}}}^{\underline{K}} \underline{\varphi}^{[p]}_{il}(b^{(j)}_{l,n}) \prod_{m = l}^{M^{(j)}_n}  \overline{\varphi}^{[p]}_{ml}(b^{(j)}_{l,n})}, &   \underline{a}^{(j)}_{kl,n} \rreq 1 \\
		1, & \underline{a}^{(j)}_{kl,n} \rreq 0.
	\end{cases} \label{eq:nu_legacy}\\
&\overline{\nu}^{[p]}_{ml}(\overline{a}^{(j)}_{ml,n})  \nn \\
&\hspace*{0mm}= \begin{cases}
	\hspace*{1mm}	\frac{\prod_{i=1}^{\underline{K}} \underline{\varphi}^{[p]}_{il}(b^{(j)}_{l,n} = m) \prod_{\substack{\scalebox{0.6}{$h=l$}\\ \scalebox{0.6}{$h\neq m$}}}^{M^{(j)}_n}  \overline{\varphi}^{[p]}_{hl}(\bkm)}{\hspace{-2mm} \sum_{\substack{\hspace{-10mm}\scalebox{0.6}{$b^{(j)}_{l,n}=0$} \\ \scalebox{0.6}{$b^{(j)}_{l,n} \notin \{m,\underline{K} + m\}$}}}^{K^{(j)}_n} \hspace{-8mm} \prod_{i=1}^{\underline{K}} \underline{\varphi}^{[p]}_{il}(\scb) \prod_{\substack{\scalebox{0.6}{$h=l$}\\ \scalebox{0.6}{$h\neq m$}}}^{M^{(j)}_n}   \overline{\varphi}^{[p]}_{hl}(\bkm)}, & \hspace*{-3mm} \overline{a}^{(j)}_{ml,n} \rreq 1 \\
		1, & \hspace*{-3mm} \overline{a}^{(j)}_{ml,n} \rreq 0.
	\end{cases} \label{eq:nu_new}  
\end{align}
Finally, by calculating the explicit summations and multiplications in \eqref{eq:nu_legacy} and \eqref{eq:nu_new}, it results in
\begin{align}
&\underline{\nu}^{[p]}_{kl}(\underline{a}^{(j)}_{kl,n}) \nn \\
&\hspace*{3mm}=	\begin{cases}
		\hspace*{-1mm} \frac{1}{1 + \sum_{\substack{\scalebox{0.6}{$i=1$} \\ \scalebox{0.6}{$i\neq k$}}}^{\underline{K}} \underline{\varphi}^{[p]}_{il}(\bi) +  \sum_{m = l}^{M^{(j)}_n}  \overline{\varphi}^{[p]}_{ml}(\bkm)}, &  \rrmv \rrmv \underline{a}^{(j)}_{kl,n} \rreq 1 \\
		1, & \rrmv \rrmv \underline{a}^{(j)}_{kl,n} \rreq 0
	\end{cases} \label{eq:nu_legacy_end}\\[-7mm]\nn
\end{align}
\begin{align}
&\overline{\nu}^{[p]}_{ml}(\overline{a}^{(j)}_{ml,n}) \nn \\ 
&\hspace*{3mm}=\begin{cases}
		\hspace*{-1mm} \frac{1}{1 + \sum_{i=1}^{\underline{K}} \underline{\varphi}^{[p]}_{il}(\bi) +  \sum_{\substack{h=l \\ h\neq m}}^{M^{(j)}_n}  \overline{\varphi}^{[p]}_{hl}(\bkm)}, &  \rrmv \rrmv \overline{a}^{(j)}_{ml,n} \rreq 1\\
		1, & \rrmv \rrmv \overline{a}^{(j)}_{ml,n} \rreq 0.
	\end{cases} \label{eq:nu_new_end}  
\end{align}

%% file: biography.tex
\begin{IEEEbiography}[{\includegraphics[width=1in,height=1.25in,clip,keepaspectratio]{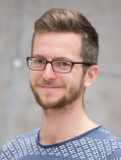}}]{Lukas~Wielandner} (S'20) received his Dipl.-Ing. (MSc.) degree in technical physics from Graz University of Technology, Austria, in 2018.  
He received his Ph.D. degree in electrical engineering at the Signal Processing and Speech Communication Laboratory (SPSC) of Graz University of Technology, Austria  in 2022. His research interests include localization and navigation, estimation/detection theory, inference on graphs and iterative message passing algorithms. 
\end{IEEEbiography}

\begin{IEEEbiography}[{\includegraphics[width=25mm,height=32.15mm,clip,keepaspectratio]{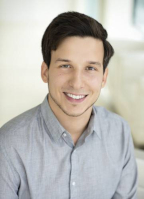}}]{Alexander~Venus} (S'20) received his B.Sc. and  Dipl.-Ing.\ (M.Sc.\ ) degrees (with highest honors) in biomedical engineering and information and communication engineering from Graz University of Technology, Austria in 2012 and 2015, respectively. He was a research and development engineer at Anton Paar GmbH, Graz from 2014 to 2019. He is currently a project assistant at Graz University of Technology, where he is pursuing his Ph.D. degree.
	
His research interests include radio-based localization and navigation, statistical signal processing, estimation/detection theory, machine learning and error bounds. 
\end{IEEEbiography}

\begin{IEEEbiography}[{\includegraphics[width=1in,height=1.25in,clip,keepaspectratio]{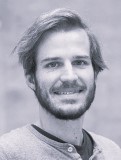}}]{Thomas Wilding}
(S'17) received his B.Sc. and  Dipl.-Ing.\ (M.Sc.) degrees in audio and electrical engineering from the University of Music and Performing Arts Graz, Austria in 2013 and 2016, respectively, and his Ph.D.\ from Graz University of Technology, Austria in 2022. He is currently a post-doctoral researcher at Graz University of Technology working on positioning, sensing and environment learning in wireless systems.
His research interests include radio localization and navigation, graphical models and data fusion. 
\end{IEEEbiography}

\begin{IEEEbiography}[{\includegraphics[width=1in,height=1.25in,clip,keepaspectratio]{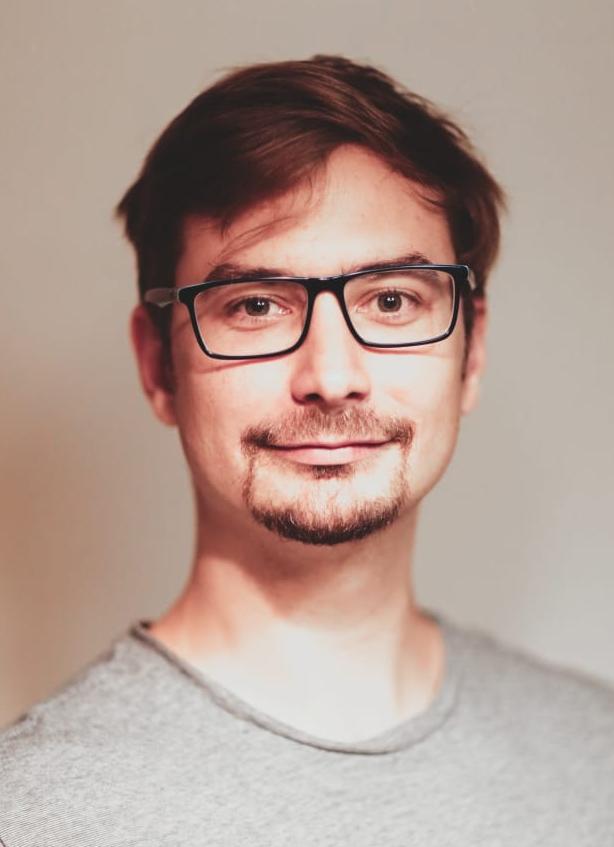}}]{Erik~Leitinger} (Member, IEEE)received his MSc and PhD degrees (with highest honors) in electrical engineering from Graz University of Technology, Austria in 2012 and 2016, respectively. He was postdoctoral researcher at the department of Electrical and Information Technology at Lund University from 2016 to 2018. He is currently a University Assistant at Graz University of Technology. Dr. Leitinger served as co-chair of the special session "Synergistic Radar Signal Processing and Tracking" at the IEEE Radar Conference in 2021. He is co-organizer of the special issue "Graph-Based Localization and Tracking" in the Journal of Advances in Information Fusion (JAIF). Dr.\ Leitinger received an Award of Excellence from the Federal Ministry of Science, Research and Economy (BMWFW) for his PhD Thesis. He is an Erwin Schr\"odinger Fellow. His research interests include inference on graphs, localization and navigation, machine learning, multiagent systems, stochastic modeling and estimation of radio channels, and estimation/detection theory. 
\end{IEEEbiography}

%% file: JAIF_2023_V7.bbl
\begin{thebibliography}{10}
\providecommand{\url}[1]{#1}
\csname url@samestyle\endcsname
\providecommand{\newblock}{\relax}
\providecommand{\bibinfo}[2]{#2}
\providecommand{\BIBentrySTDinterwordspacing}{\spaceskip=0pt\relax}
\providecommand{\BIBentryALTinterwordstretchfactor}{4}
\providecommand{\BIBentryALTinterwordspacing}{\spaceskip=\fontdimen2\font plus
\BIBentryALTinterwordstretchfactor\fontdimen3\font minus
  \fontdimen4\font\relax}
\providecommand{\BIBforeignlanguage}[2]{{%
\expandafter\ifx\csname l@#1\endcsname\relax
\typeout{** WARNING: IEEEtran.bst: No hyphenation pattern has been}%
\typeout{** loaded for the language `#1'. Using the pattern for}%
\typeout{** the default language instead.}%
\else
\language=\csname l@#1\endcsname
\fi
#2}}
\providecommand{\BIBdecl}{\relax}
\BIBdecl

\bibitem{LeitingerJSAC2015}
E.~Leitinger, P.~Meissner, C.~Rudisser, G.~Dumphart, and K.~Witrisal,
  ``Evaluation of position-related information in multipath components for
  indoor positioning,'' \emph{{IEEE} J. Sel. Areas Commun.}, vol.~33, no.~11,
  pp. 2313--2328, Nov. 2015.

\bibitem{WitrisalSPM2016}
K.~Witrisal, P.~Meissner, E.~Leitinger, Y.~Shen, C.~Gustafson, F.~Tufvesson,
  K.~Haneda, D.~Dardari, A.~F. Molisch, A.~Conti, and M.~Z. Win,
  ``High-accuracy localization for assisted living: {5G} systems will turn
  multipath channels from foe to friend,'' \emph{{IEEE} Signal Process. Mag.},
  vol.~33, no.~2, pp. 59--70, Mar. 2016.

\bibitem{LeitMeyHlaWitTufWin:TWC2019}
E.~{Leitinger}, F.~{Meyer}, F.~{Hlawatsch}, K.~{Witrisal}, F.~{Tufvesson}, and
  M.~Z. {Win}, ``A belief propagation algorithm for multipath-based {SLAM},''
  \emph{{IEEE} Trans. Wireless Commun.}, vol.~18, no.~12, pp. 5613--5629, Dec.
  2019.

\bibitem{MenMeyBauWin:JSTSP2019}
R.~{Mendrzik}, F.~{Meyer}, G.~{Bauch}, and M.~Z. {Win}, ``Enabling situational
  awareness in millimeter wave massive {MIMO} systems,'' \emph{{IEEE} J. Sel.
  Topics Signal Process.}, vol.~13, no.~5, pp. 1196--1211, Sep. 2019.

\bibitem{GentnerTWC2016}
C.~Gentner, W.~Jost, T.and~Wang, S.~Zhang, A.~Dammann, and U.~C. Fiebig,
  ``Multipath assisted positioning with simultaneous localization and
  mapping,'' \emph{{IEEE} Trans. Wireless Commun.}, vol.~15, no.~9, pp.
  6104--6117, Sept. 2016.

\bibitem{KulmerPIMRC2018}
J.~{Kulmer}, F.~{Wen}, N.~{Garcia}, H.~{Wymeersch}, and K.~{Witrisal}, ``Impact
  of rough surface scattering on stochastic multipath component models,'' in
  \emph{Proc. IEEE PIMRC 2018}, Bologna, Italy, Dec. 2018, pp. 1410--1416.

\bibitem{WenKulWitWym:TWC2021}
F.~Wen, J.~Kulmer, K.~Witrisal, and H.~Wymeersch, ``5{G} positioning and
  mapping with diffuse multipath,'' \emph{{IEEE} Trans. Wireless Commun.},
  vol.~20, no.~2, pp. 1164--1174, 2021.

\bibitem{PohZhaStaCaiDamHoe:IEEEAcess2022}
R.~P\"ohlmann, S.~Zhang, E.~Staudinger, S.~Caizzone, A.~Dammann, and P.~A.
  Hoeher, ``Bayesian in-situ calibration of multiport antennas for {DoA}
  estimation: Theory and measurements,'' \emph{IEEE Access}, vol.~10, pp.
  37\,967--37\,983, 2022.

\bibitem{DurrantWhyte2006}
H.~Durrant-Whyte and T.~Bailey, ``{Simultaneous localization and mapping: Part
  I},'' \emph{IEEE Robot. Autom. Mag.}, vol.~13, no.~2, pp. 99--110, June 2006.

\bibitem{Dissanayake2001}
M.~Dissanayake, P.~Newman, S.~Clark, H.~Durrant-Whyte, and M.~Csorba, ``A
  solution to the simultaneous localization and map building ({SLAM})
  problem,'' \emph{IEEE Trans. Robot. Autom.}, vol.~17, no.~3, pp. 229--241,
  June 2001.

\bibitem{LeiGreWit:ICC2019}
E.~{Leitinger}, S.~{Grebien}, and K.~{Witrisal}, ``Multipath-based {SLAM}
  exploiting {AoA} and amplitude information,'' in \emph{Proc. IEEE ICCW-19},
  Shanghai, China, May 2019, pp. 1--7.

\bibitem{KimGraGaoBatKimWym:TWC2020}
H.~{Kim}, K.~{Granstr{\"o}m}, L.~{Gao}, G.~{Battistelli}, S.~{Kim}, and
  H.~{Wymeersch}, ``{5G} {mmWave} cooperative positioning and mapping using
  multi-model {PHD} filter and map fusion,'' \emph{{IEEE} Trans. Wireless
  Commun.}, vol.~19, no.~6, pp. 3782--3795, Mar. 2020.

\bibitem{KimGranSveKimWym:TVT2022}
H.~Kim, K.~Granstrom, L.~Svensson, S.~Kim, and H.~Wymeersch, ``{PMBM-based
  SLAM} filters in {5G} {mmWave} vehicular networks,'' \emph{{IEEE} Trans. Veh.
  Technol.}, pp. 1--1, May 2022.

\bibitem{LeiMey:Asilomar2020_DataFusion}
E.~Leitinger and F.~Meyer, ``Data fusion for multipath-based {SLAM},'' in
  \emph{Proc. Asilomar-20}, Pacifc Grove, CA, USA, Oct. 2020, pp. 934--939.

\bibitem{LeiTeaZhaLiaMey:Fusion2022}
E.~{Leitinger}, B.~{Teague}, W.~{Zhang}, M.~{Liang}, and F.~{Meyer}, ``Data
  fusion for radio frequency {SLAM} with robust sampling,'' in \emph{Proc.
  Fusion-22}, Link\"oping, Sweden, Jul. 2022, pp. 1--6.

\bibitem{LeiVenTeaMey:TSP2023}
E.~{Leitinger}, A.~{Venus}, B.~{Teague}, and F.~{Meyer}, ``Data fusion for
  multipath-based {SLAM}: {Combining} information from multiple propagation
  paths,'' pp. 1--17, 2023.

\bibitem{RichterPhD2005}
A.~Richter, ``{Estimation of Radio Channel Parameters: Models and
  Algorithms},'' Ph.D. dissertation, Ilmenau University of Technology, 2005.

\bibitem{ShutWanJos:CSTA2013}
D.~Shutin, W.~Wang, and T.~Jost, ``Incremental sparse {B}ayesian learning for
  parameter estimation of superimposed signals,'' in \emph{Proc. SAMPTA-2013},
  no.~1, Sept. 2013, pp. 6--9.

\bibitem{HanBadFleRao:SAM2014}
T.~L. {Hansen}, M.~A. {Badiu}, B.~H. {Fleury}, and B.~D. {Rao}, ``A sparse
  {Bayesian} learning algorithm with dictionary parameter estimation,'' in
  \emph{Proc. IEEE SAM 2014}, Jun. 2014, pp. 385--388.

\bibitem{BadHanFle:TSP2017}
M.~A. Badiu, T.~L. Hansen, and B.~H. Fleury, ``Variational {Bayesian} inference
  of line spectra,'' \emph{{IEEE} Trans. Signal Process.}, vol.~65, no.~9, pp.
  2247--2261, May 2017.

\bibitem{HanFleuRao:TSP2018}
T.~L. Hansen, B.~H. Fleury, and B.~D. Rao, ``Superfast line spectral
  estimation,'' \emph{{IEEE} Trans. Signal Process.}, vol.~PP, no.~99, pp.
  1--1, Feb. 2018.

\bibitem{LiLeiVenTuf:TWC2022}
X.~Li, E.~Leitinger, A.~Venus, and F.~Tufvesson, ``Sequential detection and
  estimation of multipath channel parameters using belief propagation,''
  \emph{{IEEE} Trans. Wireless Commun.}, vol.~21, no.~10, pp. 8385--8402, Apr.
  2022.

\bibitem{GreLeiFleWit:Arxiv2023}
\BIBentryALTinterwordspacing
S.~Grebien, E.~Leitinger, B.~H. Fleury, and K.~Witrisal, ``Super-resolution
  estimation of {UWB} channels including the diffuse component --- {An} {SBL}
  inspired approach,'' \emph{ArXiv e-prints}, vol. abs/2308.01702, 2023.
  [Online]. Available: \url{https://arxiv.org/abs/2308.01702}
\BIBentrySTDinterwordspacing

\bibitem{MeyWilJ21}
F.~Meyer and J.~L. Williams, ``Scalable detection and tracking of geometric
  extended objects,'' \emph{{IEEE} Trans. Signal Process.}, vol.~69, pp.
  6283--6298, Oct. 2021.

\bibitem{WilliamsLauTAE2014}
J.~Williams and R.~Lau, ``Approximate evaluation of marginal association
  probabilities with belief propagation,'' \emph{IEEE Trans. Aerosp. Electron.
  Syst.}, vol.~50, no.~4, pp. 2942--2959, Oct. 2014.

\bibitem{MeyerProc2018}
F.~Meyer, T.~Kropfreiter, J.~L. Williams, R.~Lau, F.~Hlawatsch, P.~Braca, and
  M.~Z. Win, ``Message passing algorithms for scalable multitarget tracking,''
  \emph{Proc. {IEEE}}, vol. 106, no.~2, pp. 221--259, Feb. 2018.

\bibitem{Koc:TAES2008_EOT}
J.~W. {Koch}, ``Bayesian approach to extended object and cluster tracking using
  random matrices,'' \emph{{IEEE} Trans. Aerosp. Electron. Syst.}, vol.~44,
  no.~3, pp. 1042--1059, Jul. 2008.

\bibitem{GranstroemFatemiSvenssonTAES2020_PMBMFilterExtendedTargets}
K.~{Granstr{\"o}m}, M.~{Fatemi}, and L.~{Svensson}, ``Poisson multi-{B}ernoulli
  mixture conjugate prior for multiple extended target filtering,''
  \emph{{IEEE} Trans. Aerosp. Electron. Syst.}, vol.~56, no.~1, pp. 208--225,
  June 2020.

\bibitem{MeyerICASSP2020}
F.~{Meyer} and J.~L. {Williams}, ``Scalable detection and tracking of extended
  objects,'' in \emph{Proc. ICASSP 2020}, Barcelona, Spain, May 2020, pp.
  8916--8920.

\bibitem{GranstroemTAES2012}
K.~{Granstr{\"o}m}, C.~{Lundquist}, and O.~{Orguner}, ``Extended target
  tracking using a {Gaussian}-mixture {PHD} filter,'' \emph{{IEEE} Trans.
  Aerosp. Electron. Syst.}, vol.~48, no.~4, pp. 3268--3286, Oct. 2012.

\bibitem{GraBau:JAIF2017}
K.~Granstr{\"{o}}m and M.~Baum, ``Extended object tracking: {Introduction},
  overview and applications,'' \emph{J. Adv. Inf. Fusion}, vol.~12, Dec. 2017.

\bibitem{KolFri:PGM2009}
D.~Koller and N.~Friedmann, \emph{Probabilistic Graphical Models: {Principles}
  and Techniques}.\hskip 1em plus 0.5em minus 0.4em\relax Cambridge, MA, USA:
  MIT Press, 2009.

\bibitem{WieVenWilLei:Fusion2023}
L.~Wielandner, A.~Venus, T.~Wilding, and E.~Leitinger, ``Multipath-based {SLAM}
  with multiple-measurement data association,'' in \emph{Proc. Fusion-23},
  Charleston, USA, Jul. 2023, pp. 1--8.

\bibitem{Kay1998}
S.~M. Kay, \emph{Fundamentals of Statistical Signal Processing: {D}etection
  Theory}.\hskip 1em plus 0.5em minus 0.4em\relax Upper Saddle River, NJ, USA:
  Prentice Hall, 1998.

\bibitem{BarShalom11}
Y.~Bar-Shalom, P.~K. Willett, and X.~Tian, \emph{{Tracking and Data Fusion: A
  Handbook of Algorithms}}.\hskip 1em plus 0.5em minus 0.4em\relax Storrs, CT,
  USA: Yaakov Bar-Shalom, 2011.

\bibitem{WilLeiMueWit:PIMRC2020}
T.~Wilding, E.~Leitinger, U.~M\"uhlmann, and K.~Witrisal, ``Modeling human body
  influence on {UWB} channels,'' in \emph{Proc. IEEE PIMRC-20}, London, United
  Kingdom, Oct. 2020.

\bibitem{WilLeiMueWit:EuCAP2021}
T.~Wilding, E.~Leitinger, U.~Muehlmann, and K.~Witrisal, ``Statistical modeling
  of the human body as an extended antenna,'' in \emph{Proc. EuCAP-2021},
  D\"usseldorf, Germany, Apr. 2021, pp. 1--5.

\bibitem{SchuFle:EUCAP2010}
F.~M. Schubert, B.~H. Fleury, P.~Robertson, R.~Prieto-Cerdeirai, A.~Steingass,
  and A.~Lehner, ``Modeling of multipath propagation components caused by trees
  and forests,'' in \emph{Proceedings of the Fourth European Conference on
  Antennas and Propagation}, 2010, pp. 1--5.

\bibitem{SchuFle:EUCAP2012}
F.~M. Schubert, B.~H. Fleury, R.~Prieto-Cerdeira, A.~Steingass, and A.~Lehner,
  ``A rural channel model for satellite navigation applications,'' in
  \emph{2012 6th European Conference on Antennas and Propagation (EUCAP)},
  2012, pp. 2431--2435.

\bibitem{BarShalom2002EstimationTracking}
Y.~Bar-Shalom, T.~Kirubarajan, and X.-R. Li, \emph{Estimation with Applications
  to Tracking and Navigation}.\hskip 1em plus 0.5em minus 0.4em\relax New York,
  NY, USA: John Wiley \& Sons, Inc., 2002.

\bibitem{MerUlmKoc:TAES2016}
M.~Mertens, M.~Ulmke, and W.~Koch, ``Ground target tracking with {RCS}
  estimation based on signal strength measurements,'' \emph{{IEEE} Trans.
  Aerosp. Electron. Syst.}, vol.~52, no.~1, pp. 205--220, Feb. 2016.

\bibitem{WitrisalWCL2016}
K.~Witrisal, E.~Leitinger, S.~Hinteregger, and P.~Meissner, ``Bandwidth scaling
  and diversity gain for ranging and positioning in dense multipath channels,''
  vol.~5, no.~4, pp. 396--399, May 2016.

\bibitem{WilGreLeiMueWit:ACSSC2018}
T.~Wilding, S.~Grebien, E.~Leitinger, U.~M\"uhlmann, and K.~Witrisal,
  ``Single-anchor, multipath-assisted indoor positioning with aliased antenna
  arrays,'' in \emph{Proc. Asilomar-18}, Pacifc Grove, CA, USA, Oct. 2018, pp.
  525--531.

\bibitem{LepRabLeG:FUSION2013}
A.~Lepoutre, O.~Rabaste, and F.~Le~Gland, ``Exploiting amplitude spatial
  coherence for multi-target particle filter in track-before-detect,'' in
  \emph{Proc. FUSION 2013}, Oct. 2013, pp. 319--326.

\bibitem{LepRabLeG:TAES2016}
------, ``Multitarget likelihood computation for track-before-detect
  applications with amplitude fluctuations of type {Swerling} 0, 1, and 3,''
  vol.~52, no.~3, pp. 1089--1107, June 2016.

\bibitem{VenLeiTerWit:TWC2023}
A.~Venus, E.~Leitinger, S.~Tertinek, and K.~Witrisal, ``A graph-based algorithm
  for robust sequential localization exploiting multipath for
  obstructed-{LOS}-bias mitigation,'' \emph{{IEEE} Trans. Wireless Commun.},
  pp. 1--1, June 2023.

\bibitem{LerroACC1990}
D.~Lerro and Y.~Bar-Shalom, ``Automated tracking with target amplitude
  information,'' in \emph{1990 American Control Conference}, May 1990, pp.
  2875--2880.

\bibitem{Poo:B94}
H.~V. Poor, \emph{An Introduction to Signal Detection and Estimation},
  2nd~ed.\hskip 1em plus 0.5em minus 0.4em\relax New York: Springer-Verlag,
  1994.

\bibitem{KscFreLoe:TIT2001}
F.~Kschischang, B.~Frey, and H.-A. Loeliger, ``Factor graphs and the
  sum-product algorithm,'' \emph{{IEEE} Trans. Inf. Theory}, vol.~47, no.~2,
  pp. 498--519, Feb. 2001.

\bibitem{MeyHliHla:TSPIN2016}
F.~Meyer, O.~Hlinka, H.~Wymeersch, E.~Riegler, and F.~Hlawatsch, ``Distributed
  localization and tracking of mobile networks including noncooperative
  objects,'' vol.~2, no.~1, pp. 57--71, Mar. 2016.

\bibitem{MeyerTSP2017}
F.~Meyer, P.~Braca, P.~Willett, and F.~Hlawatsch, ``A scalable algorithm for
  tracking an unknown number of targets using multiple sensors,'' \emph{{IEEE}
  Trans. Signal Process.}, vol.~65, no.~13, pp. 3478--3493, July 2017.

\bibitem{Loe:SMP2004_FG}
H.-A. Loeliger, ``An introduction to factor graphs,'' \emph{{IEEE} Signal
  Process. Mag.}, vol.~21, no.~1, pp. 28--41, Feb. 2004.

\bibitem{Schuhmacher2008}
D.~Schuhmacher, B.-T. Vo, and B.-N. Vo, ``{A consistent metric for performance
  evaluation of multi-object filters},'' \emph{{IEEE} Trans. Signal Process.},
  vol.~56, no.~8, pp. 3447--3457, Aug. 2008.

\bibitem{SalVal:JSAC1987}
A.~Saleh and R.~Valenzuela, ``A statistical model for indoor multipath
  propagation,'' \emph{{IEEE} J. Sel. Areas Commun.}, vol.~5, no.~2, pp.
  128--137, Feb. 1987.

\bibitem{PedersenJTAP2018}
T.~Pedersen, ``Modeling of path arrival rate for in-room radio channels with
  directive antennas,'' \emph{{IEEE} Trans. Antennas Propag.}, vol.~66, no.~9,
  pp. 4791--4805, 2018.

\end{thebibliography}
